\documentclass[preprint,12pt,number]{elsarticle}
\usepackage[latin1]{inputenc}
\usepackage{amsmath}
\usepackage{amsfonts}
\usepackage{amssymb}
\usepackage{graphicx} 
\usepackage{epsfig}
\usepackage[all]{xy}
\usepackage{hyperref} 
\usepackage{epstopdf}
\usepackage{psfrag}
\usepackage{cite}       
\usepackage{bm}         
\usepackage{verbatim}   
\def\href#1#2{#2}	
\newcommand{\psib}{\ensuremath{\overline{\psi}}}
\newcommand{\phib}{\ensuremath{\overline{\phi}}}

\newcommand{\lambdab}{\ensuremath{\overline{\lambda}}}
\newcommand{\omegab}{\ensuremath{\overline{\omega}}}
\newcommand{\Wp}{\ensuremath{W^{\prime}}}
\newcommand{\Wpp}{\ensuremath{W^{\prime\prime}}}

\newcommand{\Nc}{\ensuremath{{\cal N}}}
\newcommand{\cN}{\ensuremath{{\cal N}}}
\newcommand{\beq}{\begin{equation}}
\newcommand{\eeq}{\end{equation}}
\newcommand{\KD}{Dirac-K\"{a}hler }
\newcommand{\cF}{\ensuremath{{\cal F}}}

\newcommand{\cA}{{\cal A}}
\newcommand{\cAb}{{\overline{\cal A}}}
\newcommand{\cFb}{{\overline{\cal F}}}
\newcommand{\cD}{{\cal D}}
\newcommand{\cDb}{{\overline{\cal D}}}
\newcommand{\cQ}{{\cal Q}}
\newcommand{\cU}{{\cal U}}
\newcommand{\cUb}{{\overline{\cal U}}} 
\newcommand{\Tr}{{\rm Tr\;}}
\newcommand{\bx}{{\bf x}}
\newcommand{\bmu}{{\boldsymbol \mu}}
\newcommand{\bnu}{{\boldsymbol \nu}}

\newcommand{\bzero}{{\boldsymbol 0}}
\newcommand\bal{ \begin{align}}
\newcommand\eal{\end{align} }

\newcommand\eqn[1]{\label{eq:#1}} 
\newcommand\Eq[1]{eq.~\eqref{eq:#1}} 
\newcommand\eq[1]{eq.~\eqref{eq:#1}} 
\newcommand\half{{\textstyle{\frac{1}{2}}}} 
\newcommand\fourth{{\textstyle{\frac{1}{4}}}} 
\newcommand\eight{{\textstyle{\frac{1}{8}}}} 
\newcommand\fourthi{{\textstyle{\frac{i}{4}}}}

\newcommand\vev[1]{\langle #1 \rangle}

\newcommand\bfz{\mathbf{z}}
\newcommand\bfZ{\mathbf{Z}}

\newcommand\bfmu{\boldsymbol{\mu}}

\newcommand\bfsig{\boldsymbol{\sigma}}

\newcommand{\sla}[1]{\kern .25em\raise.18ex\hbox{$/$}\kern-.80em #1}
\newcommand{\Dslash}{\hbox{\kern .25em\raise.18ex\hbox{$/$}\kern-.80em  $D$}}
\newcommand{\dslash}{\hbox{\kern .25em\raise.18ex\hbox{$/$}\kern-.55em  $\partial$}}

\newcommand{\CD}{{\cal D}}
\newcommand{\CE}{{\cal E}}
\newcommand{\CF}{{\cal F}}
\newcommand{\CO}{{\cal O}}
\newcommand{\CN}{{\cal N}}

\newcommand{\CQ}{{\cal Q}}

\newcommand{\CL}{{\cal L}}
\newcommand{\bfn}{{\bf n}}
\newcommand{\bfr}{{\bf r}}
\newcommand{\bfR}{{\bf R}}

\newcommand{\bfe}{{\bf  e}}
\newcommand{\xh}{\mathbf{\hat{x}}}
\newcommand{\yh}{\mathbf{\hat{y}}}

\newcommand{\mybar}[1]%
        {\kern 0.6pt\overline{\kern -0.6pt#1\kern -0.6pt}\kern 0.6pt}
\journal{Physics Reports}
\begin{document}

\begin{frontmatter}
\title{Exact lattice supersymmetry\tnoteref{t1}} 
\tnotetext[t1]{Preprint: INT-PUB-09-014, SLAC-PUB-13567, SU-4252-887}

\author{Simon Catterall}
\ead{smc@physics.syr.edu} 
\address{Department of Physics, Syracuse University, Syracuse, NY13244 }

\author{David B. Kaplan}
\ead{dbkaplan@phys.washington.edu}
\address{ Institute  for Nuclear Theory, University of Washington, Seattle, WA 
    98195-1550}

\author{Mithat \"Unsal}
\ead{unsal@slac.stanford.edu}
\address{ SLAC and Physics Department, Stanford University, Stanford, CA 94305 }

\begin{abstract}
We provide an  introduction to recent 
lattice formulations of supersymmetric theories which are invariant under one or more
real supersymmetries
at nonzero lattice spacing. These include the
especially interesting case of ${\cal N}=4$ SYM in 
four dimensions. 
We discuss approaches based both on twisted supersymmetry and 
orbifold-deconstruction techniques
and show their equivalence in the case of gauge theories. The presence of an 
exact supersymmetry
reduces and in some cases eliminates the need for fine tuning to
achieve a continuum limit invariant under the full supersymmetry of
the target theory. We discuss 
open problems. 
\end{abstract}

\end{frontmatter}

\tableofcontents 
\newpage

\section{Introduction}
\label{intro}

Whether or not supersymmetry
is discovered to be a symmetry of nature, strongly coupled
supersymmetric theories will always be a source of fascination  \citep{Wess:1992cp,Nilles:1983ge,Martin:1997ns,Terning:2003th}.  In
these theories one can find explicit examples of many of the basic
mechanisms and objects put forward in the early days of gauge
theories: confinement, chiral symmetry breaking, magnetic monopoles
and dyons, conformal field theories, etc.  Especially intriguing are
the connections between 
theories with sixteen supercharges and both supergravity and string
theory \citep{Banks:1996vh,Maldacena:1997re,Itzhaki:1998dd}. 

Until recently, a
nonperturbative lattice formulation for all but a few of these theories  remained
elusive despite many efforts over the years.
The problem has been that
discretization tends to completely break the supersymmetry, so that no
characteristics of the continuum theory are present without excessive
fine-tuning. In the language of the renormalization group, the
lattice theory typically flows away from any 
supersymmetric fixed point as the
cut-off is removed.  Past attempts to fix this by imposing an exact supersymmetric  subalgebra 
on the lattice action typically resulted in a loss of Poincar\'e invariance \citep{Banks:1982ut}.

In the past few years,
however, there have been significant advances in our understanding,
which have led to the construction of a number of interesting
supersymmetric lattice theories, including $\CN=4$ supersymmetric Yang-Mills (SYM)
in four dimensions, in which these fine tuning problems are under
much better control.

The new development has been the construction of lattice actions 
which possess a subset of the supersymmetries of the continuum 
theory and have a Poincar\'{e} invariant continuum limit. 
The presence of the exact supersymmetry on
the lattice provides a way to obtain the continuum limit with no fine tuning,  
or fine tuning much less than conventional lattice constructions
(in which there is no exact supersymmetry  at the cut-off scale.) 
In this review, we  introduce some of the 
ideas which lead to the construction of these supersymmetric lattice
theories. 

Two main approaches have been proposed to formulate 
such supersymmetric lattice theories, which are now understood to be closely related. 
One is based on the idea of `twisting' and Dirac-K\"{a}hler fermions 
\citep{Rabin:1981qj, Becher:1982ud}. The twisting procedure  
is based on a decomposition  
of  Lorentz  spinor supercharges into a sum of integer spin ($p$-form)
tensors under a 
diagonal subgroup of the Lorentz group and some large 
global symmetry of the theory, usually referred to as $R$-symmetry.
The twisted formulation of supersymmetry goes back to Witten \citep{Witten:1988ze}
in his seminal construction of topological field theories,
but actually had been anticipated in earlier lattice work using
\KD fields \citep{Elitzur:1982vh,Sakai:1983dg,Kostelecky:1983qu,Scott:1983ha,Aratyn:1984bc}. The precise connection between
 \KD fermions and topological twisting was found by
Kawamoto and collaborators \citep{Kawamoto:1999zn,Kato_bf,D'Adda_super}. 
The key observation is that 
the zero-form supercharge that arises after twisting
is a scalar which squares to zero, and 
constitutes a closed subalgebra of the full twisted superalgebra.  
It is this scalar supersymmetry 
that can be made manifest in the lattice action even at finite lattice spacing
\citep{Catterall:2005df,Catterall:2005eh,Catterall_topo,Catterall_wz1,
Giedt:2007hz,Giedt_wz,Sugino_sym1,Sugino_2d,D'Adda_2d}.

The second approach derives a supersymmetric
lattice theory by orbifolding a certain supersymmetric
matrix model. This `mother' matrix theory is obtained by dimensional
reduction of a SYM theory with a very
large gauge symmetry. The projection is chosen so as to induce a lattice
structure and to preserve one or more supersymmetries of the mother
theory. The resulting theory is also gauge invariant and preserves 
a discrete subgroup of the continuum Lorentz and global 
symmetries 
\citep{Kaplan:1983sk,ArkaniHamed:2001ie,Kaplan:2002wv,
Cohen:2003xe, Cohen:2003qw,Kaplan:2005ta,Endres:2006ic,Damgaard_orb,Giedt_rev1}.
The theories obtained in this fashion have a degenerate ground state 
(called a moduli space), where the distance from the origin of the moduli space
has an interpretation as the inverse lattice spacing, as first implemented in 
``deconstruction" \citep{ArkaniHamed:2001ca, ArkaniHamed:2001ie}.  
The continuum limit is thus 
defined as a particular scaling limit out to infinity in the moduli space and 
the result is a supersymmetric gauge theory where full super-Poincar\'{e}
symmetry 
is recovered\footnote{To be more precise the continuum limit of these
constructions is actually invariant under a twisted version of
the super-Poincar\'{e} group. Whether one can `untwist' the theory to obtain
a target theory with the usual super-Poincar\'{e} invariance is related
to the amount of residual fine tuning needed to obtain full
supersymmetry.}.
 
Even though these two approaches seem different at first glance, they do 
generate similar actions
and lattices. The reason behind  this is that the Dirac-K\"{a}hler 
decomposition of the fermions is indeed encoded into the charges of the 
fermions encountered in the orbifold projection\citep{Unsal:2006qp,Damgaard_orb,Damgaard:2007xi} -- 
the number of non-zero
components of the r-charge vector characterizing the
orbifold lattice field matching the degree of the $p$-form component
of the corresponding twisted \KD field.
The common 
thread of both approaches is the exact preservation of (nilpotent) scalar 
supercharges on the lattice, which automatically dictates  the 
distribution of the bosonic degrees of freedom in the lattice given the 
Dirac-K\"ahler construction. Indeed we will show that one can obtain
the supersymmetric orbifold lattices by a direct discretization of an
appropriately chosen twist of the target SYM theory
\citep{Catterall:2007kn,Damgaard:2008pa}. 

It is also useful to keep in mind the limitations of the   twisting and orbifolding formalisms.  
These  techniques   only apply to a  sub-class of supersymmetric gauge theories.
The ability to construct a manifestly supersymmetric lattice in this formalism requires the R-symmetry group to contain  $SO(d)$ -- the d-dimensional 
(Euclidean) Lorentz symmetry group -- as a subgroup. If so, one can apply the idea of twisting 
as shown in Fig.\ref{fig:sym}. Clearly this constraint excludes the formulation of some other interesting theories, such as ${\cal N}=2$ SYM (the Seiberg-Witten theory) or generic ${\cal N}=1$  supersymmetric QCD theories, or 
theories of more phenomenological interest such as the MSSM. Lattice constructions of these interesting theories  are currently  open problems. 
It is also fair to say that much theoretical work remains to be done
to understand how much fine tuning is required in order that these lattice
theories inherit the full supersymmetry of the target theory
in the continuum limit -- perturbative calculations would be very
useful in this regard as we will discuss later when describing the
$\cN=4$ construction in detail.

The first part of the review will motivate the study of lattice supersymmetry and
give an overview of some of the basic ideas:  
why it is difficult to build supersymmetric lattice
theories and why naive discretizations of continuum supersymmetric
theories lead to fine tuning problems. We argue that supersymmetry
should arise as an accidental symmetry on taking the
continuum limit of some suitable lattice model, and offer as an example  a supersymmetric theory without scalars:  the interesting $\CN=1$ super Yang-Mills theory in $d=4$ dimensions.  We then discuss why in supersymmetric theories with scalars, only an exact lattice 
supersymmetry can keep scalars massless without fine-tuning and allow for the full supersymmetry algebra to emerge
as an accidental symmetry at long distances,.
Tautology is avoided since the lattice model
need only preserve a subset of
the continuum supersymmetry to  avoid or at least ameliorate fine-tuning.  This naturally leads into a discussion of twisted supersymmetry and \KD fermions.

 Before progressing to  more complicated theories, we next consider
supersymmetric quantum mechanics and the two dimensional
Wess-Zumino and sigma models. This will allow us to illustrate the nature of the fine tuning problems
that are encountered and how realization of an exact lattice
supersymmetry enables the full supersymmetry to emerge in the continuum limit.  The connection between twisting, Nicolai maps and
topological field theories is then discussed in the context of these examples.  
We then turn to gauge theories, first presenting the twisted supersymmetry approach to $(2,2)$ SYM in two dimensions.  We then discuss the powerful  orbifold approach to lattice SYM, 
and show in detail how to obtain a
gauge invariant lattice model  invariant under one
real supersymmetry for this same $(2,2)$ SYM theory. It is shown to be
 precisely the same as
the twisted construction derived earlier. The possible
supersymmetric orbifold lattices are then classified and seen to 
include the interesting case of
$\cN=4$ SYM. We summarize the content of this lattice model and show how it can
also be generated by discretization of the Marcus twist
of $\cN=4$ SYM theory confirming, once more, the complete equivalence of the
two approaches.

We emphasize that this resultant lattice action
for $\cN=4$ SYM currently offers a promising starting point for numerical 
simulations. Indeed,
dimensional reductions of this theory are already being studied on the lattice in the context of their
conjectured equivalence to string and supergravity theories 
\citep{Catterall:2007hk,Catterall:2008yz,Catterall:2007me,
Anagnostopoulos:2007fw,Nishimura:2008ta,Kawahara:2007ib,Hanada:2008ez,
Hanada:2008gy,Hanada:2007ti,Nishimura:2008zz,Ishiki:2008te}. It is perhaps the prospect of eventually using our lattices to learn more about quantum gravity that we consider the most exciting.

\section{Supersymmetry}
\label{sec:2}

\subsection{The supersymmetry algebra}
\label{sec:2a}

Poincar\'e symmetry consists of spacetime translations, generated by $P_\mu$,
and Lorentz transformations, generated by $\Sigma_{\mu\nu}= -
\Sigma_{\nu\mu}$. The algebra has the qualitative structure
\beq
[P,P] = 0\ ,\quad [P,\Sigma]\sim P\ ,\quad [\Sigma,\Sigma]\sim \Sigma\
,
\eeq
where the meaning of the three terms are (i) translations commute with
each other; (ii) translations transform under the Lorentz group as a
4-vector; (iii) Lorentz transformations themselves transform as an antisymmetric
tensor.

Supersymmetry is the unique extension of the Poincar\'e algebra
consistent with the Coleman-Mandula theorem \citep{Coleman:1967ad}, 
where complex
spinorial generators $Q_\alpha$, $\bar Q_{\dot\alpha}$ are added with
the (anti-) commutation relations:\footnote{We have neglected
the possibility of central charge terms in this simplified
discussion}
\beq
\left\{ Q,Q\right\} = 0\ ,\quad[P,Q]=0\ ,\quad 
[Q,\Sigma] \sim Q\ ,\quad
\{ Q,\bar Q\} \sim P\ .
\eeq

These terms tell us (i) $Q$ is Grassmann; (ii) $Q$ commutes with
spacetime translations (and hence the Hamiltonian); (iii) $Q$ transforms under
Lorentz transformation as a 2-component Weyl spinor; (iv) two
successive supersymmetry transformations yields a translation.  From
(i) and (ii) it follows that there are pairs of fermion-boson states
which are degenerate (along with possible unpaired zero energy states), and from (iv) we see that in some sense a
supersymmetry charge $Q$ is a square root of the Hamiltonian (in the
same sense that the Dirac operator is a square root of the
Klein-Gordon operator).

\subsection{Counting supercharges}
\label{sec:2b}

The supersymmetry algebra is highly constrained, and in any given
number of dimensions there are typically only a few possibilities for
how many supercharges can exist.  These constraints arise
from the requirement that the theories not contain particles
with spin greater than one (or
two in the case of supergravity). This restriction on the maximal
spin stems, in turn, from the requirement that the theories
be renormalizable.

These different solutions are often
labeled $\CN=1$, $\CN=2$, etc.  What is confusing is that the number
of supercharges for $\CN=1$ supersymmetry, for example, is different
in different numbers of dimensions.  Instead, when discussing
supersymmetric theories in dimensions other than four, we will identify
a supersymmetric 
theory by the spacetime dimension $d$, and {\it the number of real
  supercharges}, $\CQ$. Thus
$\CN=1$ supersymmetry in $d=4$ has a complex pair $Q$, $\bar Q$ which
are each two-component Weyl spinors, giving $\CQ=4$.  Similarly,
$\CN=4$ supersymmetry in $d=4$ has $\CQ=16$.

\subsection{$R$ symmetries}
\label{sec:2c}
Supersymmetric theories typically have global chiral symmetries --- generically called ``$R$-symmetries" which do not commute with the supercharges, meaning that the members of the supermultiplets transform as different multiplets under the $R$-symmetry. These symmetries turn out to play a crucial role in the implementation of lattice supersymmetry.
  The bosonic and fermionic fields 
furnish a representation of the 
$R$-symmetry, as well as Euclidean Lorentz symmetry  $SO(d)_E$, 
and the same is true for 
the supercharges. For example, 
a list of the SYM theories and their Lorentz and 
$R$-symmetries (at the classical level) are given in Table~\ref{tab:tab1}. 
These symmetries are most easily 
determined by exploiting the fact that  $\CQ=4,8,16$ SYM theories are the minimal ($\CN=1$) gauge theory in $d=4,6,10$ dimensions respectively --- in those dimensions, the theory consists of only of a gauge field and a gaugino.  The  $(\CQ=4,d=4)$  SYM theory has a $U(1)$ $R$-symmetry,  the $(\CQ=8, d=6)$ possesses an  internal  $SU(2)$  $R$-symmetry, while the $(\CQ=16, d=10)$   $\CN=1$ theory has no $R$-symmetry 
  When these theories are dimensionally reduced from $d'=4,6,10$-dimensions down to $d$ dimensions   one preserves all of the supercharges while  enlarging  the $R$-symmetry by Euclidean ``Lorentz" generators acting in the reduced  dimensions. For example, the $\CN=1$ theory in $d'=10$ dimensions dimensionally reduced to $d$ dimensions has an $SO(d)_E$ Lorentz  symmetry and    $SO(10-d)$ $R$-symmetry, as shown in the last column of  
  Table~\ref{tab:tab1}.  There are also cases where these classical symmetries may reduce  in a 
  quantum theory due to anomalies or enhance to a larger $R-$symmetry 
  at long distances.   
    For a discussion of the case 
    with $\CQ=16$ where the latter may take place, 
     see \citep{Seiberg:1997ax}.

\begin{table}[t]
\centerline{
\begin{tabular}
{|c|c||c|c|c|} \hline
Theory   & Lorentz & $\CQ=4$  &  $\CQ=8$  &  $\CQ=16$ \\ \hline\hline
$d=2$  & $SO(2)$ &$SO(2) \times U(1)$  & $SO(4) \times SU(2)$ &  $SO(8)$ 
\\ \hline
$d=3$   &$SO(3)$ &$U(1)$  & $SO(3) \times SU(2)$    &  $SO(7)$     \\ \hline
$d=4$ &$SO(4)$  &$U(1)$  &   $SO(2) \times  SU(2)$  &   $SO(6)$   \\ \hline
\end{tabular} }
\caption{\sl  The $R$-symmetry groups of various  SYM 
theories.   
\label{tab:tab1}}
\end{table}

\subsection{Why study lattice supersymmetry?}
\label{sec:2d}

Supersymmetry is interesting in its own right. It is also potentially
interesting for phenomenology, as the protection it affords scalars
from additive renormalization of their masses could have something to
do with the mysterious Higgs boson of the Standard Model.  And it is
worth studying because with the extra symmetry, many interesting
results have been obtained for supersymmetric
Yang-Mills (SYM) theories, including explicit examples of many
mechanisms postulated in the early days of Yang-Mills theories,
including spontaneous chiral symmetry breaking, confinement, magnetic
monopole condensation, strong coupling - weak coupling duality,
massless composite fermions, conformal field theory, and more \citep{Intriligator:1995au,Kaplan:1997tu}.  In
addition, many fascinating connections have been made between between
SYM theories and string theory and quantum gravity \citep{Maldacena:1997re,Itzhaki:1998dd}, as well as with topology \citep{Witten:1993yc} .

Since there are so many interesting features of SYM theories,
especially at strong coupling, it would be very desirable to be able
to have a non-perturbative definition of these theories. This is important
both from a mathematical viewpoint and also as a basis for numerical simulations.

\section{Accidental supersymmetry and twisted supercharges}
\label{sec:3}

In principle, many supersymmetric theories could be studied on the lattice by choosing the right degrees of freedom, and then tuning the couplings to the critical values which yield the supersymmetric ``target theory"  in the infrared.  However, this brute-force approach is prohibitively difficult for any but possibly the simplest theories.  A more practical approach is to construct a lattice theory that respects as many of the symmetries of the target theory as possible, limiting the number of possible operators whose coefficients need to be fine-tuned.     One might think that the lattice action would have to possess {\it all} of the symmetries of the target theory, but in fact that is not necessary due to the emergence of ``accidental" symmetries.  An accidental symmetry is a symmetry that emerges in the infrared (IR)  limit of the lattice theory, even though it is not respected by the full lattice action. This typically occurs  when the exact symmetries of the action only allow irrelevant operators which could violate the accidental symmetry transformation --- such operators become unimportant in the IR, and the symmetry then emerges.  A prime example of an accidental symmetry in the  continuum is baryon number 
violation in a Grand Unified Theory (GUT) \citep{Georgi:1974sy}: $B$-violation is mediated by  gauge boson and scalar interactions at the GUT scale; but below the GUT scale the light degrees of freedom are such that the gauge symmetries of the standard model forbid baryon violating operators with dimension less than six, and as such, baryon number violation becomes ``irrelevant" in the IR.  That explains why  GUTs can be  consistent with the observed stringent lower bounds on the proton lifetime\footnote{Actually, only SUSY GUTs are consistent
with the limits on proton lifetime and gauge coupling unification. In SUSY
GUTs, there are dimension five baryon number violating operators unless
the theory is supplemented by an additional  R-parity symmetry.}. 

Accidental symmetry also explains why lattice QCD  is able to recover (Euclidean) Poincar\'e symmetry in the continuum  limit  without fine-tuning, even though the lattice action only respects a discrete hypercubic subgroup of Poincar\'e symmetry: given the field content of QCD and both the exact hypercubic and  gauge symmetries of the lattice action, the lowest dimension operators that can be added to the action which violate continuum Poincar\'e symmetry are dimension six, such as the discrete version of $\sum_\mu \bar\psi \gamma_\mu D_\mu^3\psi$.  These are irrelevant operators which become unimportant in the IR limit of the theory, and so the full Poincar\'e symmetry is recovered without fine-tuning.  As a counter-example, consider Wilson fermions with the bare mass term set to zero; in this case the lattice action appears to have an exact chiral symmetry in all the relevant and marginal operators, with chiral symmetry breaking first appearing in the Wilson term, a dimension five operator of the form $\bar \psi D^2\psi$. Although the Wilson term is an irrelevant operator, since the exact symmetries of the lattice theory allow a dimension three fermion mass term, it will be
generated radiatively, requiring  $O(1/a)$ fine-tuning of the bare mass to obtain massless fermions in the IR.

What about supersymmetry? Supersymmetry certainly cannot be an exact symmetry on the lattice, since the supersymmetry algebra dictates that the anti-commutator of supercharges yield an infinitesimal translation \citep{Wess:1992cp}, $\{Q^i_\alpha,\bar Q^j_{\dot\beta}\} = 2\sigma^\mu_{\alpha\dot\beta} P_\mu \delta_{ij}$, and such translations do not exist on a lattice.  However, as this Report documents,  supersymmetry can emerge from a  lattice action with little or no fine-tuning due to accidental symmetry. We first describe how this works in the four-dimensional theory that is simplest to simulate on the lattice: $\CN=1$ SYM theory.

\subsection{$\CN=1$ supersymmetry in $d=4$ and accidental susy 
without scalars}
\label{sec:3d}

The $\CN=1$ SYM  theory in  $d=4$ consists of gauge bosons $v_m$ (the ``gluons'',
$m=1,\ldots 4$) and a single
Weyl fermion $\lambda_\alpha$ (the ``gluinos'', $\alpha=1,2$).  The
gluino is the 
supersymmetric partner of the gluon, and like it,  transforms in the adjoint
representation of the gauge group.  Using the two-component
fermion notation (see \citep{Wess:1992cp}), the Lagrangian for the
theory is
\beq
\CL = \bar\lambda i \bar\sigma^m D_m\lambda - \fourth v_{mn} v^{mn}\
,\eqn{SYM1}
\eeq
where $\bar\sigma^m = \{1,-\boldsymbol{\sigma}\}$ ($\boldsymbol{\sigma}$ being the
three Pauli matrices and $1$ being the unit matrix) and $v_{mn}$ is
the gauge field strength.  This theory has only one independent
coupling constant (the gauge coupling $g$) and is the most general
Lagrangian one could write down without irrelevant operators --- with
the important exception that we have omitted a fermion mass term,
$(m\lambda\lambda + h.c.)$.  At the classical level, the theory
possesses a global $U(1)$ symmetry under which $\lambda \to
e^{i\alpha}\lambda$.  This does not commute with supersymmetry
(because there is no analogous phase rotation of the gluino's partner
the gluon) and for obscure historical reasons it is therefore called
an $R$-symmetry.  Now this $U(1)$ symmetry is anomalous, and if the
gauge group is $SU(N)$, only a $Z_{2N}$ subgroup of the $U(1)$ symmetry
is exact in the full quantum theory (see, for example,
\citep{Coleman:1978ae}). Note that a gluino mass term would explicitly
violate this $Z_{2N}$ $R$-symmetry. 
It is known that gluino condensation occurs in this theory
($\vev{\lambda\lambda}\ne 0$), and that
the global $Z_{2N}$ symmetry is spontaneously broken to $Z_2$, giving
rise to domain walls, where the strength of the condensate and the
domain wall tension can be analytically related. 

Can we investigate these properties on the lattice?  After all, the
theory looks simpler than QCD which has several flavors of quarks with
different masses, which is routinely simulated.

The key to accidental  supersymmetry for a lattice realization of this $\CN=1$ SYM theory is the observation that the only ``bad'' relevant operator allowed by gauge
plus Lorentz symmetries is a gaugino mass term...and this is forbidden
by the $Z_{2N}$ chiral $R$-symmetry. 
This observation can be put to use to construct a lattice theory.  Either one can use a Majorana Wilson fermion and fine-tune the gaugino mass to zero 
(see \citep{Montvay:1995ea,Feo:2004kx,Demmouche:2008ms,Demmouche:2008aq}
and references therein), or one can start with chiral lattice fermions  and obtain the supersymmetric target theory  without fine-tuning  \footnote{This scenario whereby
supersymmetry could be realized as an accidental symmetry was first
proposed in \citep{Kaplan:1983sk}, and specifically for  lattice
simulation in \citep{Curci:1986sm}.}.  Luckily, the problem of how to realize chiral fermions on the lattice
has already been solved: the two related techniques are to use domain
wall fermions (DWF) \citep{Kaplan:1992bt}, or overlap
fermions\citep{Narayanan:1994gw,Neuberger:1997fp}.  Formulations of $\CN=1$ SYM with overlap fermions were first proposed in  \citep{Neuberger:1997bg}; domain wall fermion formulations of the lattice theory are found in
 \citep{Nishimura:1997vg,Maru:1997kh, Kaplan:1999jn}.  Here we sketch the domain wall fermion (DWF) formulation of this theory \citep{Kaplan:1999jn}, before discussing the more complicated case of how accidental supersymmetry can arise in theories with scalar fields in subsequent sections.

\begin{figure}[t]
\centerline{\resizebox{0.5\columnwidth}{!}{%
 \includegraphics{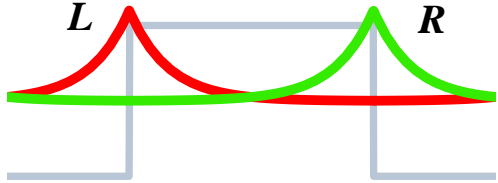} }}
\caption{The profile of the domain wall fermion mass in the fifth
  dimension, showing the chiral zero-modes  (L,R)
  bound to the two domain walls where the fermion mass
  switches sign.}
\label{fig:dwf}
\end{figure}


The DWF formulation
is formulated on a (compact) five-dimensional lattice, with a massive
fermion whose mass equals $+m_0$ on half the lattice and $-m_0$ on the
other half.  The 4-dimensional hypersurfaces where the mass changes sign are
called ``domain walls'', and on solving the free Dirac equation, one
finds two  massless 4-dimensional fermion modes, one  with $\gamma_5=+1$ bound to one domain
wall, and the other with $\gamma_5=-1$ bound to the other wall, as
shown in Fig.~\ref{fig:dwf}. (There 
is actually a small mass which vanishes exponentially in the fifth
dimensional separation between the two domain walls, which for
the purposes of this simplified discussion, we will
assume is negligible - see \citep{Giedt:2008xm,Endres:2008tz,Endres:2009yp} 
for a
discussion of practical issues 
issues arising from this non-zero {\it residual mass}). Four
dimensional gauge
fields are introduced {\it \'a la} Wilson, constant in the fifth
dimension, with the fermion transforming as an adjoint under the gauge
group. The low energy spectrum therefore looks like a $d=4$ theory
consisting of a gauged massless 
adjoint Dirac fermion and gauge bosons.  This $d=4$ (Euclidean) Dirac fermion takes the
form $\Psi = \begin{pmatrix} \alpha \cr \bar\beta \end{pmatrix}$,
$\bar\Psi = \left(\bar\alpha^T \ \ \beta^T \right)$, where
$\alpha$ and $\beta$ are the 2-component chiral spinors stuck to the
two domain walls respectively. Since the gauge fields are constant in
the fifth dimension, they are insensitive to the fact that the spinors
$\alpha$ and $\beta$ reside at different places in the extra
dimension. Imposition of the Majorana condition is
equivalent to requiring $\Psi = R_5 C \bar\Psi^T$, where $R_5$ is the
reflection in the fifth dimension which interchanges the two domain
walls, and $C$ is the $d=4$ charge conjugation matrix.  To implement a
Majorana fermion
in the Euclidean path integral then, we just replace $\bar\Psi$
everywhere by $\Psi^T R_5^T C^T$, so that the Dirac Lagrangian $\bar\Psi
\Dslash \Psi$ becomes instead $\Psi^T R_5^T C^T \Dslash\Psi$, and the 
Dirac determinant $\det\Dslash$ is  replaced by the Pfaffian ${\rm
  Pf}R_5^T C^T \Dslash$, which is real and non-negative.  Simulation of this theory is computationally challenging, but recently 
 much progress has been made with the first ab initio
 calculations of the
 chiral condensate being reported in
\citep{Fleming:2000fa,Giedt:2008cd,Endres:2008tz,Endres:2009yp,Giedt:2008xm}.
 
\subsection{Accidental SUSY with scalars?}
\label{sec:3e}

In the previous section we saw that a gauged adjoint Majorana fermion
in four dimensions was automatically supersymmetric provided that the
relevant mass term $m \lambda\lambda$ vanished.  Since this mass term
violates a $Z_{2N}$ chiral symmetry as well as supersymmetry, it
follows that in a lattice theory that correctly implements the chiral
symmetry, supersymmetry will automatically emerge as an accidental
symmetry in the continuum limit, despite the fact that the lattice
action is not supersymmetric at all.  

Unfortunately, this simple reasoning does not extend readily to other
supersymmetric theories, which all contain scalars as well as
fermions, and possibly gauge fields.  The problem is that
supersymmetry is broken by the relevant operator responsible for
scalar masses $m^2 |\phi|^2$ (among others), which breaks the
fermion-boson degeneracy.  Following the example of $\CN=1$ SYM, we
would like to identify some symmetry ($other$ than supersymmetry)
which is broken by a scalar mass
term, and which can be implemented exactly on the
lattice. Unfortunately, unlike fermions, there is no chiral
symmetry which can be invoked to forbid a scalar mass; the only
symmetry that can do that is a shift symmetry $\phi\to \phi + f$, and
this shift symmetry is too restrictive, dictating only derivative
interactions for the scalar,  and hence applicable only to Goldstone
bosons (furthermore sigma models are not thought
to be renormalizable in four dimensions).
Thus the only useful symmetry that can forbid the undesirable
scalar mass term is supersymmetry itself.

We are apparently left with a paradox: implementing supersymmetry
exactly on the lattice seems impossible, and so we would like it to
emerge as an accidental symmetry; but in order for supersymmetry to
emerge as an accidental 
symmetry, we are forced to suppress scalar mass renormalization, which
requires implementing supersymmetry exactly on the lattice!

Perhaps we don't have to find an exact lattice
implementation of  $all$ of the
supersymmetry of the target theory, but only realize part of the
supersymmetric algebra?  After all, the full Poincar\'e group is not
realized on the lattice, but only the finite subgroup generated by
finite translations and  rotations by $\pi/2$, yet the Poincar\'e
group emerges as an accidental symmetry.  It is natural then to ask
whether there could exist a ``subgroup'' of supersymmetry on the
lattice? But the answer is no:  whereas rotations, for example, are
parameterized by a bosonic angle which can be large (e.g., $\pi/2$)
supersymmetric transformations are characterized by a Grassmann
parameter, which is necessarily infinitesimal, just as
there exist classical bosonic fields (such as the electric field) but
not classical fermionic fields \footnote{There are
  such things as ``supergroups'' defined with Grassmann generators, but they do
  not seem to be of any practical use for constructing
   supersymmetric lattices.}.

Instead one must ask whether it is possible to preserve a subalgebra
of the full extended supersymmetric algebra
\beq
\{Q^i_\alpha,Q^j_\beta\}=0\ ,\quad\{\bar Q^i_{\dot\alpha},\bar Q^j_{\dot
  \beta}\}=0\ ,\quad \{Q^i_\alpha,\bar Q^j_{\dot\alpha}\} = 2 P_m
\sigma^m_{\alpha\dot\beta}\delta_{ij}\ ,
\eqn{susyalg}\eeq
where $i,j=1,\ldots,\CN$ run over different supercharges (for $\CN=1,2,4$ supersymmetry respectively in $d=4$). A number of potential obstacles are immediately obvious:
\begin{enumerate}[i.]
\item
The same old problem we keep returning to: how can a subalgebra of
\eq{susyalg} be chosen given that the $P_m$, 
the generator of infinitesimal translations, does not exist on the
lattice? (An early attempt at lattice SUSY was to work in a Hamiltonian formulation, so that infinitesimal time translations $P_0=H$ were maintained; however while this enabled exact lattice supersymmetry, it precluded a Lorentz invariant continuum limit \citep{Banks:1982ut}.)
\item
How can one isolate part of the algebra without destroying the hypercubic
lattice symmetry, and thereby making it impossible to recover
Poincar\'e symmetry in the continuum, let alone supersymmetry?
\item
Less abstractly, how is it possible to implement scalars, fermions and
gauge bosons in a symmetric fashion on the lattice?  For example, $\CN=4$
SYM has one gauge field, four Weyl gauginos, and six real scalars in
the same supersymmetric multiplet.  If we put the gauge fields on
links, surely their scalar superpartners have to be on links too! But
then the scalars will transform nontrivially under lattice rotations,
which suggests they  can't transform as scalars (rotationally
invariant objects) in the continuum limit. 
\item
SYM theories have $R$-symmetries ($U(1)$, $U(2)$ and $SU(4)$
respectively for $\CN=1,2,4$ theories in $d=4$; larger symmetries in lower
dimensions) which are $chiral$ symmetries; how are we to implement
chiral fermions in a way that makes them look symmetric with their
gauge and scalar partners?!
\end{enumerate}

These arguments would incorrectly seem to rule out implementing accidental lattice supersymmetry with scalars. The way out of this cul-de-sac is to recognize that in Euclidean space, $Q^i$ and $\bar Q^i$ are independent; calling them all $q^i$,   a subset of the supercharges $\{q^i\}$ can be preserved that are nilpotent: $\{q^i,q^j\}=0$ (up to a gauge transformation).   Keeping such charges exact on the lattice solves the problem of not having infinitesimal $P_\mu$ generators, but does not solve the issue of Lorentz invariance since the supercharges being kept belong to incomplete spinor representations of the Lorentz group.  An answer to this  conundrum can be found  in the peculiar formulation of staggered or \KD fermions, where the discrete point symmetry of the lattice is not just embedded in the full Lorentz group (or its Euclidean analogue), but in the combined (Lorentz)$\times$(flavor) group.  Under the discrete lattice symmetry, the fermions then naturally decompose as $n$-index antisymmetric tensors, instead of spinors. Furthermore,
these antisymmetric tensor components generically
include one or more scalars.   If the supercharges singled out for preservation on the lattice are scalars under this lattice symmetry, it becomes plausible that Lorentz symmetry could be preserved in the continuum limit.  

While providing a clue for how to realize supersymmetric lattices, the above discussion leaves obscure how to create ``staggered scalars" and ``staggered gauge fields" so that supersymmetric multiplets composed of (gauge boson, gaugino, gauge scalar) could appear on the lattice.  As we show below, the Gordian knot is cut by formulating twisted supersymmetry, or  by following the orbifold/deconstruction procedure.  By means of these related techniques we are able to solve the above conundrums in a new
and unanticipated way.

\subsection{Twisted supersymmetry} 
\label{sec:3f}

In this section, we first briefly review  the  concept of twisting in 
extended supersymmetric gauge theories in the continuum  formulation on 
${\mathbb R^d}$  \citep{Witten:1988ze} and sketch its relation to orbifold 
projections of supersymmetric matrix models, which we will discuss next.
 
As we have discussed,
extended supersymmetric gauge theories usually possess large 
chiral symmetries, called $R$-symmetries.  The bosonic and fermionic fields 
furnish a representation of the relevant
$R$-symmetry, as well as Euclidean Lorentz symmetry  $SO(d)_E$, 
and the same is true for 
the supercharges. For example, 
a list of the SYM theories and their Lorentz and 
$R$-symmetries (at the classical level) are given in Table~\ref{tab:tab1}. 
For  six of the  theories shown in 
Table~\ref{tab:tab1},  the $R$-symmetry group possess an $SO(d)_R$ subgroup.   
Hence, the full global  symmetry of the supersymmetric  theory has a 
subgroup   
$SO(d)_E \times SO(d)_R \subset SO(d)_E \times G_R$.  
To construct the twisted theory, we identify the diagonal  $SO(d)'$ subgroup
in  $SO(d)_E \times SO(d)_R $,   
and declare  and treat it  as the new  Lorentz 
symmetry of the  theory \footnote{ We will not distinguish spin groups 
$Spin(n)$ from   $SO(n)$ unless otherwise specified.}.\beq
SO(d)'= {\rm Diag} (SO(d)_E \times SO(d)_R) 
\eeq
 In particular, when we eventually create a lattice theory, the point group of the lattice will be a discrete subgroup of this  $SO(d)'$, and {\it not} of $ SO(d)_E$, as one might have supposed.
Since the details of each such construction are slightly different, 
let us restrict to generalities first (later we discuss in some detail how twisting works in
the context of $(2,2)$ SYM in two dimensions). In every case we will consider, fermionic fields transform as spinor representations under both $SO(d)_E$ and $SO(d)_R$.
Since the product of two half-integer spins always has integer 
spin, {\it all} fermionic degrees of freedom will be in
integer spin representations of $SO(d)'$, 
direct sums of scalars, vectors, and general $p$-form tensors. Let us label a 
$p$-form fermion as $\psi^{(p)}$. 
In  all of our applications, the $\CF$ different fermions of a target field theory
in $d$ dimensions    are distributed in multiplets of
$SO(d)'$ as 
\beq 
{\rm fermions} \rightarrow  \;  \frac{\CF}{2^d} 
( \psi^{(0)} \oplus\psi^{(1)}  \oplus  \ldots  \psi^{(d)}) 
\eeq
where the multiplicative factor up front is one, two, four or eight.  
For a given 
$p$-form, there are $  \frac{\CF}{2^d} \binom{d}{ p}$ fermions. 
Summing over all $p$,  we obtain 
the total number of fermions in the target theory:
$\frac{\CF}{2^d} \sum_{p=0}^{d}  \binom{d}{p}= \CF $

Turning to the bosonic fields, 
the gauge bosons  $V_{\mu}$ transform as $(d, 1)$ under $SO(d)_E \times SO(d)_R$, while the scalars typically include a subset we can label  as   $S_{\mu}$, transforming
as $(1,d)$. 
Thus both $S_{\mu}$  and $V_{\mu}$  transform as $d$-vectors  (1-forms) under the diagonal
$SO(d)'$ symmetry.  In theories with more than 
$d$ scalars in the  untwisted theory, these remnants become either $0$-forms or 
$d$-forms under $SO(d)'$.

The $\CQ$ supercharges  also 
decompose into a sum of $p$-forms under the diagonal group: 
\beq 
{\rm Supercharges} \rightarrow  \;  \frac{\CQ}{2^d} 
( Q^{(0)} \oplus Q^{(1)}  \oplus  \ldots  Q^{(d)}) 
\eqn{DK2}
\eeq
We may then write the supersymmetry algebra without using any spinor indices just in
terms of $p$-forms. What is important is the fact that there exists
one or more spin-0 
nilpotent supercharges $Q^{(0)}\equiv Q$. That means, the twisted formulation
of the supersymmetry  algebra contains a subalgebra 
\begin{equation}
Q^2 \; \cdot = 0
\label{Eq:nilpotent}
\end{equation}
This nilpotent supercharge is then insensitive to the background 
geometry. In fact, if the base space of the theory is an arbitrary 
$d$-dimensional curved  manifold $M^{d}$, then there exist no covariantly 
constant spinors. However, there may exist covariantly constant, spin-0 fields. 
Hence, globally, only the  scalar  supercharges 
are  preserved when the theory is carried on curved spacetime.  
Furthermore, if $M^d$ is flat, this transition from integer-spin, $p$-form   supercharges 
to spinor supercharges is a simple change of basis, a redefinition. 
In flat spacetime, so long as the scalar supercharge 
is not declared as a BRST operator, there is no 
physical distinction between the twisted and untwisted theories.
We can construct true
topological 
field theories if we additionally require that the charge $Q$ be interpreted as
a BRST operator. This is discussed throughly in the context of the string 
theory and the theory of four manifolds in \citep{Birmingham:1991ty}. 

\begin{figure}[t]
\centerline{\resizebox{0.5\columnwidth}{!}{%
 \includegraphics{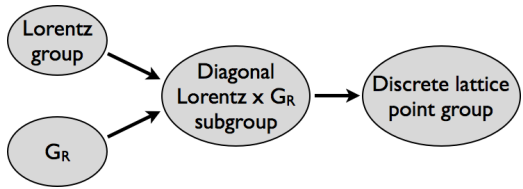
 } }}
\caption{The lattice point group in supersymmetric lattices cannot be
  considered to be a subgroup of just the Lorentz group, but rather it
  is a subgroup of
the product of the Lorentz group and and the $R$-symmetry group, $G_R$.}
\label{fig:sym}
\end{figure}

The application 
to topological field theories leads to the appearance of the term 
``topological twisting''. 
In fact, the twisting operation can be thought of as conceptually unrelated to
topological field theory or 
supersymmetry. As we will discuss, the well known staggered
fermion formulation of lattice 
fermions (or Kogut-Susskind fermions) is in fact 
an example of twisting, with the lattice fermions living in a  
diagonal subspace of the flavor and real spaces as shown in Fig.~\ref{fig:sym}.

The action  expressed in terms of fields which form representations  of the  
twisted Lorentz  group $SO(d)'$   instead of the usual 
Lorentz symmetry, is called the twisted action.  Typically, 
the twisted action
can be expressed as a sum of $Q$-exact and $Q$-closed terms, 
where $Q$ refers to one or more of the
scalar supercharges.  The arguments for this follow directly from the structure of the
twisted subalgebra as we explain later.

A key feature of this twisting process (shown in Fig.\ref{fig:sym})  is that 
none of the  degrees of freedom are 
spinors under $SO(d)'$. Both   bosons and 
fermions are in integer spin representations. They are $p$-form tensors 
of $SO(d)'$. 
This particular form of the twisted theory is the bridge to lattice 
supersymmetry and orbifold lattices.   
A  $p$-form continuum
field  may naturally be associated with a $p$-cell  on the hypercubic 
lattice \citep{Rabin:1981qj}. We will see that 
this is exactly what an orbifold lattice does. The orbifold projection 
places the fermions on sites, links, faces, or more
generally to $p$-cells. 
Moreover, since scalars of the Lorentz symmetry are 1-forms of the twisted 
theory, they can naturally be amalgamated with the gauge bosons  into 
complex bosons -- the two degrees of freedom being
associated with oppositely oriented links. This
complexification of gauge fields had been noted earlier in
continuum twists of the $\cN=4$ theory \citep{Marcus}. 

We now see how twisting allows us to circumvent  the  first three obstacles 
listed in section \ref{sec:3e}
\begin{enumerate}[i.]
\item
 By focusing on the nilpotent supercharge, the connection between supersymmetry and infinitesimal translations is sidestepped;

 \item By identifying the lattice symmetry with a discrete subgroup of $SO(d)'$ and implementing the supercharges which are $SO(d)'$-scalars, the supersymmetric subalgebra we have selected does not interfere with obtaining a Poincar\'e invariant continuum limit;
 
 \item
 With fermions and bosons falling into similar $SO(d)'$ representations, it becomes possible to imagine that they could be treated similarly on the lattice as would befit members of the same supersymmetry multiplet.
 \end{enumerate}
 In relation to the fourth point in section \ref{sec:3e}, we will see that the $R$-symmetries are not exact on the lattice, but emerge as accidental symmetries in the continuum limit, along with Poincar\'e invariance and full supersymmetry.

One may ask how do these orbifold projections
know about the representations of the twisted group ? We will come back to 
this point later, after discussing several applications.  The punchline is 
that the point group symmetry of the supersymmetric lattice is {\it not} 
a subgroup of the 
Euclidean Lorentz  group, but in fact a discrete subgroup of the twisted 
rotation group $SO(d)'$. 
In the continuum, the orbifold lattice 
theory becomes 
the twisted version of the desired target field theory. And in flat space, 
the change of variables which takes the twisted form to the canonical 
form  essentially  undoes the twist.

The type of twist discussed in this section is 
sometimes referred as {\it maximal twist} as it
involves the twisting of the full Lorentz symmetry group as opposed 
to twisting one of its subgroups.  In this sense,  
the  four dimensional  $\CN=2$ theory  can only admit a {\it half twisting}
as its $R$-symmetry 
group is not as large as $SO(4)_E$ \citep{Witten:1988ze}.
The other two
theories, $\CN=1$ in $d=4$ and  $\CN=1$ in  $d=3$ shown in 
Table~\ref{tab:tab1} do not admit a nontrivial twisting 
as there is no nontrivial homomorphism from their Euclidean rotation group
to their $R$-symmetry group. Thus the methods described in this
Report cannot be used in those cases.

After this general discussion we turn now to a pedagogical
discussion of these problems and their solution in the
context of simpler models -- namely supersymmetric quantum mechanics
and the Wess-Zumino model.

\section{Supersymmetric quantum mechanics on the lattice }
\label{sec:6}

Supersymmetric quantum mechanics constitutes a good toy model for both
understanding some of the problems encountered when trying to
study supersymmetry on the lattice and some of the ways to circumvent
these problems. Specifically even in this simple model we will
see the issue of fine tuning arising in naive discretizations
of the continuum theory and how this can be handled in low
dimensions by performing perturbative lattice calculations
to subtract off the dangerous radiative
corrections. Furthermore, we will also see 
how we can make a change of variables which
exposes a nilpotent supersymmetry and allows us to write down a lattice
action which is explicitly invariant under this supersymmetry. This
change of variables is just the twisting procedure we have already
alluded to but restricted to the case of one (Euclidean) dimension.
We also show that the exact supersymmetry ensures that these dangerous
radiative corrections then cancel automatically and the resulting lattice
theory does not suffer from fine tuning problems.

The continuum theory was first written down by Witten as a toy
model for understanding supersymmetry breaking \citep{Witten_qm}. 
The model comprises a single
commuting bosonic coordinate $\phi(t)$ and two anticommuting
fermionic coordinates $\psi_1(t)$, $\psi_2(t)$. We will be working
in the language of path integrals in Euclidean space which here
means we treat the time coordinate $t$ as Euclidean. The continuum action reads
\begin{equation}S=\int dt\;\frac{1}{2}\left(\frac{d \phi}{dt}\right)^2+\frac{1}{2}P^\prime(\phi)^2
+\frac{1}{2}\psi_i\frac{d \psi_i}{dt}+i\psi_1\psi_2P^{\prime\prime}(\phi)
\end{equation}
where $P^\prime$ is an arbitrary polynomial in $\phi$ and
$P^{\prime\prime}$ its derivative. The function $P(\phi)$ is often
called the superpotential.

\subsection{Algebra - two supersymmetries}
\label{sec:6a}
This action is invariant under the two supersymmetries given below
where $\epsilon_A$, $\epsilon_B$ are infinitesimal Grassmann
parameters.
\begin{align}
\delta_A \phi & = \psi_1\epsilon_A & \delta_B \phi &= \psi_2\epsilon_B\nonumber\\
\delta_A \psi_1 &=  \frac{d \phi}{dt}\epsilon_A & \delta_B \psi_1&= -iP^\prime\epsilon_B\\
\delta_A \psi_2 &= iP^\prime\epsilon_A &\delta_B \psi_2&= \frac{d\phi}{dt}\epsilon_B\ .\nonumber
\end{align}

It is a simple exercise to verify these invariances. Simply
carry out the variation of the fields and use the Grassmann
property e.g. $\{\epsilon_A,\psi_i\}=0$. In both cases the
only slightly nontrivial terms encountered take, in the former
case, the form
\begin{equation}
\delta_A S=\int dt\; i\epsilon\left(P^\prime\frac{d\psi_2}{dt}+\frac{d\phi}{dt}P^{\prime\prime}\psi_2\right)
\end{equation}
In this case a simple integration by parts sets the term inside the
brackets equal to zero. From an operational point of view this
is what ruins supersymmetry on the lattice -- since this operation
requires the use of the Leibniz rule which does not hold for lattice difference operators \citep{Fujikawa_leib,Bruckmann_nc}.
Notice that $\delta_A^2=\delta_B^2=\frac{d}{dt}$ when acting
on any field\footnote{We need to use the equations of motion to show
this for $\psib$}. Since $H\equiv \frac{d}{dt}$ in Euclidean
space this corresponds
to the usual supersymmetry algebra reduced to the
quantum mechanics case of one dimension. 

\subsection{Naive discretization}
\label{sec:6b}
Let us now proceed to discretize this theory initially in a naive
manner \citep{Giedt_qm,Catterall_qm}. Define the fields on lattice sites 
$x=na,\;n=0\ldots L-1$
and replace integrals by sums using periodic
boundary conditions on all fields.
To eliminate fermion doubling problems it is
sufficient in one dimension 
to replace the continuum derivative with a backward
(or forward) difference operator. 
\begin{equation}
\Delta^-f_x=f(x)-f(x-a)
\end{equation}
Upon this replacement and carrying out 
supersymmetry variation,  one  finds a non-vanishing
variation in the A case of the form (B is similar)
\begin{equation}
\delta_A S_L=\sum_x\; i\epsilon\left(P^\prime\Delta^-\psi_2+
\Delta^-\phi P^{\prime\prime}\psi_2\right) \, .
\label{variation}
\end{equation}
Using lattice integration by parts we find
\begin{equation}
\delta_A S_L=i\sum_x\; \epsilon\psi_2\left(-\Delta^+P^\prime+\Delta^-\phi P^{\prime\prime}\right)
\end{equation}
Since $\Delta^-\to \Delta^+\to\frac{d}{dt}$ in the naive continuum limit
it is clear that this term is O(a) and vanishes in the naive
continuum limit. However it clearly does not vanishing at finite
lattice spacing and thus the naive lattice action is {\it not}
invariant under supersymmetry transformations.

\begin{figure}
\begin{center}
\resizebox{9.0cm}{!}{\includegraphics{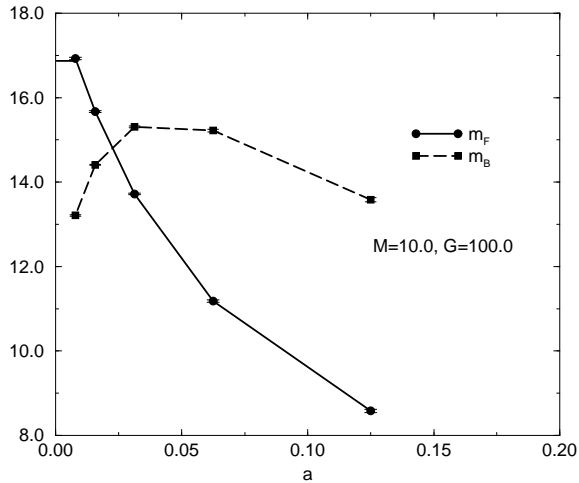}}
\caption{Boson and fermion masses vs lattice spacing for naive action}
\label{qm_naive}
\end{center}
\end{figure}

As we emphasized earlier, the absence of
an exact classical supersymmetry allows the quantum effective action to
develop further SUSY violating interactions. 
This problem can be seen explicitly when we simulate this
naive lattice theory. 
Fig.~\ref{qm_naive} shows a plot of the boson and fermion
masses $m_BL$, $m_FL$  extracted from a simulation of the naively discretized
action with $P(\phi)=m\phi+g\phi^3$ with $mL=10.0$ and
$gL^2=100.0$ shown as a function of
the lattice spacing. Clearly they are not equal and indeed
the problem worsens as $a\to 0$ consistent with the
existence of relevant SUSY breaking counterterms. The plot also
shows the expected result in the continuum limit $m_BL=m_FL=16.87$ which
can be computed straightforwardly using Hamiltonian methods.

The mismatch  arises through
radiative corrections and hence we  are led to an analysis
of loop corrections in the lattice theory in the next section.

\subsection{Feynman diagrams and power counting}
\label{sec:6c}

In practice the only diagrams we need to be concerned about 
are ones with a positive superficial degree of divergence. Only
these will generate { relevant} SUSY violating interactions in
the lattice effective action. Reisz's theorem \citep{Reisz} guarantees that all
Feynman diagrams with a negative degree of divergence converge to
their continuum counterparts as $a\to 0$ -- and hence cannot 
contribute new supersymmetry breaking terms.

The good thing is that
since this quantum mechanical model is  super-renormalizable theory,
there are only a finite number of such U.V sensitive
diagrams and they
occur at low orders in perturbation theory
\citep{Giedt_qm,Golterman_wz,Elliott:2005bd,Kaplan_rev,Elliott:2007bt,Suzuki_reduc}. Only these diagrams need
to be examined carefully when we go to
the lattice. We will see
that it is possible that the contribution of such graphs in lattice
perturbation theory {\it do not} converge to the continuum result
as $a\to 0$
and hence can yield SUSY breaking effects. The reason lies with the
would-be fermion doublers in the lattice description -- it is possible
for these high momentum states to contribute additional effects
at the cut-off scale. This is similar to the classic calculation of
Karsten and Smit showing how the chiral anomaly arises in lattice QCD
\citep{Smit}.
 
\begin{figure}
\begin{center}
\resizebox{9.0cm}{!}{\includegraphics{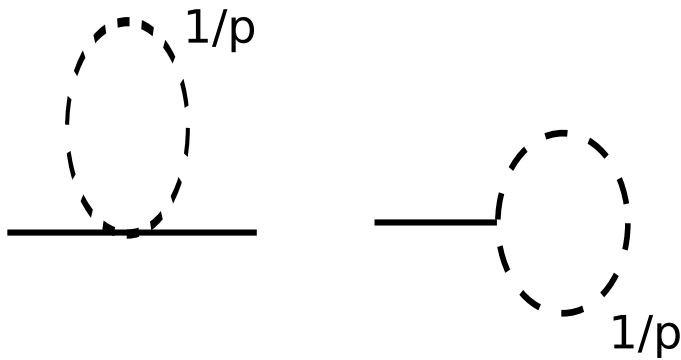}}
\caption{One loop Feynman graphs for SUSYQM}
\label{loopsWitten}
\end{center}
\end{figure}

Let us take as an example once again
the potential $P^\prime=m\phi+g\phi^3$. In this
case it is a simple exercise to show that the only dangerous
Feynman graph is the one-loop fermion contribution to the boson
propagator (the  diagram on the left in Fig.~\ref{loopsWitten})
\citep{Giedt_qm}. 

In the continuum this contributes:
\begin{equation}
\sigma_{\rm cont}=6g\int_{-\frac{\pi}{a}}^{\frac{\pi}{a}}\frac{dp}{2\pi}
\frac{-ip+m}{p^2+m^2}
\end{equation}
where we use $\pi/a$ as the effective continuum momentum cut-off.
The divergent piece of the integral is zero by the symmetry $p\to -p$ and
we find 
\begin{equation}
\Sigma_{\rm cont}=6g\left(\frac{1}{\pi}\tan^{-1}{\frac{\pi}{2ma}}\right)\sim 
6g\left(\frac{1}{2}+{\cal O}(ma)\right)
\end{equation}
The same diagram on a lattice of size $L$ (and using a backward difference operator) yields
\begin{equation}
\Sigma_{\rm latt}=\frac{6g}{L}\sum_{k=0}^{L-1}\frac{-2i\sin{({\pi k}{L})}
e^{i(\frac{\pi k}{L})}+ma}{\sin^2{(\frac{\pi k}{L})}+m^2}
\end{equation}
Notice that the phase factor breaks the $k\to -k$ symmetry. Indeed,
the lattice yields {\it twice} the continuum result when the
limit $a\to 0$ is taken {\it after} doing the sum. 
In order to understand this effect, use 
$\Delta^-=\Delta^S+\frac{1}{2}m_W$  where  $\Delta^S$ is the symmetric difference operator 
having the same symmetry $k\to -k$ as the continuum, and $m_W$  is the Wilson operator i.e difference between forward and backward difference operators or equivalently the discrete laplacian.  
We can understand the lack of convergence to
the continuum result as resulting from the
additional contribution of a doubler state with $k\sim \frac{\pi}{a}$
and mass determined
by the Wilson term $m \sim {\cal O}(\frac{1}{a})$.

\begin{figure}
\begin{center}
\includegraphics[height=8cm,width=6cm,angle=90]{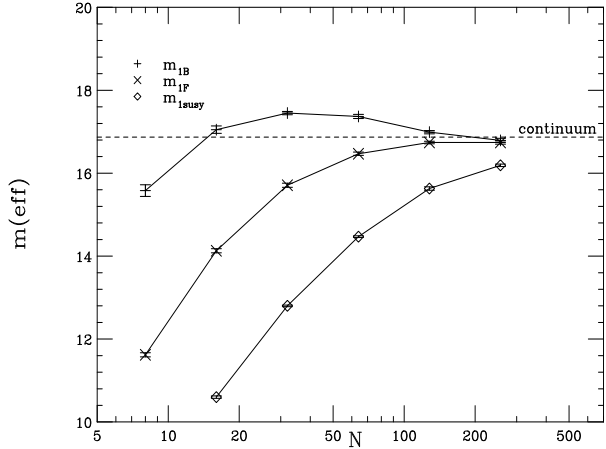}
\caption{Counterterm corrected boson/fermion masses
(From  Giedt et al. \citep{Giedt_qm})} \label{qm_naive_corr}
\end{center}
\end{figure}

This additional contribution shifts the mass squared
of the boson by an additional
$3g$ which breaks supersymmetry. To restore SUSY one needs only add
a new boson mass counterterm to the lattice action 
\begin{equation}
S_L\to S_L+\sum_x 3g\phi^2
\end{equation}
The resultant lattice theory does {\it not} manifest exact supersymmetry
at finite lattice spacing but will nevertheless flow to the
correct supersymmetric continuum theory without further fine
tuning as $a\to 0$. This can be seen in Fig.~\ref{qm_naive_corr} which shows the boson and
fermion masses derived from a simulation of the naive action
corrected by this one-loop counter term. The x-axis shows
the number of lattice points $N=L_{\rm phys}/a$. The upper two curves
correspond to the boson and fermion masses. The lower
curve corresponds to the result from the exactly supersymmetric
action which we discuss next. The dotted line is the expected continuum
value (the parameters coincide with those used earlier for the
naive lattice action)

Clearly the counterterm corrected lattice action generates boson and
fermion masses which approach each other as the lattice spacing is
reduced.

We now turn to a lattice discretization which preserves supersymmetry
{\it exactly} at non-zero lattice spacing. Such a theory will necessarily
contain the counterterm we have just discovered plus additional
irrelevant terms which will keep the boson and fermion masses equal
to all orders in the lattice spacing.
 
\subsection{Exact lattice supersymmetry}
\label{sec:6d}

In the last section we discussed how to diagnose the counterterms which must be added to the lattice action to avoid fine tuning.
In general for $d=1,2,3$ dimensions,   the coefficients to these may be computed
in perturbation theory. However, in the case of this supersymmetric
quantum mechanics model we can do
better -- it is possible to find a linear combination of the
supersymmetries which can be transferred intact to the 
lattice. To uncover this let us go back to eq.~(\ref{variation})
which shows the variation of the lattice action under the A-type
supersymmetry
and recognize that this term may be rewritten in terms of a variation of another operator under the B-type supersymmetry
\begin{equation}
\delta_A S_L=-i\delta_B \sum_x P^\prime\Delta^-\phi
\end{equation}
Similarly it is not hard to show that
\begin{equation}
\delta_B S_L=i\delta_A\sum_x P^\prime\Delta^-\phi
\end{equation}
Thus the linear combination 
\begin{equation}
\left(\delta_A+i\delta_B\right)S_L=-\left(\delta_A+i\delta_B\right)O
\end{equation}
where $O=\sum_x P^\prime\Delta^-\phi$. Notice, again, that $O$
would vanish in a continuum theory as it would correspond to
a total derivative term. 
Thus a lattice action invariant under supersymmetry $\delta S_{\rm exact}=0$
can be
constructed by adding $O$ to the original naive action
\begin{equation}
S_L^{\rm exact}=
\sum_x \frac{1}{2}(\Delta^- \phi)^2+\frac{1}{2}{P^\prime}^2+P^\prime\Delta^-\phi+\psib(\Delta^-+
P^{\prime\prime})\psi
\label{exact_act}
\end{equation}
where we have defined new fermion fields 
\begin{eqnarray}
\psi &=&\frac{1}{\sqrt{2}}(\psi_1+i\psi_2)\nonumber\\
\psib&=&\frac{1}{\sqrt{2}}(\psi_1-i\psi_2)
\end{eqnarray}
and the
exact supersymmetry $\delta=\frac{1}{\sqrt{2}}(\delta_A+i\delta_B)$.
The latter symmetry acts on the fields as follows
\begin{eqnarray}
\delta x&=&\psi\epsilon\nonumber\\
\delta \psi&=&0\nonumber\\
\delta \psib&=&(\Delta^-\phi+P^\prime)\epsilon
\end{eqnarray}
Notice that this derived supersymmetry is no longer the
``square root'' of a translation but instead  is {\it nilpotent} (using the
equations of motion). This fact is at the heart of how we are
able to build an exact supersymmetry -- the algebra of
the corresponding supercharge is simply $\{Q,Q\}=0$ and does not involve the energy or momentum. 
Equivalently, the invariance $\delta S_L=0$ 
does not require use of the Leibniz rule.

Notice also that {\it two}
supersymmetries were required to find such a nilpotent
supercharge -- the continuum theory has {\it extended} supersymmetry.
This will be seen to be a general property of lattice models
with exact supersymmetry.
It is not hard to show that the other supersymmetry
$\delta^\prime=\frac{1}{\sqrt{2}}\left(\delta_1-i\delta_2\right)$ is
still broken on the lattice
\begin{equation}
\delta^\prime S_L=2\delta^\prime O
\end{equation}
Finally, if we examine the form of the lattice action
 in eq.~(\ref{exact_act}) we see the bosonic piece can be rewritten
\begin{equation}
S_B=\sum_x \left(\Delta^- \phi+P^\prime(\phi)\right)^2
\end{equation}
The cross term that appears after the square is expanded is the
correction needed to ensure exact supersymmetry. It vanishes
in the continuum as a total derivative. While it is identically zero
in the continuum it constitutes a new relevant operator by lattice
power counting. It generates an additional bosonic
1-loop Feynman graph which
cancels the corresponding fermion loop (due to the derivative interaction)
restoring supersymmetry.

\subsection{Nicolai map}
\label{sec:6e}

The partition function\footnote{In the path integral formulation, we impose 
supersymmetry preserving  periodic boundary conditions for all fields. In operator formalism, 
this corresponds to the  partition function $Z= {\rm tr} [ e^{-\beta H} (-1)^F]$ which, for supersymmetric theories, is just the  Witten index of the theory.}
governing the quantum theory takes the
form
\begin{equation}
Z=\int D\phi D\psi D\psib e^{-S}=\int D\phi {\rm det}
\left(\Delta^-+P^{\prime\prime}\right)e^{-S_B}
\end{equation}
Notice, however, a curious fact; if we imagine changing variables
from $\phi$ to new variables ${\Nc}=\Delta^-\phi+P^\prime(\phi)$ we
will encounter a Jacobian which is just 
${\rm det}\left(\frac{\delta\Nc}{\delta \phi}\right)$. This Jacobian
{\it cancels} the fermionic action and yields a very
simple expression
for the partition function \citep{Nicolai}
\begin{equation}
Z=\int  D {\Nc} \;  e^{-\sum_x {\Nc}^2}
\end{equation}
corresponding to a set of simple uncoupled bosonic oscillators.
This change of variables is called a {\it Nicolai map} and
the existence of a {\it local} Nicolai map is at the heart of
why these models may be discretized in a SUSY preserving manner
\citep{Nicolai,Cecotti:1982ad}.

It has immediate consequences; the detailed form of the
superpotential $P(\phi)$ has disappeared in this final form of
$Z$ -- hence the latter cannot depend on any coupling constants
in the model -- it is a {\it topological invariant} \citep{Catterall_topo}. We may use this fact
to derive an {\it exact} value for the expectation value of the
bosonic action which holds for all interaction couplings. Replacing
$S_L$ by $\mu S_L$ it is clear that $Z$ does not change.
The statement $\frac{\partial\ln{Z}}{\partial\mu}=0$ then
implies that $<S_L>=0$ which in turn implies that the bosonic
action $S_B$ is given by the expectation value of the fermionic action.
Since the latter is quadratic in the fermion fields a simple 
scaling argument shows the expectation value of the
latter simply counts the number of
degrees of freedom. We thus find that
\begin{equation}
<S_B>=\frac{1}{2}N_{\rm d.o.f}
\end{equation}
a result which usually only applies to a free theory but here is
valid as a consequence of supersymmetry at all values of the
coupling.
As an example of this we quote the measured value $<S_B>=1.99985(20)$
obtained from a Monte Carlo calculation using 
a lattice with
$L=4$ points and corresponding to the superpotential
parameters $mL_{\rm phys}=2.5$ and $gL_{\rm phys}^2=6.25$. This
is to be compared with the exact result expected on the
basis of exact supersymmetry $<S_B>=2$.

\subsection{Ward identities}
\label{sec:6f}

The classical invariance of the action is manifested in the
quantum theory by exact relationships between different
correlation functions. Consider the expectation value of
some operator $\cal O$. If we make a change of variables
$\phi\to\phi^\prime=\phi+\delta\phi$ we find
\begin{equation}
<{\cal O}>=\frac{1}{Z}\int D\phi^\prime O(\phi^\prime) e^{-S(\phi^\prime)}
\end{equation}
If $\delta\phi$ corresponds to some symmetry $S(\phi^\prime)=S(\phi)$ and, assuming the measure is also invariant
under this shift, we deduce a corresponding Ward identity
\begin{equation}
<\delta {\cal O}>=0
\end{equation}
Notice, that this result can be reinterpreted in the language of
canonical quantization as the usual
statement that the supercharge annihilates the
vacuum in the absence of spontaneous symmetry breaking.

\begin{figure}
\begin{center}
\resizebox{9.0cm}{!}{\includegraphics{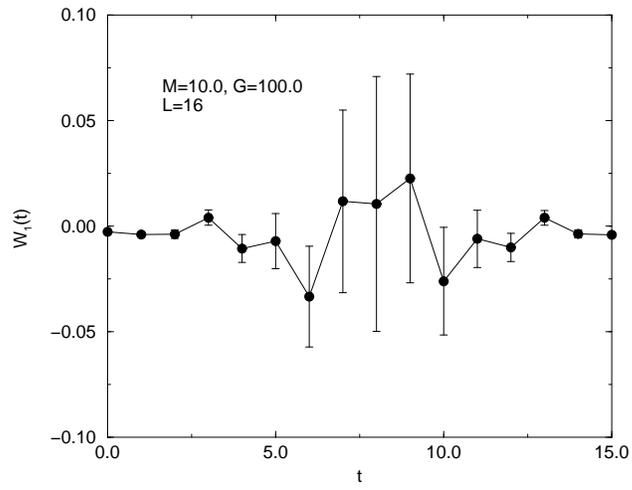}}
\caption{Ward identity for supersymmetric quantum mechanics}
\label{qm_ward}
\end{center}
\end{figure}

This can be exemplified in the case of supersymmetry by choosing
the operator ${\cal O}=\psib_x\phi_y$ yielding the Ward
identity
\begin{equation}
<\psib_x\psi_y>+<(\Delta^-\phi+P^\prime)_x\phi_y>=0
\end{equation}
relating the fermion 2pt function to a bosonic correlator.
A numerical calculation of this
Ward identity is shown in Fig.~\ref{qm_ward}
for the lattice parameters $L=16$, $mL=10$ and $gL^2=100$ 
corresponding to a strongly interacting theory with
dimensionless coupling $g/m^2=1$. 

\begin{figure}
\begin{center}
\resizebox{9.0cm}{!}{\includegraphics{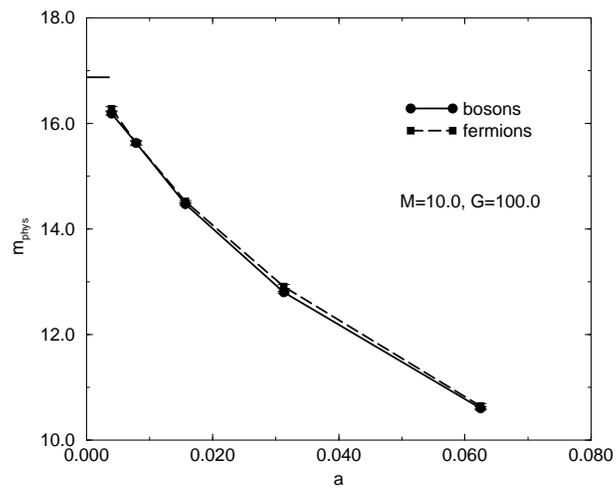}}
\caption{Boson and fermion masses vs lattice spacing for
supersymmetric action } \label{qm_mass}
\end{center}
\end{figure}

This result already ensures that the masses of the lowest
lying fermionic and bosonic excitations must be equal -- one
of the most obvious predictions of a supersymmetric theory.
Further evidence to this effect is shown in Fig.~\ref{qm_mass}
which plots the lowest lying bosonic and fermionic masses
versus lattice spacing. Evidently
the masses are degenerate within rather small statistical errors.
Additionally note that they appear to extrapolate rather nicely
to the exact continuum answer obtained independently
by Hamiltonian
methods\footnote{Additional work on lattice implementations of this model including some high statistics, fine
lattice spacing results can be found in the recent work \citep{Giedt_qm, Bergner:2007pu}.}.

Contrast this with the result obtained for the naive action shown
in Fig.~\ref{qm_naive} for the same lattice
parameters (again the continuum result is shown as a tick mark on
the y axis). Clearly the boson and fermion masses
are very different for small lattice spacing reflecting the
necessity of introducing the appropriate counter term to
ensure supersymmetry as $a\to 0$. 

\subsection{Topological field theory form - twisting}
\label{sec:6g}

The nilpotent character of the exact supersymmetry can be
rendered true off-shell by adding an auxiliary field $B$
\citep{Catterall_topo}. The off-shell
algebra reads
\begin{eqnarray}
Q\phi&=&\psi\nonumber\\
Q\psi&=&0\nonumber\\
Q\psib&=&B\nonumber\\
QB&=&0
\end{eqnarray}
where we absorbed the anticommuting parameter $\epsilon$
into the variation $\delta$ and introduced a corresponding fermionic variation $Q$ which closely corresponds to the
exact supercharge of the canonical formalism. It is trivial
to verify that $Q^2=0$ now on all fields without use of the
equations of motion.
Using this field $B$ the bosonic action can now be rewritten as
\begin{equation}
S_B=\sum_x -B(D^-\phi+P^\prime)-\frac{1}{2}B^2
\end{equation}
The original action is recovered after integration over $B$. It
is easy to verify that the supersymmetry variation of the new 
action is still zero.

Furthermore, it is now trivial to show that
entire action is nothing but the
$Q$ variation of a particular function -- it is said to be $Q$-exact.
\begin{equation}
S_L=Q  \sum_x\psib(-D^-\phi-P^\prime-\frac{1}{2}B) 
\end{equation} 
In obtaining the action, one must treat the fermionic variation $Q$ as a 
Grassmann
and anticommute it through the other Grassmann fields. 

In this form the invariance of the lattice action is manifest -- it simply
relies on the nilpotent property of $Q$ and the fact that the
action is the $Q$-variation of something. This $Q$-exact
structure should remind one of BRST gauge fixing and it turns
out that this is not a coincidence -- theories with local Nicolai
maps and nilpotent supercharges such as the quantum
mechanics model discussed here can be obtained by a process
of gauge fixing. However, unlike the usual gauge fixing procedure
the gauge fixing employed in this context
arises during quantization of a bosonic
model with a classical topological shift symmetry. 

Consider a theory comprising a bosonic field $\phi$ living on a
lattice in Euclidean time. Take as classical action the trivial
function $S(\phi)=0$. To construct the quantum theory we
must integrate over the field $\phi$. But the classical theory
is invariant under the topological symmetry
\begin{equation}
x\to x+\epsilon
\end{equation}
where $\epsilon$ is an {\it arbitrary} smooth function.
Thus, to quantize this model we must fix a gauge. Let us
employ the gauge condition
${\cal N}(\phi)=0$. The correct quantum partition function is
then given by
\begin{equation}
Z=\int D\phi\; {\rm det}(\frac{\partial {\cal N}}{\partial \phi})
e^{-\frac{1}{2\alpha}{\cal N}^2(\phi)}
\end{equation}
where we have included the usual Fadeev-Popov determinant
and inserted an arbitrary gauge fixing parameter $\alpha$.
If we represent the determinant in terms of ghost fields $\psi$
and $\psib$ and
choose the specific gauge fixing function
${\cal N}=D^-\phi+P^\prime$ (with Feynman gauge parameter $\alpha=1$)
we recover our original quantum mechanics model!
In this case the nilpotent supersymmetry we uncovered
is nothing
more than the usual BRST symmetry arising from quantizing the
underlying topological symmetry and the Nicolai map is just the
gauge fixing function! Thus, we see in our simple
quantum mechanics example, that
the nilpotent twisted supersymmetry we have constructed by
taking linear combinations of the original supercharges has
an alternative interpretation as a BRST charge arising in
quantizing an underlying bosonic theory. 

In BRST gauge fixing we would then
go on to impose the physical state
condition that \beq
Q|{\rm physical\; state}>=0\eeq
To construct
a true topological quantum field theory this is what is done.

However, in the context of the lattice SUSY constructions
described in this Report this is {\it not} what is done.
Such a restriction would be equivalent to a projection to
the vacuum states of the target supersymmetric theory 
-- which is much too restrictive. Hence
we will {\it not} impose this condition here and 
instead merely use the
topologically twisted reformulations of these
supersymmetric theories as simply more convenient starting
points for constructing lattice actions which retain a degree
of supersymmetry.

As we have seen they simply correspond to
an exotic change of variables in the original theory -- one
that exposes a nilpotent supersymmetry explicitly.

\subsection{Semiclassical exactness}
\label{sec:6h}

The topological structure that we have exposed in this
quantum mechanics model, which is at the heart of our
ability to discretize it, has many important consequences. 
Consider a set of $Q$-invariant operators such as 
${\cal O}_1(x_1), \ldots, {\cal O}_n(x_n) $. 
Most importantly, the expectation values and connected correlators  
 of such operators  in  a theory with a $Q$-exact action 
$S=Q\Lambda$ may
be computed {\it exactly} in the semi-classical limit. It is easy to
see this -- replacing $S$ by $\mu S$ we can write down an expression
for the expectation value
\begin{equation}
\langle{\cal O} \rangle_{\mu}= \frac{1}{Z} \int {\cal O} \; e^{-\mu S}
\end{equation}
Differentiating this expression with respect to $\mu$ leads to
\begin{equation}
\frac{\partial}{\partial \mu} \langle{\cal O}\rangle_{\mu}=  -  \langle {\cal O} {Q \Lambda} \rangle_{\mu} +  \langle {\cal O} \rangle_{\mu}  \langle{Q \Lambda} \rangle_{\mu} = 0 
\end{equation}
where the equality just follows from recognizing that ${\cal O}Q\Lambda=Q({\cal
O}\Lambda)$ which can be recognized
as a supersymmetric Ward identity.
Thus expectation values of $Q$-invariant observables
are independent of $\mu$ {\it in the absence of spontaneous breaking of
the $Q$-symmetry} and hence can be computed exactly
in the semiclassical limit $\mu\to\infty$. In this limit we need
only do 1-loop calculations around the classical vacua\footnote{This is the
basis of the argument used by Matsuura to show that the vacuum energy
of certain supersymmetric orbifold theories remains zero to all orders
in the coupling \citep{Matsuura:2007ec}}.
Generalizations of
this argument allow one to show that $Q$-invariant observables
in the continuum twisted theories
are independent of any background metric and, in the case
where they are not $Q$-exact, possess expectation values that
correspond to topological invariants of the background space.

\subsection{Supersymmetry breaking}
\label{sec:6i}

In this section we discuss, in the context of our lattice
quantum mechanics model, a mechanism by which $Q$-supersymmetry
may be spontaneously broken. This offers a prototype for the kind
of dynamical susy breaking 
mechanism we would ultimately like to examine in lattice
simulations of more realistic models.

The partition function for the system with periodic boundary  conditions on all fields is a topological invariant called the
Witten index. We may evaluate it easily from the Nicolai map
formulation. We may deform the map
\begin{equation}
{\cal N}(\phi)=\Delta^-\phi+P^\prime(\phi)
\end{equation}
in such a way as keep only the highest power of $\phi$ in the
potential ${\cal N}(\phi)\to \phi^n$. Consider the expression for $Z$
in this limit (here $S$ contains both
the bosonic action $S_B$ and the effective interaction coming from
the fermions $\ln{\rm det}(\frac{\partial{\cal N}}{\partial\phi})$) 
\begin{equation}
Z=\int_{-\infty}^{\infty} D\phi e^{-S}=\int_{-\infty}^0D\phi e^{-S}+\int_0^\infty D\phi e^{-S}
\end{equation}
In the case where $n$ is odd the map $\phi\to\cN=\phi^n$ is single
valued and the previous expression is just equivalent to 
\begin{equation}
Z=\int_{-\infty}^{\infty} D{\cal N}e^{-{\cal N}^2}
\end{equation}
But for $n$ even the map is not single valued and the limits 
on the first integral change leading to the result
\begin{equation}
Z=\int_{\infty}^0 D{\cal N}e^{-{\cal N}^2}+\int_0^{\infty}D{\cal N}e^{-{\cal N}^2}=0
\end{equation}
Thus the Witten index is zero for superpotentials where
the highest power of $\phi$ in $P^\prime(\phi)$ is even. 
A non-zero Witten index implies that supersymmetry is unbroken. 
A vanishing Witten index allows for supersymmetry breaking.
To see this recall that the Witten index is simply the difference
between the number of fermionic and bosonic vacua. If this
is non-zero supersymmetry {\it cannot} break since any vacuum
state that is lifted to positive energy necessarily occurs with
a superpartner state of opposite statistics which is not possible
if $W$ is non-zero. However, if $W=0$ supersymmetry breaking
can (and often does) occur. 

However, powerful non-renormalization theorems
guarantee that supersymmetry cannot break
in any finite order of perturbation theory \citep{Weinberg}. 
If it is to occur
it must proceed through a non-perturbative mechanism
- for example instantons. In the quantum mechanics model
these correspond to non-trivial field configurations
satisfying
\begin{equation}
\frac{d\phi}{dt}+P^\prime(\phi)=0
\end{equation}
Such non-trivial field configurations can occur
when the asymptotic values of $\phi$ tend to two different
classical vacua as $t\to \pm\infty$. The instantons are then
just
kink solutions.

When $P^\prime(\phi)=0$ has only one solution they cannot
occur but it is easy to construct examples where there are
two solutions eg. $P^\prime(\phi)=(\phi^2-a^2)$ which
clearly corresponds to a theory with vanishing Witten index $W=0$. 
The instanton solution is then
\begin{equation}
\phi=a\tanh{\frac{1}{2}(t-c )}
\end{equation} 
where $c$ is an arbitrary constant corresponding to the
center of the instanton. For the bosonic action based on
${\cal N}^2$ the action of this configuration is zero. Furthermore,
associated with the center coordinate $c$ is an exact bosonic
zero mode of the form
\begin{equation}
\phi_0=a{\rm sech}^2{\frac{1}{2}(t-c)}
\end{equation}
which is exponentially localized on the instanton. Supersymmetry
then dictates that there is a corresponding fermionic
zero mode. One can understand the vanishing of
$W$ as a consequence of the
Grassmann integration over this
fermionic zero mode.
Indeed, to get nonzero expectation values we need observables
that absorb this zero mode. They are of the form
\begin{equation}
{\cal O}=f^\prime(\phi)\psi
\end{equation}
But notice that this observable is itself the $Q$-variation of
something. Hence, if instantons condense in the
vacuum of this theory we will find $<Q{\cal O}>\ne 0$ signaling 
supersymmetry breaking.

It is instructive to see how one might see this breaking within
a Monte Carlo simulation.
On the lattice an isolated instanton cannot be realized because of
the periodic boundary conditions on the bosonic field. The lowest
energy configuration then consists of an instanton anti-instanton
pair. If the pair are widely separated this is an approximate
solution of the classical equations of motion with a
classical action $S_{\rm I\overline{I}}=\frac{4}{3}a^3$
Associated with this configuration there will be a low lying, localized
{\it and hence normalizable} fermion mode which is a superpartner to the approximate
bosonic zero mode corresponding to motion of the instanton center. In the thermodynamic limit
this mode can induce supersymmetry breaking. The cleanest way to see this is to
consider the system at non-zero temperature by employing antiperiodic boundary conditions
for the fermions. These boundary conditions break supersymmetry explicitly and the
question of whether supersymmetry
breaks spontaneously is then reduced
to a computation of the expectation value of the energy 
(written in $Q$-exact form) in the thermodynamic limit as the temperature
is sent to zero.  
Numerical simulations of this theory indicate that the energy, extrapolated to
first to
zero lattice spacing, and subsequently to zero temperature, is indeed
non-zero \citep{Kanamori:2007yx,Kanamori:2007ye}.

\subsection{Generalizations}
\label{sec:6j}

\subsubsection{Supersymmetric Yang Mills Quantum Mechanics}
\label{sec:6j.1}

These models arise as dimensional reductions of
${\cal N}=1$ SYM theory in $d=4,6,10$ dimensions down to
one (Euclidean) dimension.
The $d=10$ theory with $\cQ=16$ supercharges and
$N$ colors
is especially interesting as it 
is conjectured to be dual to type IIa string theory containing $N$
D0-branes. 
The type IIA string theory reduces to a supergravity theory for  
low energies
compared to the string scale $(\alpha^\prime)^{-1/2}$. In this limit
the thermal theory contains black holes with $N$ units of D0-charge. Their energy, $E$, is a function of their Hawking temperature, $T$.
Defining $\lambda=N g_s\alpha^{\prime -3/2}$ where $g_s$ is the
string coupling, we may write a dimensionless energy and temperature $ 
\epsilon = E \lambda^{-1/3}$ and $t = T \lambda^{-1/3}$. One finds  
provided we take $N$ large and $t \ll 1$ the black hole is weakly  
curved on string scales and the quantum string corrections are  
suppressed. The energy of this black hole can
be precisely computed by standard methods \citep{Itzhaki:1998dd} giving,
\begin{eqnarray}
\label{blackholeE}
\epsilon = c\; N^2 t^{14/5} \qquad c = \left( \frac{2^{21} 3^{12} 5^
{2}}{7^{19}} \pi^{14} \right)^{1/5} \simeq 7.41 .
\end{eqnarray}
Duality posits that the thermodynamics of this black hole
should be reproduced by the dual Yang-Mills quantum mechanics
at the same temperature with $g_s\alpha^{\prime -3/2} = g_{YM}^2$
so that $\lambda$ is to be identified with the 't Hooft coupling.

A one dimensional  lattice action with $\CQ=8$ exact supersymmetries 
has been constructed for the target $\CQ=16$ 
matrix quantum mechanics \citep{Kaplan:2005ta}.  
However, as our previous analysis
shows, the renormalization of supersymmetric quantum
mechanics is sufficiently simple that a naive latticization may also 
be employed. The only diagram
exhibiting a superficial degree of divergence $D\ge 0$ corresponds
to the one loop fermion contribution to the tadpole diagram for
the field $\psi$ which has $d=0$.  However an easy calculation
shows that the log divergent part of
this amplitude contains a factor
\begin{equation}
f_{abc}\delta_{ab}
\end{equation} 
which vanishes on account of the antisymmetry of the
structure constants $f_{abc}$. Thus {\it any} naive discretization
of this theory (in which there are no doubles) will flow
automatically to the supersymmetric theory in the continuum
limit as was explicitly shown in \citep{Catterall:2007fp}. 

A suitable lattice action takes the form $S=S_B-\ln{\rm Pf(\mathcal{O})}$
where
\begin{equation}
S_{B}=\frac{NL^3}{\lambda R^3}  \sum_{a=0}^{L-1} \mathrm
{Tr}\left[ \frac{1}{2}  \left( D_- X_i \right) - [ X_{i,a}, X_{j,a} ]
^2  \right]
\end{equation}
and the fermion operator $\mathcal{O}$ is defined by
\begin{equation}
\mathcal{O}_{ab}=\left( \begin{matrix} 0 & (D_-)_{ab} \\ -(D_-)_
{ba} & 0 \end{matrix} \right) - \gamma^i \left[ X_{i,a}, \cdot
\right]  \mathrm{Id}_{ab}
\end{equation}
and we have rescaled the continuum fields $X_{i,a}$ and $\Psi_{i,
\alpha}$
by powers of the lattice spacing $R/L$ to render them dimensionless
($L$ is the number of lattice points).
The covariant derivatives are given by
$(D_- W_i)_a = W_{i,a} - U_{a-1}^{\dagger}W_{i,a-1} U_{a-1}$
and we have introduced a Wilson gauge link field $U_a$. Notice that  
the fermionic operator is
free of doublers and is manifestly antisymmetric in this basis.

\begin{figure}
\begin{center}
\resizebox{9.0cm}{!}{\includegraphics{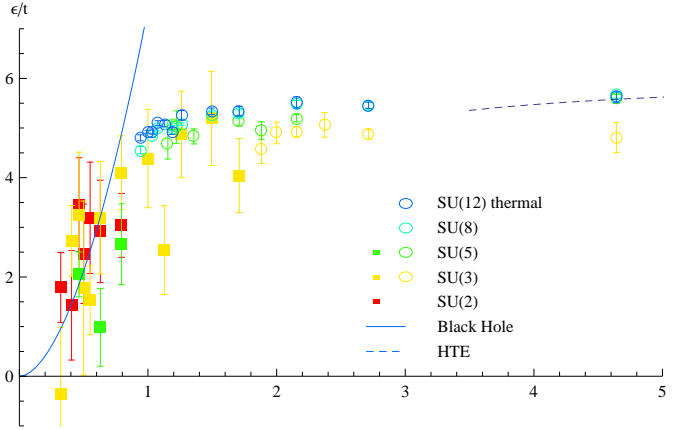}}
\caption{Mean energy vs temperature for SYMQM (points). Solid line 
corresponds to the prediction from the black hole solution in supergravity}
\label{black}
\end{center}
\end{figure}

Studies of the 16 supercharge
model using this action have have been initiated in 
\citep{Catterall:2007me,Catterall:2008yz}. Work using a gauge fixed
approach momentum space approach have been reported in \citep{Anagnostopoulos:2007fw,Hanada:2008ez,Hanada:2008gy,Nishimura:2008ta,
Kawahara:2007ib,Hanada:2007ti,Nishimura:2008zz}. The results from the two
methods are in agreement and lend strong support to the conjectured
duality of strong coupling Yang-Mills quantum mechanics and
type II string theory in $\rm AdS$ space. A comparison between the Yang-Mills
system and its gravity dual is shown in Fig.~\ref{black}

The plots shows the mean Yang-Mills energy (in units of $\lambda^{\frac{1}{3}}$)
versus dimensionless
temperature $t=T/\lambda^{\frac{1}{3}}$ for a number of colors $N$ in
the range $N=3-12$
obtained from a Monte Carlo simulation of the model
\citep{Catterall:2008yz}. The solid line shows the
corresponding result obtained for the charge $N$ black hole solution of
type IIa supergravity as given in earlier in eq.~(\ref{blackholeE}).
The dashed line gives the high temperature result.
It should be noted that the agreement seen in the figure was
obtained without any fitting of the data.

\subsubsection{Sigma models}
\label{sec:6j.2}

The simple quantum mechanics model may be generalized by
replacing the field $\phi$ by a set of fields $\phi^i$
to be considered as coordinates
in some target space with metric $g_{ij}$.
The fermions and auxiliary field $B$ also pick up
target space indices and the nilpotent symmetry is
naturally generalized to \citep{Catterall_sig1}
\begin{eqnarray}
Q\phi^i&=&\psi^i\nonumber\\
Q\psi^i&=&0\nonumber\\
Q\psib_i&=&B_i\psib_j\Gamma^j_{ik}\psi_k\nonumber\\
QB_i&=&B_j\Gamma^j_{ik}\psi_k-\frac{1}{2}\psib_jR^j_{ilk}\psi_l\psi_k
\end{eqnarray}
The quantities $\Gamma^i_{jk}$ and $R^j_{jkl}$ are the usual
Riemann connection and curvature.  It is straightforward
to show that $Q^2=0$ on all fields.
As before the supersymmetric lattice action takes the form
\begin{equation}
S=\beta Q \sum_x\psib_i\left({\cal N}^i-\frac{1}{2}g^{ij}B_j\right)
\end{equation}
Carrying out the variation and integrating over the field $B$
yields
\begin{equation}
S=\beta\sum_x \left(\frac{1}{2}{\cal N}^i{\cal N}_i-\psib_i\nabla_k N^i\psi_k+\frac{1}{4}R_{jklm}\psib^j\psib^l\psi^m\psi^k\right)
\end{equation}
which is invariant under the on-shell SUSY
\begin{eqnarray}
Q\phi^i&=&\psi^i\nonumber\\
Q\psi^i&=&0\nonumber\\
Q\psib_i&=&{\cal N}_i-\psib_j\Gamma^j_{ik}\psi_k
\end{eqnarray}
The curved space lattice Nicolai map is simply
\begin{equation}
{\cal N}^i=\Delta^+\phi^i
\end{equation}
which allows us to rewrite the fermion kinetic term in the form
\begin{equation}
\psib_i\left(\Delta^+\psi^i+
\Gamma^i_{kj}\Delta^+\phi^k\psi^j\right)
\end{equation}
Notice that while the lattice action is now manifestly supersymmetric (and free of doubles) it is no longer general coordinate invariant  in the
target space -- the term $\Delta^+\phi^k$ is {\it not}
a target space vector unlike its continuum cousin involving
derivatives. However, one  might hope that this symmetry
along with the other supersymmetry 
is restored in the continuum limit (large $\beta$) where the
fields vary slowly over the lattice. There is some
evidence for this in the analogous two dimensional
sigma models \citep{Catterall_sig2}. It would be very interesting to
investigate this in more detail in this simpler one dimensional
case.
 
\subsection{Summary}
\label{sec:6k}

There are several conclusions we can draw from the discussion
of these quantum
mechanical models. The first is
that low dimension supersymmetric theories can sometimes be handled
using discretizations that break all the continuum supersymmetries --
the super-renormalizable nature of the theories ensures that only
a finite number of
dangerous Feynman diagrams exist and they
occur in low orders of perturbation theory. These diagrams can be
computed using lattice perturbation theory and counter-terms 
can be used to  subtract off the susy breaking effects. 

This has been verified in great detail in the work of the Jena group
who have conducted
high precision simulations
of the non-gauge quantum mechanics theory
with very convincing results \citep{Bergner:2007pu}. But we have seen that
things are even 
easier with exact supersymmetry -- linear combinations of the
original supersymmetries can be found
which are nilpotent and a lattice
action constructed that is exactly invariant under one of these
new supersymmetries. This is the simplest example of the more general
twisting procedure discussed in the introduction
to this Report. In one dimension there is
no notion of Lorentz symmetry and the procedure reduces to the simple
observation that from the two real supersymmetries obeying $Q_1^2=Q_2^2=H$
one can construct two complex conjugate symmetries $Q=Q_1+ iQ_2, 
{\overline Q} = Q_1- iQ_2, $
which are nilpotent. One of these can then be chosen to
survive the transition to
the lattice.

We now turn to the simplest quantum field theory where a similar
construction is possible and where Lorentz symmetry plays a less trivial
role -- the two dimensional Wess-Zumino model.

\section{Two dimensional theories without gauge symmetry}
\label{sec:7}

\subsection{Wess-Zumino Model}
\label{sec:7a}

Having discussed the case of quantum mechanics the next task
is to see how to generalize these ideas to field theory. The simplest
place to start is in two dimensions and perhaps the simplest
example of a  lattice theory which exhibits an exact supersymmetry is 
gotten by lifting the (non-gauge) quantum
mechanics to two dimensions. This will lead to a Wess-Zumino
model
\citep{Catterall_wz1,Catterall_wz2,Giedt_wz,Giedt:2004qs,
Beccaria_break,Fujikawa_wz,Kastner:2008zc}.  (An alternative is to fine-tune a finite number of counter-terms as in the 2+1 dimensional Wess-Zumino model \citep{Lee:2006if}.)
We saw that for quantum mechanics a minimum of
two supersymmetries was necessary to build an exact
lattice supersymmetry so we are led to consider the ${\cal N}=2$
Wess-Zumino model with continuum action
\begin{eqnarray}
S_{\rm WZ}&=&\int d^2x\, \partial_\mu\phi\partial_\mu\phib+
W^\prime(\phi)W^\prime(\phib)+
\psib\gamma_\mu\partial_\mu\psi+\nonumber\\
&+&\psib\left(\frac{1}{2}\left(1+\gamma_5\right)\Wpp(\phi)+
\frac{1}{2}\left(1-\gamma_5\right)\Wpp(\phib)\right)\psi
\label{cont_wzaction}
\end{eqnarray}
which contains a complex scalar field $\phi$ 
with analytic
superpotential $W(\phi)$, coupled to a Dirac fermion
$\psi$. It will turn out to be easier to rewrite the fermion
action. Choosing a chiral basis
\begin{equation}
\begin{array}{cc}
\gamma_1=\left(\begin{array}{cc}
0&1\\
1&0\end{array}\right)
&
\gamma_2=\left(\begin{array}{cc}
0&-i\\
i&0\end{array}\right)
\\
\end{array}
\end{equation}
allows us to rewrite the fermion operator as
\begin{equation}
M_{\rm F}=\left(\begin{array}{cc}
i\Wpp(\phib)&\partial_{\overline{z}}\\
\partial_z&-i\Wpp(\phi)
\end{array}\right)
\end{equation}
To build a lattice action which preserves supersymmetry we
will require a local Nicolai map as for quantum mechanics
\citep{Sakai:1983dg}. This
means that the fermion operator should be realizable as
a Jacobian representing the change of variables
from $\phi\to {\cal N}(\phi)$. Analogy with one dimension 
together with the above form of the fermion
operator suggests the form
\begin{equation}
{\cal N}(\phi)=\partial_{\overline{z}}\phi+i\Wp(\phib)
\end{equation}
Indeed in the continuum the bosonic action derived from
${\cal N}\overline{\cal N}$ differs from the continuum one given by eq.~(\ref{cont_wzaction}) 
by a cross term
$\int dzd\overline{z} \partial_z\phib\Wp(\phib)+{\rm h.c}$, similar to that
encountered in supersymmetric quantum mechanics. Again in the continuum
this term is a total derivative and can be discarded.
Discretizing this Nicolai map then leads to
a lattice action $S_L$ which is invariant under the following
supersymmetry \citep{Catterall_wz1,Sakai:1983dg,Giedt_wz}
\begin{equation}
S_L=\sum_x {\cal N}\overline{\cal N}+\omegab\left(
D^s_{\overline{z}}\lambda+
i\Wpp_L(\phib)\omega\right)+
\lambdab\left(D^s_z\omega-i\Wpp_L(\phib)\lambda\right)
\end{equation}
where
\begin{eqnarray}
Q\phi&=&\lambda\nonumber\\
Q\phib&=&\omega\nonumber\\
Q\lambda&=&0\nonumber\\
Q\omega&=&0\nonumber\\
Q\omegab&=&\overline{{\cal N}}\nonumber\\
Q\lambdab&=&{\cal N}
\end{eqnarray}
where $\psi=\left(\begin{array}{c}\omega\\ \lambda\end{array}\right)$,
$\psib=\left(\begin{array}{c}\omegab \\ \lambdab\end{array}\right)$.
In this expression we have replaced the
continuum derivative by a {\it symmetric} difference
operator. To remove the would be doublers from this
expression one can add a Wilson mass term to the
superpotential
\begin{equation}
\Wp_L(\phi)=\Wp(\phi)+\frac{1}{2}D^+_zD^-_{\overline{z}}\phi
\end{equation} 
Again it is easy to see that $Q^2=0$ using the equations of
motion. By introducing fields $B,\overline{B}$ we can again
write the action in $Q$-exact form and
render the symmetry $Q$ nilpotent off-shell \citep{Catterall_topo}.
\begin{equation}
S_L=Q\sum_x \left[  \omegab\left({\cal N}+\frac{1}{2}B\right)+
                       \lambdab\left(\overline{\cal N}+\frac{1}{2}\overline{B}\right) \right]
\end{equation}
where the action of the supersymmetry $Q$ contains the new
elements
\begin{eqnarray}
Q\omegab&=&\overline{B}\nonumber\\
Q\lambdab&=&B\nonumber\\
QB&=&0\nonumber\\
Q\overline{B}&=&0
\end{eqnarray} 
The existence of an exact supersymmetry results 
in a set of exact lattice Ward identities
constraining the form of the quantum theory.
Perhaps the simplest of these identities is given by 
$<S_L>=<Q\Lambda>=0 $ leading to
a prediction for the mean bosonic action for {\it any} superpotential. 
This is illustrated in table~\ref{qm_act} which shows the
expectation value of the bosonic action as a function of the number
of lattice sites.
\begin{table}
\begin{center}
\begin{tabular}{||c|c|c||}
\hline
L & $<S_B>$ & $\frac{1}{2}N_{dof}$\\\hline
4 & 31.93(6) & 32\\\hline
8 & 127.97(7) & 128\\\hline
16 & 512.0(3) & 512\\\hline
32 & 2046(3) & 2048\\\hline
\end{tabular}
\caption{Mean bosonic action $<S_B>$ versus exact SUSY value, equal to half the number of degrees of freedom.}
\label{qm_act}
\end{center}
\end{table}  

\begin{figure}
\begin{center}
\resizebox{9.0cm}{!}{\includegraphics{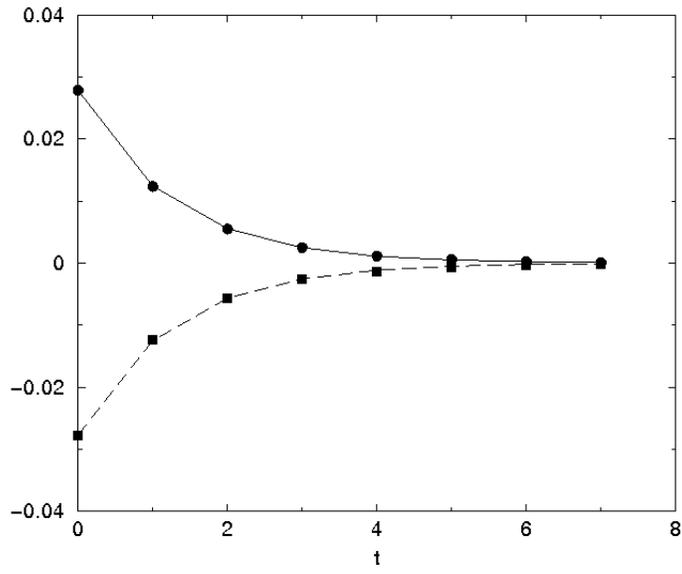}}
\caption{Real parts of bosonic $<\phi_x{\cal N}_y>$ and fermionic
contributions $<\lambda_x\lambdab_y>$ to $Q$-ward identity for
Wess-Zumino model with superpotential parameters $mL=10.0$
and $gL=3.0$}\label{wz_ward}
\end{center}
\end{figure}

Again, the restoration of full supersymmetry
occurs without fine tuning as a consequence of exact supersymmetry
\citep{Catterall_wz1,Giedt_wz,Kastner:2008zc,Bergner:2007pu,Wozar:2008jb}.
Figure~\ref{wz_ward} shows the bosonic
and fermionic contributions to a particular $Q$-supersymmetric
Ward identity resulting from
the $Q$-variation of the operator $O=\phi_x\lambdab_y$ in
the theory with $W(\phi)=m\phi+g\phi^2$. The parameters
correspond to $mL=10.0$
and $gL=3.0$ on a lattice with $L=8\times 8$ sites. Clearly the
two curves add to zero as would be expected for a theory
which realizes exact supersymmetry.

\subsection{Sigma models}
\label{sec:7b}

The one dimensional sigma  model may be lifted to two dimensions
if sufficient constraints are placed on the target space. In practice
this means that the manifold should be K\"{a}hler. In the case of
a curved target space we must add the Wilson term in the form
of a {\it twisted} mass term rather than a superpotential.  
We refer to \citep{Catterall_sig2} for details. This
breaks the $Q$ symmetry softly although numerical results
still favor the restoration of full supersymmetry without fine
tuning in the continuum limit
\citep{Catterall_sig2}.

\subsection{Twisting in two dimensions}
\label{sec:7c}

The process of exposing a linear combination of supercharges
that is nilpotent can be understood in a systematic way. This will furnish an explicit example of the
twisting process discussed in the introduction.

Consider a two dimensional theory with 4 (real) supercharges 
(${\cal N}=2$ supersymmetry). Such a theory contains 2 
degenerate Majorana spinors which transform into each other
under an internal $SO(2)_I$ symmetry. The spinors also transform under
an independent $SO(2)_E$ (Euclidean)
Lorentz symmetry. Following on from our general
discussion of twisting in the introduction it is
possible to decompose the fields of the theory under the
{\it diagonal subgroup} corresponding to making equal rotations
in the base and internal spaces
\begin{equation}
SO(2)^\prime={\rm Diag}\left(SO(2)_I\times SO(2)_L\right)
\end{equation}
In practice this means that the supercharges
$q_\alpha^i$ ($i$ corresponds to internal space, 
$\alpha$ to rotations) are to be treated as a $2\times 2$ matrix
\citep{D'Adda_super,Kato_bf}. It is
then natural to expand the
supercharge matrix on products of two
dimensional Dirac gamma matrices  (Pauli matrices) ($\mu=1,2$)
\begin{equation}
q=QI+Q_\mu\gamma_\mu+
Q_{12}\gamma_1\gamma_2
\end{equation} 
In this process we see that all the supercharges are decomposed
in terms of geometrical quantities -- scalars, vectors and
antisymmetric tensors. 
Furthermore the original supersymmetry algebra
becomes
\begin{eqnarray}
\{Q,Q\}&=&\{Q_{12},Q_{12}\}=\{Q,Q_{12}\}=
\{Q_\mu,Q_\nu\}=0\nonumber\\
\{Q,Q_\mu\}&=&p_\mu\nonumber\\ \{Q_{12},Q_\mu\}&=&\epsilon_{\mu\nu}p_\nu
\end{eqnarray}
showing that indeed the scalar component is nilpotent as
required. Actually we see that the momentum is $Q$-exact
which makes it plausible that the entire energy-momentum
tensor is nilpotent. Thus it should be no surprise that 
supersymmetric theories reformulated in this {\it twisted}
basis often have $Q$-exact actions 
as we have already seen \citep{Catterall:2005eh}\footnote{ Two known exceptions are the $(\CQ=8, d=3)$ and 
 $(\CQ=16, d=4)$ target SYM theories. These theories,  in their twisted and lattice regularized forms,  
contain a $Q$-closed operator in their action \citep{Cohen:2003qw, Kaplan:2005ta}.}.

\subsection{\KD fermions in two dimensions}
\label{sec:7d}

If the supercharges are tensorial we would expect the
same to be true for the fermions themselves.
Taking the standard free fermion action for a theory with 
2 degenerate Majorana species and replacing the fermions
$\psi_\alpha^i$ by matrices we find that it can be
trivially rewritten as \citep{D'Adda_super, Catterall_n=2}
\begin{equation}
S_{\rm F}={\rm Tr}\left[\Psi^\dagger \gamma_\mu\partial_\mu \Psi\right]
\end{equation}
Expanding the matrices into (real) components $(\frac{1}{2}\eta,\psi_\mu,\chi_{12})$ and doing the trace yields
\begin{equation}
S_{\rm F}=\frac{1}{2}\eta \partial_\mu\psi_\mu+\chi_{12}\left(\partial_1\psi_2-\partial_2\psi_1\right)
\end{equation}
This geometrical rewriting of the fermionic action yields the so-called
K\"{a}hler-Dirac action which is most naturally rewritten using the
language of differential forms as
\begin{equation}
\Psi .\left(d-d^\dagger\right)\Psi
\end{equation}
where the \KD field $\Psi$ is now just the set of
components $(\frac{1}{2}\eta,\psi_\mu,\chi_{12})$ and
$d$ is the exterior derivative whose action on general
rank $p$ antisymmetric tensors (forms) $\omega_{\left[\mu_1\ldots\mu_p\right]}$ yields a rank $p+1$
tensor with components $\omega_{\left[\mu_1\ldots\mu_p\mu_{p+1}\right]}$ and the
square bracket notation indicates complete antisymmetrization
between all indices. The dot notation just indicates that corresponding tensor components are multiplied and integrated
over space. The operator $d^\dagger$ is the corresponding
adjoint operator mapping rank $p$ tensors to rank $p-1$.

This recasting of the action in geometrical terms not only yields
a nilpotent supersymmetry but allows us to discretize the action
{\it without} inducing fermion doubles \citep{Rabin:1981qj}. The prescription is simple. 
\begin{itemize}
\item{Replace a continuum derivative by a {\it forward difference}
operator if it derives from the exterior derivative, a curl-like
operation on the component fields. } 
\item{Replace a continuum derivative by the {\it backward} difference
if it comes from the adjoint operator $d^\dagger$, which implements a divergence-like operation on the component fields.}
\end{itemize}
We can see this explicitly if we take the preceding fermion
action and write it in the form
\begin{equation}
(\frac{1}{2}\eta,\chi_{12}).\left(\begin{array}{cc}
\Delta^-_1&\Delta^-_2\\
\Delta^+_2&-\Delta^+_1
\end{array}\right)\left(\begin{array}{c}
\psi_1\\
\psi_2
\end{array}\right)
\end{equation}
It is clear that the determinant of the matrix operator is equal to
the usual double free bosonic determinant precluding the
existence of additional zeros of the fermion operator.
Thus, discretizations based on \KD fermions
do not require additional ad hoc Wilson mass terms to be added.
This  turns out to be particularly useful for models with massless fermions
such as the extended supersymmetric Yang Mills theories we consider later.

Finally, notice that the action for free \KD fermions
can be mapped into the action for staggered quarks.
In two dimensions simply introduce a lattice with half the
original lattice spacing. Place the link fields $\psi_1(x)$, $\psi_2(x)$
and the plaquette field $\chi_{12}(x)$ on the sites of this new lattice
with the scalar $\eta(x)$ being placed at the original site $x$.
It is straightforward to verify that all the backward and forward
differences now become symmetric differences acting now
on these new site fields on the doubled lattice. The usual
staggered phases arise as a consequence of the antisymmetrization of
the derivatives.

Indeed this latter construction offers yet another way to
see that \KD fields have no doubling -- they
are equivalent at the free field level (all that matters for
doubling) to staggered fermions. However, unlike the
usual situation in QCD the supersymmetric theories we
are studying automatically contain the correct number
of degenerate fermion flavors represented by the full
staggered fermion determinant and the usual
rooting problem that plagues staggered fermion formulations
of QCD is avoided.

Finally, it is possible to recast our previous Wess Zumino construction in the language of these \KD fields. Consider just
the kinetic term and let
\begin{eqnarray}
\omega&=&\frac{1}{2}\eta+i\chi_{12}\nonumber\\
\lambda&=&\psi_1+i\psi_2
\end{eqnarray}
Then the expression
\begin{equation}
\omega^\dagger D_{\overline{z}}\lambda
\end{equation}
yields the previous \KD action in two dimensions. The scalar
field and its complex conjugate are formed from the bosonic
partners $(C,A_\mu,B_{12})$ in the same way.

Actually the appearance of these complex field combinations 
can be understood from within a \KD perspective. From a given \KD
field $\Psi=(\frac{1}{2}\eta,\psi_\mu,\chi_{12})$ we can 
construct a dual field $\tilde{\Psi}$ with
dual tensor components. This duality operation
takes
a rank $p$ tensor and replaces it by a $(d-p)$ rank tensor, and is a realization of 
 Hodge duality on the lattice.  
Schematically
\begin{equation}
f_p\stackrel{*}{\to} f_{d-p}
\end{equation} 
where the dual components are given by
\begin{equation}
\tilde{f}_{\mu_1\ldots\mu_{d-p}}=\epsilon_{\mu_1\ldots\mu_d}
f_{\mu_{d-p+1}\ldots\mu_d}
\end{equation}
Notice that two applications of the duality operation yields {\it minus} the identity. Thus a  projector onto self-dual \KD fields
would take the form
\begin{equation}
P^+=\frac{1}{2}\left(I+i*\right)
\end{equation}
with a corresponding projector equipped with a minus sign for
projection on anti-self dual fields.
If we decompose the original \KD field on its self-dual and
anti-self-dual parts we can verify that the coontinuum \KD action 
separates into two independent parts. This allows us to
restrict attention to say the self-dual component. This is what
happens in the Wess-Zumino and sigma model cases.

\section{Two dimensional gauge field theories -- twisted ${\cal N}= (2,2)$ SYM }
\label{sec:8}

\subsection{Continuum formulation}
\label{sec:8a}

In section~\ref{sec:7c} we argued that 
any two dimensional supersymmetric theory with four (or integer multiples of four) supercharges can be reformulated in twisted variables. 
Furthermore we have an uncovered explicit
examples of this in the case of the $\cN=2$ Wess-Zumino model
and sigma models. 
However, while the existence of a twisted formulation
of a given continuum field theory 
is certainly a necessary condition for constructing a lattice
model with exact supersymmetry, it does not guarantee that one
exists. In general it is necessary for any lattice theory to
satisfy additional constraints. Perhaps the most important
of these is seen when we try to implement
the procedure for gauge theories. In this case
the requirement that the discretization procedure
maintaining exact gauge introduces additional difficulties
which we have not considered up to this point. In this section we
examine this issue in some detail concentrating on perhaps the
simplest canonical case of

${\cal N}=(2,2)$ SYM target theory in $d=2$ 
 dimensions.\footnote{In two dimensions supercharges can be
specified as ``left-handed'' or ``right-handed'', and this theory has
two of each, so it is often called $(2,2)$ SYM.} The first Euclidean lattice formulation for this theory was given in Ref.\citep{Cohen:2003xe} 
using the orbifold approach, which will be discussed in \S \ref{sec:9b}.  
Lattice formulations based on the concept of twisting
were then proposed in \citep{Sugino_sym1,Catterall_n=2}. These latter
approaches start from a particular twist of the continuum
theory given by the action
\beq
S=\beta Q{\rm Tr}\int
d^2x\left(
\frac{1}{4}\eta[\phi,\phib]+2\chi_{12}F_{12}+\chi_{12}B_{12}+\psi_\mu D_\mu \phib\right)
\label{gfermion}
\eeq
Here all fields $f(x)$
are in the adjoint representation of $SU(N)$ with
$f(x)=\sum_{a=1}^{N^2-1} f^a(x)T^a$ with {\it antihermitian}
generators $T^a$ satisfying ${\rm Tr}T^aT^b=-\delta^{ab}$ \footnote{
In \S \ref{sec:8} and \S \ref{sec:8d}, we use anti-hermitian generators with ${\rm Tr}T^aT^b=-\delta^{ab}$. In the rest, our Lie algebra generators and their normalization convention  is 
${\rm Tr}T^aT^b=+ \delta^{ab}$.}.
The covariant derivatives act as
\begin{equation}
D_\mu f=\partial_\mu f+[A_\mu,f]
\end{equation}
while the action of $Q$ on the twisted fields is given by
\begin{eqnarray}
QA_\mu&=&\psi_{\mu}\nonumber\\
Q\psi_\mu&=&-D_\mu\phi\nonumber\\
Q\phib&=&\eta\nonumber\\
Q\eta&=&[\phi,\phib]\nonumber\\
QB_{12}&=&[\phi,\chi_{12}]\nonumber\\
Q\chi_{12}&=&B_{12}\nonumber\\
Q\phi&=&0
\end{eqnarray}
Notice that $Q^2=\delta^\phi$ an infinitesimal gauge transformation on the fields.
Carrying out the $Q$-variation on eq.~(\ref{gfermion})
and subsequently integrating over the field $B_{12}$ leads to the
action
\begin{eqnarray}
S&=&\beta {\rm Tr}\int d^2x\left(
\frac{1}{4}[\phi,\phib]^2-\frac{1}{4}\eta [\phi,\eta]-F_{12}^2-D_\mu \phi D_\mu \phib \right. \nonumber\\
&-&\left.\chi_{12} [\phi,\chi_{12}]-
2\chi_{12}\left(D_1\psi_2-D_2\psi_1\right)-\psi_\mu D_\mu\eta+\psi_\mu [\phib,\psi_\mu]\right)
\label{twist_sym_action}
\end{eqnarray}
The bosonic sector of this action is precisely the usual Yang-Mills
action while the fermionic sector constitutes, as expected,
a \KD
representation of the usual spinorial action \citep{Catterall:2005eh}.

It is worth pointing out that
the twisted theory possesses an additional $U(1)$ symmetry inherited from
the remaining $R$-symmetry of the model which is given by
\begin{eqnarray}
\psi_\mu\to e^{i\alpha}\psi_\mu\nonumber\\
\chi_{12}\to e^{-i\alpha}\chi_{12}\nonumber\\
\eta\to e^{-i\alpha}\eta\nonumber\\
\phi\to e^{2i\alpha}\phi\nonumber\\
\phib\to e^{-2i\alpha}\phib
\label{u1}
\end{eqnarray}

Two different discretization schemes have been
proposed to generate a lattice model from this
continuum theory \citep{Sugino_sym2,Sugino_sym2,Catterall_n=2} and preliminary
simulations have already been done
\citep{Catterall_sims,Catterall_rest,Kanamori:2007yx,Suzuki_sims}. However, the lattice formulation
described in \citep{Catterall_n=2} and \citep{Aratyn} suffers from 
a doubling of degrees of freedom with
respect to the continuum theory which has been discussed in
\citep{Damgaard:2007xi,Takimi:2007nn}. The lattice formulation
introduced by Sugino in \citep{Sugino_sym2} is also problematic
since the vacuum state turns out to be infinitely
degenerate at least in dimensions greater than two.

Both sets of problems can be evaded using a
alternative twist based on the strictly nilpotent supercharge
$Q+iQ_{12}$ introduced later \citep{Catterall:2007kn,Damgaard:2008pa}. 
As we will see
the resultant lattice actions then
reproduce {\it precisely} the corresponding
orbifold actions \citep{Cohen:2003xe},  which we discuss in \S \ref{sec:9b}. 
Before describing this alternative twist 
we show how to construct the action of the
additional (non-scalar) twisted supersymmetries on the twisted
fields. This is an important issue as it allows us to construct
Ward identities for all the broken supersymmetries in the
discretized theory. The question of whether the full supersymmetry
of the continuum target theory is
regained in the continuum limit 
can then be examined 
by examining whether these Ward identities are satisfied
as the lattice spacing is sent to zero. 
Preliminary work in this direction is reported in
\citep{Kanamori:2008bk, Catterall_rest}.

\subsection{Additional twisted supersymmetries}
\label{sec:8b}

It is straightforward to construct the 
additional twisted supersymmetry transformations of
the component fields \citep{Catterall_rest}. As we have described in
section~\ref{sec:7d} the fermion
kinetic term can be written in the matrix form
\beq
S_F=\int d^2x{\rm Tr} \Psi^\dagger \gamma .D\Psi\eeq
where $\Psi$
corresponds to the
matrix form of the \KD field
\beq	
\Psi=\frac{\eta}{2}I+\psi_\mu\gamma_\mu+\chi_{12}\gamma_1\gamma_2\eeq
This term is clearly invariant under $\Psi\to \Psi\Gamma^i, i=1\ldots 4$ and
$\Gamma^i$ corresponds to one of the set $(I,\gamma_1,\gamma_2,\gamma_1\gamma_2)$
Consider first the case $\Gamma^4=\gamma_1\gamma_2$.
In terms of the component fields the transformation $\Psi\to\Psi\Gamma^4$
effects a duality map
\begin{eqnarray}
\frac{\eta}{2}&\to& -\chi_{12}\nonumber\\
\chi_{12}&\to& \frac{\eta}{2}\nonumber\\
\psi_\mu&\to& -\epsilon_{\mu\nu}\psi_\nu
\end{eqnarray}
Such an operation clearly leaves the Yukawa terms invariant and trivially
all bosonic terms. It is thus a symmetry of the continuum action.
By combining such a transformation with the original 
action of the scalar
supercharge one derives an additional supersymmetry of the theory -- that
corresponding to the twisted supercharge $Q_{12}$.
Explicitly this supersymmetry will transform the component fields of
the continuum theory in
the following way
\begin{eqnarray}
Q_{12}A_\mu &=& -\epsilon_{\mu\nu}\psi_\nu\nonumber\\
Q_{12}\psi_\mu&=& -\epsilon_{\mu\nu} D_\nu \phi\nonumber\\
Q_{12}\chi_{12}&=&-\frac{1}{2}[\phi,\phib]\nonumber\\
Q_{12}B_{12}&=&[\phi,\frac{\eta}{2}]\nonumber\\
Q_{12}\phib&=&-2\chi_{12}\nonumber\\
Q_{12}\frac{\eta}{2}&=&B_{12}\nonumber\\
Q_{12}\phi&=&0
\end{eqnarray}
From the $Q$ and $Q_{12}$ transformations
it is straightforward to verify the following algebra holds 
\begin{eqnarray}
\{Q,Q\}&=&\{Q_{12},Q_{12}\}=\delta_\phi\nonumber\\
\{Q,Q_{12}\}&=&0
\end{eqnarray}
where $\delta_\phi$ denotes an infinitesimal gauge transformation with
parameter $\phi$. This allows us to construct strictly nilpotent
symmetries $\hat{Q}_{\pm}=Q\pm iQ_{12}$ in the continuum theory
corresponding to using the (anti)self-dual components of the
original \KD field.

In the same way we can try to build an additional supersymmetry
by combining the invariance of the fermion kinetic term 
under $\Psi\to\Psi\Gamma^1$ with the existing scalar supersymmetry.
This effects the following transformation of fermion fields:
\begin{eqnarray}
\frac{\eta}{2}&\to&\psi_1\nonumber\\
\chi_{12}&\to& -\psi_2\nonumber\\
\psi_1&\to& \frac{\eta}{2}\nonumber\\
\psi_2&\to&-\chi_{12}
\end{eqnarray}
However, the Yukawas and bosonic terms are only invariant under
such a transformation if we simultaneously make the
transformation $\phi\to -\phib$.
The resultant explicit action of $Q_1$ and $Q_2$ on the component fields is given
by

\begin{align}
Q_{1}A_1 &= \frac{\eta}{2}&Q_{2} A_1&=\chi_{12}\nonumber\\
Q_{1}A_2&=-\chi_{12}&Q_{2} A_2&=\frac{\eta}{2}\nonumber\\
Q_{1}\psi_1&=-\frac{1}{2}[\phi,\phib]&Q_{2} \psi_1&=B_{12}\nonumber\\
Q_{1}\psi_2&=-B_{12}&Q_{2} \psi_2&=-\frac{1}{2}[\phi,\phib]\nonumber\\
Q_{1}\chi_{12}&=-D_2\phib &Q_{2} \chi_{12}&=D_1\phib \\
Q_{1} B_{12}&=[\phib,\psi_2]&Q_{2} B_{12}&=-[\phib,\psi_1]\nonumber\\
Q_{1}\phib&=0 &Q_{2} \phib&=0\nonumber\\
Q_{1}\frac{\eta}{2}&= D_1\phib & Q_{2} \frac{\eta}{2}&=D_2\phib\nonumber\\
Q_{1}\phi&=-2\psi_1&Q_{2} \phi&=-2\psi_2\nonumber
\end{align}

Again, we can verify the following algebra holds
\begin{eqnarray}
\{Q_1,Q_1\}&=&\{Q_2,Q_2\}=\delta_{-\phib}\nonumber\\
\{Q_1,Q_2\}&=&0
\end{eqnarray}
with $\delta_{-\phib}$ a corresponding gauge transformation with parameter
$-\phib$. This allows us to construct yet another pair of nilpotent
supercharges in the continuum theory $\overline{Q}_\pm=Q_1\pm iQ_2$. 

It is interesting to check also the anticommutators of these new charges
$\hat{Q}_\pm$ and $\overline{Q}_\pm$. It is a straightforward exercise to
verify the following algebra holds {\it on-shell} 
\begin{eqnarray}
\{\hat{Q}_+,\overline{Q}_-\}&=&\{\hat{Q}_-,\overline{Q}_+\}=0\nonumber\\
\{\hat{Q}_+,\overline{Q}_+\}&=&4(D_1+iD_2)\nonumber\\
\{\hat{Q}_-,\overline{Q}_-\}&=&4(D_1-iD_2)
\label{anti}
\end{eqnarray}
As an example consider $\{\hat{Q}_+,\overline{Q}_+\}\psi_1$
\beq
\{\hat{Q}_+,\overline{Q}_+\}=\{Q,Q_1\}-\{Q_{12},Q_2\}+i\left(\{Q_{12},Q_1\}+\{Q,Q_2\} \right)
\eeq
Using the component transformations listed above
the relevant anticommutators are
\begin{eqnarray}
\{Q,Q_1\}\psi_1&=&2D_1\psi_1\nonumber\\
\{Q,Q_2\}\psi_1&=&2D_1\psi_2+2[\phi,\chi_{12}]\nonumber\\
\{Q_{12},Q_1\}\psi_1&=&2D_2\psi_1\nonumber\\
\{Q_{12},Q_2\}\psi_1&=&2D_2\psi_2+[\phi,\eta]
\end{eqnarray}
Thus we find
\beq
\{\hat{Q}_+,\overline{Q}_+\} \psi_1=2D_1\psi_1-2D_2\psi_2-[\phi,\eta]+
                           i(2D_1\psi_2+2D_2\psi_1+2[\phi,\chi_{12}])\eeq
Using the equations of motion
\begin{eqnarray}
-2D_1\psi_1-2D_2\psi_2-[\phi,\eta]&=&0\nonumber\\
-2D_2\psi_1+2D_1\psi_2+2[\phi,\chi_{12}]&=&0
\end{eqnarray}
we can easily verify the second line of eq.~(\ref{anti}). Notice that
from these new charges $\hat{Q}_{\pm}$, $\overline{Q}_\pm$ 
we can build spinorial supercharges of the form
\beq\left(\begin{array}{c}
\hat{Q}_+\\
\overline{Q}_-
\end{array}\right)
\eeq
in which case the algebra given in eq.~(\ref{anti}) represents
the usual supersymmetry algebra in a chiral basis (up to a gauge
transformation).

\subsection{Self-dual twist}
\label{sec:8c}

As we have seen it is possible to derive an additional {\it strictly
nilpotent} supersymmetry in the gauge theory
case by combining the scalar charge $Q$ with
its dual pseudoscalar charge $Q_{12}$ in the form
$\cQ=Q-iQ_{12}$. Furthermore, using the transformations derived
in the previous section we can easily show that 
\begin{eqnarray}
\cQ\cA=\cQ\left(A_1+iA_2\right)&=&2\left(\psi_1+i\psi_2\right)\nonumber\\
\cQ\left(\psi_1+i\psi_2\right)&=&0\nonumber\\
\cQ\cAb=\cQ\left(A_1-iA_2\right)&=&0
\end{eqnarray}
This new supercharge is associated to a alternative
twist of the Yang-Mills theory which we may call the
{\it self-dual twist}. It will turn out that this twist is
intimately connected to the orbifold lattice constructions
we discuss next. The key observation is that the original 4 on-shell
bosonic degrees of freedom can be realized in terms of the {\it complex}
gauge fields $\cA_\mu$ and $\cAb_\mu$ together with
a new set of twisted supersymmetry
transformations  
\begin{eqnarray}
\cQ\; \cA_\mu&=&\psi_\mu\nonumber\\
\cQ\; \psi_\mu&=&0\nonumber\\
\cQ\; \cAb_\mu&=&0\nonumber\\
\cQ\; \chi_{\mu\nu}&=&-\cFb_{\mu\nu}\nonumber\\
\cQ\; \eta&=&d\nonumber\\
\cQ\; d&=&0
\end{eqnarray}
Notice that this supersymmetry implies that the fermions are to be treated
as complex
which is natural in a Euclidean theory.
As in previous constructions the twisted action in two dimensions can be written
in $\cQ$-exact form $S=\beta\cQ\; \Lambda$ where $\Lambda$ now is given by
the expression
\beq
\Lambda=\int
\Tr\left(\chi_{\mu\nu}\cF_{\mu\nu}+\eta [ \cDb_\mu,\cD_\mu ]-\frac{1}{2}\eta
d\right)
\eeq
and we have introduced the complexified covariant derivatives (we again employ
an antihermitian basis for the generators of $U(N)$)
\begin{eqnarray}
\cD_\mu&=&\partial_\mu+\cA_\mu=\partial_\mu+A_\mu+iB_\mu\nonumber\\
\cDb_\mu&=&\partial_\mu+\cAb_\mu=\partial_\mu+A_\mu-iB_\mu
\end{eqnarray}
Doing the $\cQ$-variation and integrating out the field $d$ yields
\beq
S=\int\Tr \left(-\cFb_{\mu\nu}\cF_{\mu\nu}+\frac{1}{2}[ \cDb_\mu, \cD_\mu]^2-
\chi_{\mu\nu}\cD_{\left[\mu\right.}\psi_{\left.\nu\right]}-\eta \cDb_\mu\psi_\mu\right)\eeq
The bosonic terms can be written
\begin{eqnarray}
\cFb_{\mu\nu}\cF_{\mu\nu}&=&\left(F_{\mu\nu}-[B_\mu,B_\nu]\right)^2+
\left(D_{\left[\mu\right.}B_{\left.\nu\right]}\right)^2\nonumber\\
\frac{1}{2}\left[\cDb_\mu,\cD_\mu\right]^2 &=& -2\left(D_\mu B_\mu\right)^2
\end{eqnarray}
where $F_{\mu\nu}$ and $D_\mu$ denote the usual field strength and
covariant derivative depending on the real part of the connection $A_\mu$.
After integrating by parts the term linear in $F_{\mu\nu}$ cancels
and the final bosonic action reads\footnote{The bosonic action is real positive
definite on account of the antihermitian basis that we have chosen.}
\beq
S_B=\int\Tr \left(-F^2_{\mu\nu}+2B_\mu D_\nu D_\nu B_\mu-[B_\mu,B_\nu]^2\right)\eeq
Notice that the imaginary parts of the gauge field have transformed into
the two scalars of the SYM theory! This is further confirmed
by looking at the fermionic part of the action which can be rewritten
in $2\times 2$ block form as
\beq
\left(\begin{array}{cc}\chi_{12}&\frac{\eta}{2}\end{array}\right)
\left(\begin{array}{cc}-D_2-iB_2&D_1+iB_1\\
                        D_1-iB_1&D_2-iB_2\end{array}\right)
\left(\begin{array}{c}\psi_1\\ \psi_2\end{array}\right) \; . 
\eeq
which is easily recognized as the dimensional reduction of ${\cal N}=1$
SYM theory in four dimensions in which a chiral representation
is employed for the fermions. As usual the scalar fields $B_\mu$ arise from
the gauge fields in the reduced directions.

\subsection{Lattice theory for $(2,2)$ SYM}
\label{sec:8d}

In this section we show how to discretize this self-dual twist of the two
dimensional Yang-Mills model with $\cQ=4$ supercharges. 
To do this 
we employ the
geometrical discretization scheme proposed in \citep{Catterall_n=2}. 
In general continuum p-form fields are mapped
to lattice fields defined on $p$-subsimplices of a general
simplicial lattice. In the case of hypercubic lattices
this assignment is equivalent to placing a $p$-form with
indices $\mu_1\ldots\mu_p$ on the link connecting $\bx$ with
$(\bx+\bmu_1+\ldots+\bmu_p)$ where $\bmu_i,i=1\ldots p$ corresponds to
a unit vector in the lattice. 
Actually this is not quite the full story; each link has two
possible orientations and we must also specify which orientation
is to be used for a given field. A positively oriented field
corresponds to one in which the link vector has positive components with
respect to this coordinate basis.

Continuum derivatives on such a hypercubic
lattice are represented by lattice difference operators acting on these
link fields. Specifically,
covariant derivatives appearing in curl-like operations 
and {\it acting on positively oriented fields} are replaced by
a lattice gauge covariant forward difference operator whose action on
lattice scalar and vector fields is given by
\begin{eqnarray}
\cD^{(+)}_\mu f(\bx)=\cU_\mu(\bx)f(\bx+\bmu)-f(\bx)\cU_\mu(\bx)\nonumber\\
\cD^{(+)}_\mu f_\nu(\bx)=\cU_\mu(\bx)f_\nu(\bx+\bmu)-f_\nu(\bx)\cU_\mu(\bx+\bnu)
\label{derivs}
\end{eqnarray}
where $\bx$ denotes a two dimensional lattice vector and
$\bmu=(1,0)$, $\bnu=(0,1)$ unit
vectors in the two coordinate directions.
Here, we have replaced the continuum {\it complex}
gauge fields $\cA_\mu$ by non-unitary link fields $\cU_\mu=e^{i\cA_\mu}$. 
The backward difference
operator $\cDb^-_\mu$ replaces the continuum covariant derivative
in divergence-like operations and its action on (positively oriented)
lattice vector fields can be
gotten by requiring  that it to be the adjoint to $\cD^+_\mu$. Specifically
its action on lattice vectors is
\beq
\cDb^{(-)}_\mu f_\mu(\bx)=f_\mu(\bx)\cUb_\mu(\bx)-
\cUb_\mu(\bx-\bmu)f_\mu(\bx-\bmu)
\eeq
The nilpotent scalar supersymmetry now acts on the lattice fields
as
\begin{eqnarray}
\cQ\; \cU_\mu&=&\psi_\mu\nonumber\\
\cQ\; \psi_\mu&=&0\nonumber\\
\cQ\; \cUb_\mu&=&0\nonumber\\
\cQ\; \chi_{\mu\nu}&=&\cF^{L\dagger}_{\mu\nu}\nonumber\\
\cQ\; \eta&=&d\nonumber\\
\cQ\; d&=&0
\end{eqnarray}
Here we written the lattice field strength as 
\beq
\cF^L_{\mu\nu}=\cD^{(+)}_\mu
\cU_\nu(\bx)=\cU_\mu(\bx)\cU_\nu(\bx+\bmu)-
\cU_\nu(\bx)\cU_\mu(\bx+\bnu)\label{field}\eeq
which
reduces to the continuum (complex) field strength in the naive continuum
limit and is automatically antisymmetric in the indices $(\mu,\nu)$.

Notice that this supersymmetry transformation
implies that the fermion fields $\psi_\mu$ have 
the same orientation as their superpartners
the gauge links $\cU_\mu$ and run from $\bx$ to $(\bx+\bmu)$. However, the
field $\chi_{\mu\nu}$ must have
the same orientation as $\cF^{L\dagger}_{\mu\nu}$ and
hence is to be assigned to the 
negatively oriented link running from $(x+\bmu+\bnu)$ down to $\bx$ 
i.e parallel to the vector $(-1,-1)$. This link choice also follows naturally from
the matrix representation of the \KD field $\Psi$
\beq
\Psi=\eta I+\psi_\mu \gamma_\mu +\chi_{12}\gamma_1\gamma_2\eeq
which associates the field $\chi_{12}$ with the
lattice vector $\bmu_1+\bmu_2=\bmu+\bnu$. We will see that
the negative orientation is crucial for allowing us
to write down gauge invariant expressions for the fermion kinetic
term.
Finally, it should be clear that the scalar fields
$\eta$ and $d$ can be taken to transform simply as site fields.

These link mappings and orientations are conveniently summarized by
giving the gauge transformation properties of the lattice fields
\begin{eqnarray}
\eta(\bx)&\to&G(\bx)\eta(\bx)G^\dagger(\bx)\nonumber\\
\psi_\mu(\bx)&\to&G(\bx)\psi_\mu(\bx)G^\dagger(\bx+\bmu)\nonumber\\
\chi_{\mu\nu}(\bx)&\to&G(\bx+\bmu+\bnu)\chi_{\mu\nu}G^\dagger(\bx)\nonumber\\
\cU_\mu(\bx)&\to&G(\bx)\eta(\bx)G^\dagger(\bx)\nonumber\\
\cUb_\mu(\bx)&\to&G(\bx+\bmu)\cUb_\mu(\bx)G^\dagger(\bx)
\end{eqnarray}
We will see shortly that this decomposition of the fermionic degrees
of freedom over the lattice is identical to that encountered in
the orbifolding approach to lattice supersymmetry
\citep{Cohen:2003xe}. Furthermore,
the above $\cQ$-variations and field assignments are equivalent to the
formulation described in \citep{D'Adda_2d} {\it provided} that we set the fermionic
shift parameter $a$ in that formulation to zero and consider only
the corresponding scalar supersymmetry. 

The lattice gauge fermion now takes the form
\beq
\Lambda=\sum_{\bx}\Tr\left( \chi_{\mu\nu}\cD^{(+)}_\mu\cU_\nu+\eta 
\cDb^{(-)}_\mu \cU_\mu-\frac{1}{2}\eta d\right)\eeq
It is easy
to see that in the naive continuum limit the lattice
divergence $\cDb^{(-)}_\mu \cU_\mu$ equals 
$[\cDb_\mu,\cD_\mu]$. Notice that with the
previous choice of orientation for the various fermionic link fields
this gauge fermion is automatically invariant under lattice gauge
transformations. There is no need for the doubling of degrees of freedom
encountered in
\citep{Catterall_n=2,Catterall_n=4}. Those constructions utilize
the twist described earlier in section ~\ref{sec:8a}
in which the
nature of the gauge fermion and the scalar supercharge led to the
presence of explicit Yukawa interactions in the theory. These,
in turn, required the lattice theory to contain fermion link fields
of both orientations and hence led to a doubling of degrees of
freedom with respect to the continuum theory. For the self-dual
twist the Yukawa interactions are embedded
into the complexified covariant derivatives and
successive
components of the \KD field representing the fermions
can be chosen with alternating orientations leading to a \KD action
which is automatically gauge invariant without these extra degrees of
freedom.

Acting with the $\cQ$-transformation shown above and again integrating out
the auxiliary field $d$ we derive the gauge and $\cQ$-invariant lattice
action
\beq
S=\sum_{\bx}\Tr\left(\cF^{L\dagger}_{\mu\nu}\cF^L_{\mu\nu}+
\frac{1}{2}\left(\cDb^{(-)}_\mu \cU_\mu\right)^2-
\chi_{\mu\nu}\cD^{(+)}_{\left[\mu\right.}\psi_{\left.\nu\right]}-
\eta \cDb^{(-)}_\mu\psi_\mu\right)
\eeq
But this is precisely the orbifold action arising in \citep{Cohen:2003xe}
with the modified deconstruction step described in \citep{Unsal_comp} and
\citep{Damgaard:2007xi} which we will describe in detail in the next
section.
The two approaches are thus entirely equivalent. 

We can use this geometrical formulation to show very easily 
that the lattice theory exhibits no fermion doubling problems. The simplest
way to do this 
is merely to notice that the lattice action at zero coupling $\cU\to I$
conforms to the
canonical form required for no doubling by the theorem of 
Rabin \citep{Rabin:1981qj}.
Explicitly, discretization of continuum actions written in terms of
p-forms
will not encounter doubling problems if
continuum derivatives acting in curl-like operations are replaced by forward
differences in the lattice theory while
continuum derivatives appearing in divergence-like operations are represented
by backward differences on the lattice. More precisely the continuum
exterior derivative $d$ is mapped to a forward difference while its adjoint
$d^\dagger$ is represented by a backward difference.

An alternative way to see this is to  examine the 
the form of the fermion operator arising in this construction.
\beq
\left(\begin{array}{cc}\chi_{12}&\frac{\eta}{2}\end{array}\right)
\left(\begin{array}{cc}-\cD^{(+)}_2&\cD^{(+)}_1\\
                        \cD^{(-)}_1&\cD^{(-)}\end{array}\right)
\left(\begin{array}{c}\psi_1\\ \psi_2\end{array}\right)
\eeq
Clearly the determinant of this operator in the free limit is nothing more
than the usual determinant encountered for scalars in two dimensions and
hence possesses no extraneous zeroes that survive the continuum limit.

Numerical simulations of this and related models are just beginning. They
rely on using the RHMC algorithm \citep{rhmc} to handle the non-local
Pfaffian that arises after integration over the twisted fermions. 
Table~\ref{tab:orb} shows results from a simulation of the zero dimensional
$SU(2)$
matrix system which arises by dimensional reduction of the $\cQ=4$ twisted
action we have described. As for quantum mechanics, exact supersymmetry
allows us to predict a value for the expectation value of the bosonic
action $\kappa S_B$ 
which is independent of coupling $\kappa$ and this result
is strongly borne out by the Monte Carlo data\footnote{In this case the
Pfaffian phase is exactly zero and so poses no problem in the simulation.
In two dimensions large phase fluctuations are
encountered for the model with $\cQ=4$ supersymmetries  
\citep{Giedt_det1,Giedt_det2,Catterall:2008dv}. These phase
fluctuations seem much smaller for
the model with $\cQ=16$ supersymmetries and can be handled
by standard re-weighting techniques.}.

\begin{figure}
\begin{center}
\includegraphics[height=60mm]{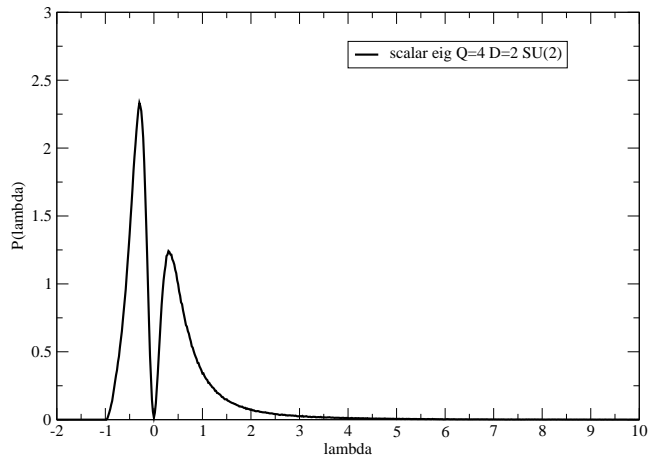}
\caption{Probability distribution for the eigenvalues of scalar
fields in $SU(2)$ theory with $\cQ=4$ supersymmetries 
in two dimensions} \label{vac}
\end{center}
\end{figure}

\begin{table}[t]
\begin{center}
\begin{tabular}{||c|c|c||}\hline
$\kappa$ & $\kappa S_B$& exact\\ \hline
1.0 & 4.39(2) & 4.5\\ \hline
10.0 & 4.49(1)& 4.5\\ \hline
100.0 & 4.49(1) & 4.5\\\hline
1000.0 & 4.52(2) & 4.5\\\hline
\end{tabular}
\end{center}
\caption{Bosonic action versus exact SUSY value for the  $\cQ=4$, $SU(2)$ matrix model.}
\label{tab:orb}
\end{table}
One of the most interesting questions that arises in these models concerns
the nature of the vacuum. Classically the models possess a continuous
infinity of vacua corresponding to taking the complex bosonic link fields
to be diagonal matrices which are constant over the lattice. Integration
over these flat directions may then  lead to IR divergences. This issue
has already been examined via numerical simulations (see fig.~\ref{vac}) where it is found that
contrary to naive expectation the eigenvalues of the
scalar fields (imaginary parts of
the complex link field) remain localized close to the origin in field
space \citep{Catterall:2008dv}.
These preliminary simulations are encouraging as they show that
these new lattice actions may indeed be very useful starting points for
numerical explorations of strongly coupled supersymmetric systems.

Our discussion thus far has taken us from simple quantum
mechanics models to genuine field theories with non-abelian
gauge symmetry. In all cases the approach we have followed is to rewrite the
theory in terms of so-called twisted variables which naturally
exposes a scalar supercharge which, with care, may then be
transfered to the lattice. The twisted constructions are elegant
and physically well motivated but the precise discretization
prescription has been arrived at in a somewhat ad hoc manner. We will
now turn to the orbifold constructions for gauge theories and show
how these discretization rules re-emerge in an essentially unique
way, being determined only by the global symmetry of the continuum
theory and the requirement of one or more exact supersymmetries.

\section{Supersymmetric lattices from orbifold projection}
\label{sec:9}

In this section we turn to an alternative construction of supersymmetric
lattice actions based on the ideas of deconstruction and orbifolding. On the
face of it this seems quite independent of the discretized twisted
constructions we have discussed in the last couple of sections.
Nevertheless we will see on closer analysis that both approaches are
intimately connected and lead to similar lattice actions in
the case of Yang-Mills theories. As we will show, the orbifold approach is a very powerful way to generate all the known SYM lattices, and in fact was how they were first derived for spatial lattices \citep{Kaplan:2002wv} and Euclidean spacetime lattices \citep{Cohen:2003xe,Cohen:2003qw,Kaplan:2005ta,Endres:2006ic}.

\subsection{Deconstruction:The AHCG model}
\label{sec:9a}

\begin{figure}[t]
\centerline{\resizebox{5.0cm}{!}{%
 \includegraphics{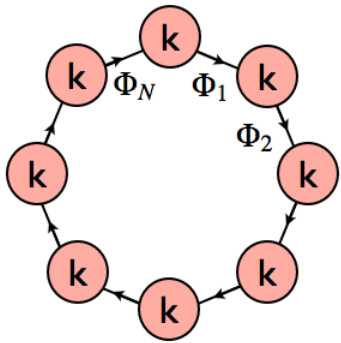} }}
\caption{A diagram for the AHCG deconstruction model, which is a
  $d=4$, $\CN=1$ supersymmetric gauge theory. Each node
  corresponds to an independent $U(k)$ gauge symmetry, with the
  associated vector supermultiplet $V_n$.  The links represent chiral
  superfields $\Phi_n$ which transform as  bifundamentals  under the
  gauge symmetries of the nodes they connect.}
\label{fig:moose}
\end{figure}

The starting point is the deconstruction method of Arkani-Hamed, Cohen and Georgi (AHCG).  
In reference \citep{ArkaniHamed:2001ca} the authors were not concerned
with latticizing supersymmetry;  instead they wanted a precise field
theoretic way to
examine claims about the phenomenology of certain field theories in
five dimensions.  In order to 
avoid ill-defined problems with renormalization in five dimensions,
they constructed a theory with four continuous dimensions, and a
latticized fifth dimension.  This can be viewed as a $d=4$ field theory
with many ``flavors'' of fields, corresponding to the discrete values
of the fifth coordinate. A diagram of the theory of interest is given in
Fig.~\ref{fig:moose}; it is an $\CN=1$ supersymmetric field theory in
$d=4$ with gauge group $U(k)^N$ with a single gauge coupling $g$, where
each $U(k)$ factor appears as a 
node in the picture.  The $n^{th}$ node has a vector multiplet
associated with it ---  a gauge field $v_m^{(n)}$ and a gaugino
$\lambda^{(n)}$.  In addition there are matter fields in the form of
chiral supermultiplets $\Phi_n$ which appear in the figure as directed links
between nodes $n$ and $(n+1)$; they transform as bifundamentals
$(\Box, \overline \Box)$ under the $U(k)\times U(k)$ gauge symmetry
associated with those two nodes, and are neutral under the rest of the
gauge symmetry; they represent the scalar and fermion component fields
$(\phi^{(n)}, \psi^{(n)})$. All the interactions in this model are
supersymmetric gauge interactions (which include certain Yukawa and $\phi^4$
couplings). Note that since all the fields transform as either adjoints
of $U(k)$ or bifundamentals of $U(k)\times U(k)$, they can all be
represented as $k\times k$ matrices with non-zero trace.

So far, this model doesn't look at all like a lattice for a 5D theory;
although there are interactions between nearest neighbors the fifth
direction, there are no bilinear ``hopping terms'' corresponding to
kinetic energy operators for motion in this extra dimension.
However, the authors noted that the theory has a ``flat direction''
corresponding to 
\beq
\vev{\phi^{(n)}} = \frac{1}{a\sqrt{2}}\,{\mathbf 1}_k
\eqn{flat}\eeq
where ${\mathbf 1}_k$ represents that $k\times k$ unit matrix, and $a$
is a length scale.  By flat direction, we mean that the theory has a
degenerate ground state, where the vacuum energy is unaffected by the
simultaneous shift of all the scalar link fields $\phi^{(n)}$ as in
\eq{flat}.  Furthermore, as we  will elaborate on below, AHCG noted that
the parameter $a$ behaves like a lattice spacing, and that in
the limit 
\beq  
N\to\infty\ ,\quad
a\to 0\ ,
\quad 
g\to 0\ ,
\quad
aN\equiv L_5\ ({\rm fixed})\ ,\quad
g^2/a\equiv g_5^2\ ({\rm fixed})\ ,
\eqn{decon}
\eeq
the model of
Fig.~\ref{fig:moose} 
has two amazing properties: 
\begin{itemize}
\item it possesses $d=5$ Poincar\'e invariance;
\item it possesses $\CQ=8$ supercharges, even though the $d=4$ model in
  Fig.~\ref{fig:moose} only respected $\CQ=4$ exact supersymmetries.
\end{itemize}
This is exactly the type of phenomenon we were looking for! Both
Poincar\'e symmetry and supersymmetry are enhanced in the continuum
limit without any fine tuning of the theory. 

We now sketch out how the 5D kinetic terms emerge in the AHCG model in
the $a\to 0$ limit, and then
discuss how to generalize their procedure to generate true lattices
where every spacetime dimension is discretized, a method called ``orbifolding''.

\subsection{Continuum limit of the AHCG model}
\label{sec:9b}

The  Lagrangian for the AHCG model possesses four types of terms:
\begin{enumerate}
\item The Yang-Mills action for the gauge fields $v_m^{(n)}$;
\item Gauge interactions for the adjoint gauginos $\lambda^{(n)}$ and
  the bifundamental matter fields $\phi^{(n)}$ and $\psi^{(n)}$, the
  latter involving both $v_m^{(n)}$ and $v_m^{(n-1)}$;
\item Yukawa interactions for the form $\sum_n\Tr
  \lambda^{(n)}\left(\psi^{(n)}\bar\phi^{(n)} -
    \bar\phi^{(n-1)}\psi^{(n-1)}\right)$;
\item A $\phi^4$ interaction (called the ``D-term'') proportional to
  $\sum_n\Tr\left( \phi^{(n+1)}\bar\phi^{(n+1)}-\bar\phi^{(n)}\phi^{(n)}\right)^2$
\end{enumerate}
It is easy to see then that indeed \eq{flat} is a flat direction of
the theory, since the D-term vanishes if each field $\phi^{(n)}$
equals the same diagonal matrix.  To see how the continuum limit emerges, we expand the
$\phi$ fields about their vacuum value as 
\beq
\phi^{(n)}(x) = \frac{1}{a\sqrt{2}}\,{\mathbf 1_k} + \frac{s^{(n)}(x)
  + i v_5^{(n)}(x)}{\sqrt{2}}
  \eqn{exp1orb}
\eeq
where $s$ and $v_5$ are hermitean matrices. Then, for example,  the
($d=4$) kinetic term for $\phi$ in the AHCG action is
\begin{eqnarray}
&&\frac{1}{g^2}\,\sum_n\int d^{4}x\, \Tr\vert D_\mu\phi^{(n)}\vert^2 =
\frac{1}{g^2}\,\sum_n\int d^{4}x\,\Tr\vert \partial_\mu\phi^{(n)} + i
v_\mu^{(n)} \phi^{(n)} - i \phi^{(n)}v_\mu^{(n+1)}\vert^2\cr
&=&\frac{1}{2g^2}\, \sum_n\int d^{4}x\, \Tr\left\vert\left( \partial_\mu s^{(n)} + i
v_\mu^{(n)}  s^{(n)} - i
s^{(n)}v_\mu^{(n+1)}\right)\right.\cr&&\qquad\left.+i\left(\partial_\mu v_5^{(n)} + i
v_\mu^{(n)}v_5^{(n)} - i v_5^{(n)} v_\mu^{(n+1)}\right) + i
\left(v_\mu^{(n)} - v_\mu^{(n+1)}\right)/a \right\vert^2\cr
&&
\xrightarrow[a\to 0]{}\,\frac{1}{2g_5^2}\,\int d^{5}x\, \Tr
(D_\mu s)^2 - \Tr v_{\mu 5}v^{\mu 5} \ ,
\end{eqnarray}
where $v_{mn}$ is the $d=5$ gauge field strength.  Note that the 5D
kinetic term for the gauge field has emerged in this limit.

The scalar ``D-term'' in the AHCG model provided the 5D kinetic term
for the field $s$ in the same limit:
\beq
\frac{1}{2g^2}\,\sum_n\int d^{4}x\,
\Tr\left(\phi^{(n+1)}\bar\phi^{(n+1)} -
  \bar\phi^{(n)}\phi^{(n)}\right){\xrightarrow[a\to 0]{}\,} \frac{1}{2
  g_5^2}\int d^5x\,\Tr \left(D_5 s\right)^2\ .
\eeq
Note that this $5D$ term is normalized the same way as the $(D_\mu s)^2$
term in the previous equation, as required by 5D Lorentz invariance.

In the AHCG model, the two Weyl fermions---the gaugino $\lambda$ and
the matter field $\psi$---combine to form one, 4-component, $d=5$
fermion
\beq
\Psi = \begin{pmatrix} \lambda\cr\bar\psi\end{pmatrix}\ ,\qquad
\bar\Psi = \left( \psi\ \ \lambda\right)
\eeq
in the $\gamma$-matrix basis
\beq
\gamma_\mu = \begin{pmatrix}  & \bar\sigma_\mu\cr \sigma_\mu &
  \end{pmatrix}\ ,\qquad \gamma_5 = \begin{pmatrix} 1 & \cr &
    -1\end{pmatrix}
\eeq
The fifth dimensional part of the fermion kinetic term (and the
$\Psi-s$ interaction)  arises from the
Yukawa interaction in the $d=4$ theory:
\begin{eqnarray}
&&\frac{1}{g^2}\,\sum_n\int d^{4}x\,
i\sqrt{2}\Tr\lambda^{(n)}\left( \psi^{(n)} \bar\phi^{(n)} -
    \bar\phi^{(n-1)}\psi^{(n-1)}\right) + {\rm h.c.}\cr&&
{\xrightarrow[a\to 0]{}\,} \frac{1}{2
  g_5^2}\int d^5x\,\Tr \left(\bar\Psi i\gamma_5 D_5\Psi -\bar\Psi
  \gamma_5[s,\Psi]\right)\ .
\end{eqnarray}

It is easy to figure out the limit of the remaining terms.  The
conclusion is that a 5D supersymmetric gauge theory emerges in the
continuum limit, consisting of the scalar $s$ arising as the real part
of the link scalar $\phi$, the fermion $\Psi,\bar\Psi$ arising both from the
gauginos $\lambda$ living at the sites of Fig.~\ref{fig:moose}, as well as the
link fermions $\psi$; and the 5D gauge field consisting of the four
components of $v_\mu$ living on the sites, and $v_5$ arising as the
imaginary part of the link scalar $\phi$. It is fascinating to see how
these 5D multiplets form by combining both site and link variables.
Most importantly for our purposes, recall the claim that this 5D gauge
theory possesses $\CQ=8$ supersymmetries, which has somehow emerged in
the $a\to 0$ limit
from the original $\CQ=4$ theory, without any fine tuning.

The mechanism by which enhanced supersymmetry emerges in the continuum
limit of the AHCG model \citep{ArkaniHamed:2001ca}
is what has been long sought for in a lattice theory --- but it is itself
still a theory in four continuous dimensions and not on a lattice.  To
construct a true 
supersymmetric lattice, we must 
``reverse engineer''  the AHCG model to find general principles for
how it is constructed, and then apply those principles 
to constructing true spacetime lattices. 

\subsection{The AHCG model via orbifolding}
\label{sec:9c}

A simple procedure exists for producing the theory represented by
Fig.~\ref{fig:moose} with $N$ sites and a $U(k)^N$ gauge symmetry.
The idea is to start with a ``mother theory'' 
which has the following properties:
\begin{itemize}
\item it is a $d=4$ field theory like the AHCG model;
\item it possesses the huge gauge group $U(N k)$;
\item it respects the number of supersymmetries of the $target$
  theory, namely $\CQ=8$.
\end{itemize}
In other words, it is a $d=4$, $\CQ=8$ gauge theory with gauge group
$U(Nk)$; such a theory is known as an $\CN=2$ SYM theory.

What we will then do is project out a $Z_N$ symmetry (which means:
identify a $Z_N$ symmetry in the theory, and set to zero all fields
which aren't neutral under that symmetry). This projection (called an
orbifold projection)  breaks the
gauge symmetry from $U(N k) \to U(k)^N$, and it breaks half the
supersymmetries of the theory, from
$\CQ=8$ to $\CQ=4$.  That leaves us with the AHCG model.

\begin{figure}[t]
\centerline{\resizebox{9.0cm}{!}{%
 \includegraphics{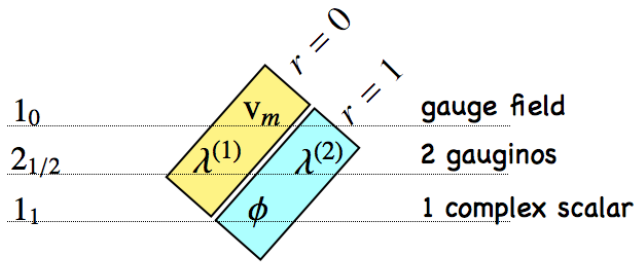} }}
\caption{The fields for $\CN=2$ SYM theory, along with their
  $SU(2)\times U(1)$ $R$-symmetry quantum numbers.  The charge $r=(Y-T_3)$
   distinguishes which fields become site
  variables in the AHCG model and which  become link variables ($r=0$
  and $r=1$ respectively).}
\label{fig:AHCGorbi}
\end{figure}

To see how this works, consider the field content of an $\CN=2$ SYM theory. The
gauge multiplet consists of a gauge field $v_\mu$, two Weyl gauginos
$\lambda^{(1,2)}$, and a complex scalar $\phi$.  It is also useful to decode the structure of the 
 $\CN=2$ supersymmetry in terms of  $\CN=1$ supersymmetry multiplet as we eventually 
 want to know which supersymmetries survive the projection. 
The  $\CN=2$  matter content  diamond  shown in Fig. \ref{fig:AHCGorbi} can be decomposed in terms of    $\CN=1$  multiplets  
$V= (v_m, \lambda^{(1)})$,  $\Phi= (\phi, \lambda^{(2)})$  and   $\CN=1'$ multiplets 
$V'= (v_m,  \lambda^{(2)}) $,  $\Phi'= (\phi,  \lambda^{(1)})$ 
as  shown below:
\begin{equation}
 \xymatrix{
& v_{m} \ar@{<->}[dl]_{\CN=1}  \ar@{<.>}[dr]^{\CN=1'}  & \\ 
\lambda^{(1)} \ar@{<.>}[dr]_{\CN=1'}  &  &  \lambda^{(2)} \ar@{<->}[dl]^{\CN=1}    \\
& \phi &
 }  
\end{equation}
Note the similarity
between this multiplet and the field content appearing in
Fig.~\ref{fig:moose}. Each of the fields transforms as the adjoint
representation of the gauge group, which in our case is $U(Nk)$; that
means we can represent the fields as $Nk\times Nk$ matrices, acted
upon by the gauge transformation $U$ as $\phi\to U\phi U^\dagger$
(except for the gauge field, which has the usual inhomogeneous
transformation).

But how to define the $Z_N$ symmetry which tells some fields to become
site variables and others to become link variables in the AHCG model?
The $\CN=2$ SYM action possesses an $SU(2)\times U(1)$ $R$-symmetry,
under which the fields transform as shown in Fig.~\ref{fig:AHCGorbi}.  We
can find a symmetry which distinguishes between fields destined to
become site variables ($v_m$ and $\lambda^{(1)}$) and link variables
($\lambda^{(2)}$ and $\phi$) by defining a $U(1)$ charge $r$ which
lives in the $SU(2)\times U(1)$ $R$-symmetry:
$r = Y-T_3$, where $Y$ is the $U(1)$ charge and $T_3$ is the third
$SU(2)$ generator.  Then as shown in Fig.~\ref{fig:AHCGorbi}, site variables
have $r=0$ and link variables have $r=1$.

\begin{figure}[t]
\centerline{\resizebox{7.0cm}{!}{%
 \includegraphics{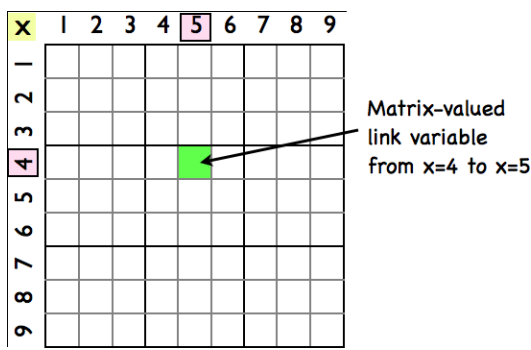} }}
\caption{Illustration of how a $9k\times 9k$ matrix can represent
  a 1D lattice with $N=9$ sites.  The highlighted $k\times k$ block
  represents a $k\times k$ matrix-valued field residing on the
  directed link from site $x=4$ to site $x=5$.}
\label{fig:1Dmat}
\end{figure}

Each of the different types of fields of the AHCG model ---
each of the  $N$  ``flavors'' of $k\times k$ matrices --- can be represented
as a single sparse $Nk
\times Nk$ matrix, as illustrated in Fig.~\ref{fig:1Dmat}. We think of
the big  $Nk
\times Nk$ matrix as being made of $N^2$ $k\times k$ blocks, each
labeled by a row number $n_i$ and a column number $n_f$; then that
block can be thought of as living on a 1D lattice as a link running
from site $n_i$ to site $n_f$.  Thus for the site variables ($r=0$) we
want to have an $Nk
\times Nk$   matrix with only diagonal $k\times k$ blocks surviving;
the link variables ($r=1$) in
Fig.~\ref{fig:moose} should become  sparse  $Nk
\times Nk$  matrices with nonzero blocks only appearing one row above
the diagonal.

We can attain the desired result by defining a $Z_N$ symmetry which
combines the $r$ symmetry with a particular $U(Nk)$ transformation:
\beq
Z_N:  \Phi \to \hat\gamma \Phi \equiv \omega^r \Omega \Phi \Omega^\dagger\ ,\qquad
\Omega = \begin{pmatrix} \omega &&& \cr &\omega&&\cr
  &&\ddots&\cr&&&\omega^N \end{pmatrix}\ ,\qquad \omega = e^{2\pi
  i/N}\ ,
\eeq
where $r$ is the particular $r$-charge for that field
$\Phi$, and each entry in $\Omega$ is proportional to a $k\times k$
unit matrix.  We then define the orbifold projection operator $\hat P \Phi =
\frac{1}{N}\sum_{i=p}^N \hat\gamma^p \Phi$ which annihilates any
sub--block in the matrix $\Phi$ which is not invariant (this follows
from the fact that $[\omega + \omega^2 +\ldots+\omega^N]=0$). Note
that this projection does not commute with the full $U(Nk)$ gauge
symmetry of the mother theory and leaves intact only the $U(k)^N$
subgroup which commutes with $\Omega$.   The
result of this projection is shown in Fig.~\ref{fig:1Dproj} and can be depicted 
as in  Fig.\ref{fig:moose}, a segment of which is shown below:
\begin{equation} 
 \xy 
(-16,0)*{V_{n-1}}; 
(-16,0)*\xycircle(4,4){-}="V_{n-1}";
(-28,0)**\dir{-} ?(.75)*\dir{<}+(0,5)*{\scriptstyle \Phi_{n-2}}; 
(0,0)*{V_n}; 
(0,0)*\xycircle(4,4){-}="V_n"; 
(-12,0)**\dir{-} ?(.75)*\dir{<}+(0,5)*{\scriptstyle \Phi_{n-1}}; 
"V_n";(12,0)**\dir{-} ?(.75)*\dir{>}+(0,5)*{\scriptstyle \Phi_{n}}; 
(16,0)*{V_{n+1}}; 
(16,0)*\xycircle(4,4){-}="V_{n+1}"; 
"V_{n+1}";(28,0)**\dir{-} ?(.75)*\dir{>}+(0,5)*{\scriptstyle \Phi_{n+1}};
\endxy 
\end{equation}
 Note
that evidently $\hat P$ also breaks the $\CN=2$ supersymmetry, since
it treats the different members of the gauge multiplet differently.
It does, however, preserve an $\CN=1$ supersymmetry, with
$V_n= (A_{\mu,n}, \lambda_{n}^{(1)})$   being an $\CN=1$ vector supermultiplet, and
  $\Phi_n= (\phi_{n}, \lambda_{n}^{(2)})\equiv(\phi_{n, n+1}, \lambda_{n,n+1}^{(2)}) $  forming an $\CN=1$ chiral matter multiplet.
The vector multiplets 
$V_n$ transform as  adjoint under the
gauge group factor $G_n$ and   chiral multiplets  $\Phi_n$
 transform as 
bi-fundamental  $(\Box, \overline \Box)$ under $G_n \times G_{n+1}$.  Thus, in the quiver, the  $\CN=1'$ is explicitly 
violated since the gauge rotation properties of used-to-be  $\CN=1'$ multiplet no longer matches as shown below:
\begin{equation}
\xymatrix{
& v_{m,n} \ar@{<->}[dl]_{\CN=1}   \ar@{<.>}[dr]|{\rm nothing }   & \\ 
\lambda^{(1)}_n    \ar@{<.>}[dr]|{\rm nothing}  &  & \lambda^{(2)}_{n, n+1} \ar@{<->}[dl]^{\CN=1}    \\
& \phi_{n,n+1} &
 }
\end{equation}
The action of a global supersymmetry transformation of an 
adjoint cannot produce
a bi-fundamental.

\begin{figure}[t]
\centerline{\resizebox{7.0cm}{!}{%
 \includegraphics{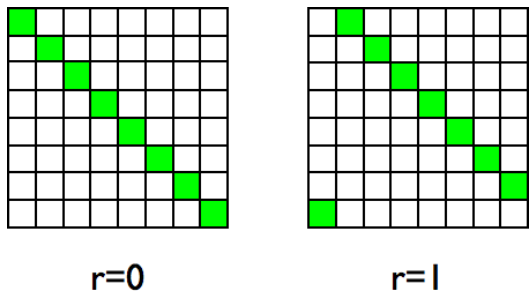} }}
\caption{The result of the $Z_N$ orbifold projection: For the
  fields $v_m$ and $\lambda^{(1)}$ with $r=0$, only the diagonal
  $k\times k$ blocks survive, and these can be interpreted as site
  variables, transforming as adjoints under the unbroken $U(k)^N$
  gauge symmetry.  The $\lambda^{(2)}$ and $\phi$ fields with $r=1$
  have only the superdiagonal blocks survive; these transform as
  bifundamentals under the $U(k)^N$ gauge symmetry, and represent the
  link variables in Fig~\ref{fig:moose} (with
  $\lambda^{(2)}\equiv \psi$).}
\label{fig:1Dproj}
\end{figure}


The punchline: by plugging the sparse matrices obtained after
projection back into the 
$\CN=2$ action, one recovers the full action of the AHCG model!  (Also see 
\citep{Giedt:2003xr,Rothstein:2001tu})

\begin{figure}[t]
\centerline{\resizebox{8.0cm}{!}{%
 \includegraphics{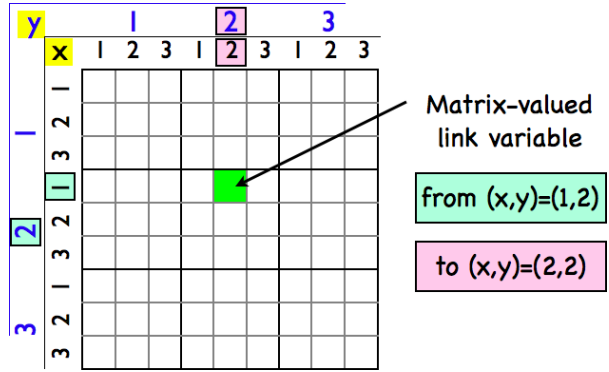} }}
\caption{Representing a two-dimensional $3\times 3$ lattice by a
  sparse $9\times 9$ matrix: each sub-block can be identified as a
  site or link variable on the 2D lattice.}
\label{fig:2Dorbi}
\end{figure}

It is straightforward now to generalize our orbifold projection
prescription in order to construct true lattices, of varying
dimensions. For example, to produce a $d=2$ lattice, we need to start
with a mother theory with a $U(N^2 k)$ gauge symmetry, and project out
a $Z_N\times Z_N$ symmetry.  The idea is that we take the $N^2k\times
N^2k$ matrices in the mother theory, divide them into $N^2$ $NK\times
Nk$ blocks, and then subdivide those into $N^2$ $k\times k$
sub-blocks.  The location of each $k\times k$ sub-block can then be
specified by four integers; the interpretation is that this is a link
variable going from one site on a 2D lattice (specified by two
integers) to another 
(specified by another two integers); see Fig.~\ref{fig:2Dorbi}.

\goodbreak
We now have a method for generating supersymmetric lattice actions:
\begin{enumerate}[i.]
\item Start with a mother theory which is an SYM with the same number of
  supercharges $\CQ$ as the target theory in the continuum;
\item This mother theory should be formulated in $zero$ dimensions (in
  other words: it is a matrix model, not a field theory),
  since we don't want any continuous dimensions, unlike the AHCG model
  which was formulated in $d=4$;
\item  For a target theory with $d$ continuous dimensions, make
  the gauge group of the mother theory $U(N^dk)$, identify the appropriate
  $Z_N^d$ symmetry that resides partly in the gauge group and partly
  in the $R$-symmetry group of the mother theory, and project it out;
\item   Travel out along the flat direction in the degenerate
  vacua  as in \eq{decon}, in order to recover the continuum
  limit of the target theory.
\end{enumerate}

Oddly enough, this diabolical recipe really works!  And in fact, it has
shown that all the different constructions of lattice
SYM theories in the literature can be shown to be equivalent to ones obtained through
orbifold projection \citep{Damgaard:2007xi}.  
As with all pacts with the devil there
is a price:  item (i) and item (iii) above are not in general compatible,
since a theory with a small number of supercharges will have a small
$R$-symmetry which will not contain a $Z_N^d$ subgroup for large
$d$. Equivalently, since each dimension requires a $Z_N$ projection
which breaks half of the remaining supercharges of the mother theory
(and since we want the lattice theory to possess at least one unbroken
supercharge) we require $\CQ \ge 2^{d}$. 
Thus to go to higher dimension $d$, one needs to consider highly
supersymmetric theories with large $\CQ$. For $d=4$, the only
supersymmetric lattice that can be constructed via this method must
have $\CQ\ge 16$, leaving $\CN=4$ SYM theory as the only possibility.

\subsection{Orbifold Lattice Theory for $\CN= (2,2)$ SYM}
\label{sec:9d}

We now briefly describe the construction of the
four supercharge theory in two dimensions which was
previously discussed from within the
twisted approach. 
The action for this
theory is easy to write down:  start with the familiar $\CN=1$ SYM
theory in $d=4$ dimensions (a gauge theory with a massless Weyl
adjoint fermion), and erase two of the space dimensions.  The gaugino
becomes a 2-component Dirac fermion $\psi$ (since $\gamma$ matrices in
$d=2$ are 
just Pauli matrices,  Dirac spinors have only two components). The
four component gauge boson in $d=4$ becomes a two component gauge boson plus
one complex scalar field $s$.  The gluon and gaugino interactions in the
$d=4$ action become 2D gauge interactions, plus Yukawa and $s^4$
interactions. The result is the action (in Euclidean spacetime)
\begin{eqnarray}
 \CL =  \frac{1}{g_2^2}  \, \Tr
  \Biggl(\bigl\vert D_m s\bigr\vert^2 + \bar \psi \,i  D_m
  \gamma_m 
  \psi + \fourth v_{mn} v_{mn} +i\sqrt{2}(\bar \psi_L [ s, \psi_R]
  +\bar\psi_R [ s^\dagger, \psi_L]) + \half
  [s^\dagger,s\,]^2\Biggr) \cr
  \eqn{targ2}
\end{eqnarray}
where $m,n=1,2$, $\psi_R$ and $\psi_L$ are the right- and left-chiral components
of a two-component Dirac field $\psi$, $D_m = \partial_m + i
[v_m,\;\cdot\;]$ is the covariant derivative, and $v_{mn} =-
i[D_m,D_n]$ is the field strength. All fields are rank-$k$ matrices
transforming as the adjoint representation of $U(k)$. This is the
target theory for which we want to construct a lattice.

To construct a lattice for this target theory, we need to start with a
matrix theory with a $U(N^2 k)$ gauge symmetry with $\CQ=4$
supersymmetries.  The way to obtain the $\CQ=4$ matrix theory is simple:
Start with the same $\CN=1$ SYM theory in
$d=4$, which we know has $\CQ=4$ supersymmetries...and then erase
$all$ spacetime coordinates from the action (and therefore, all
derivatives).  The result is a very simple  action which will
serve as our mother theory:
\begin{equation}
  S = \frac{1}{g^2}\left(\frac{1}{4} \Tr v_{mn} v_{mn} + \Tr
    \bar\psi\,  \bar\sigma_m [v_m, 
    \psi]\right)\ ,
  \eqn{momiv}
\end{equation}
where $m,n=0,\ldots,3$, $\psi$ and $\bar\psi$ are independent
complex two-component spinors, $v_m$ is the 4-vector of constant gauge
potentials, and
\begin{equation}
  v_{mn} = i[v_m,v_n]\ , \quad \sigma_m = \{ 1,\, -i{\bfsig}\}\  ,\quad
  \bar\sigma_m = \{ 1,\, i\bfsig\}\ ,
\eqn{mom}
\end{equation}
This mother theory is invariant under four independent
supersymmetries, characterized by 
the transformations
\begin{gather}
  \delta v_m = -i\bar \psi\, \bar\sigma_m\kappa + i \bar\kappa \,\bar
  \sigma_m\psi\ ,\quad
  \delta \psi = -i v_{mn}\sigma_{mn}\kappa \ ,\quad
  \delta \bar \psi = i v_{mn} \,\bar\kappa \,\bar\sigma_{mn}\ ,
  \eqn{trans}
\intertext{where}
  \sigma_{mn} \equiv
  {\textstyle{\frac{i}{4}}}
  \left(\sigma_m\bar\sigma_n-\sigma_n\bar\sigma_m\right)\ ,\qquad\qquad
  \bar\sigma_{mn} \equiv
  {\textstyle{\frac{i}{4}}}
  \left(\bar\sigma_m\sigma_n-\bar\sigma_n\sigma_m\right)\ . 
\end{gather}
where $\kappa$ and $\bar \kappa$ are independent two-component
Grassmann parameters.

The $R$-symmetry of
the mother theory is $SO(4)\times U(1) = SU(2)\times SU(2)\times
U(1)$. This result is not very mysterious:  the $U(1)$
factor is just the $U(1)$ $R$-symmetry associated with the $d=4$ $\CN=1$
SYM theory we started with to derive the mother theory.  The $SO(4)=SU(2)\times SU(2)$ factor is nothing but
what remains of the (Euclidean) Lorentz symmetry that remains even
after all spacetime coordinates are removed from the $d=4$ theory. Therefore
$v_m$ transforms as a 4-vector $=(2,2)$ under this $SU(2)\times SU(2)$, while $\psi$
transforms as a $(2,1)$ and $\bar\psi$ as a $(1,2)$.

\begin{figure}[t]
\centerline{\resizebox{11.0cm}{!}{%
 \includegraphics{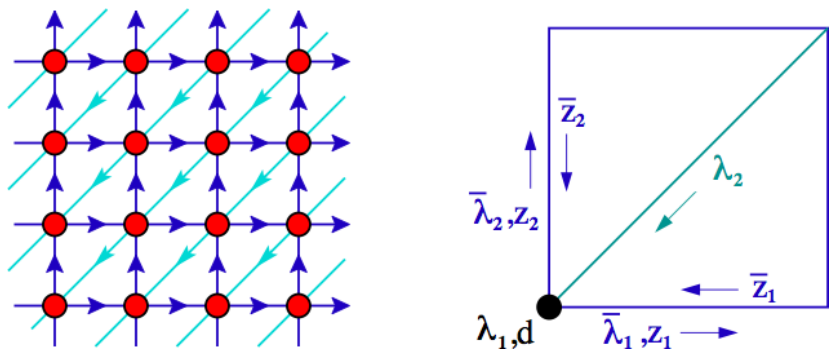} }}
\caption{The  lattice structure and the unit cell for the target
theory of \eq{targ2}, $(2,2)$ SYM in two dimensions.  The ``d''
variable is an auxiliary field you can ignore; it proves to be
convenient when developing a superfield formulation for the lattice theory. }
\label{fig:2dlat}
\end{figure}

The ``daughter theory'' we will derive from this mother theory
by orbifolding will 
be a two-dimensional lattice with $N^2$ sites and a $U(k)$ symmetry
associated with each site (the conventional way to realize a $U(k)$
gauge symmetry on a lattice). To obtain this daughter theory we must
 identify the
correct $Z_N\times Z_N$ symmetry to project out.
The trick is to define two independent analogues of the $r$-charge
from the previous section --- We will call them $\vec r
=\{r_1,r_2\}$. This vector $\vec r$ is interpreted as the directed
link in the unit cell on which a given variable resides. For example,
$\vec r=\{0,0\}$, $\vec r=\{1,0\}$ and $\vec r = \{1,1\}$ are
interpreted as a site, an $x$-link, and a diagonal link respectively.
As such, we need to define the  $Z_N\times Z_N$ symmetry so that $\vec
r$ components only take on the values $0,\pm 1$.  Furthermore, one can
show that the number of unbroken supercharges on the lattice equals
the number of fermions with $\vec r=\{0,0\}$ (e.g., living on the
sites), and so we want to choose the  $Z_N\times Z_N$ symmetry to
maximize this number.
  With a little work, it is possible to show that
a suitable choice yields the charge assignments displayed in Table~\ref{table:1}
\citep{Cohen:2003xe}, where we have written the fermion components as
\beq
\psi=\begin{pmatrix}\lambda_1\cr\lambda_2\end{pmatrix}\ ,\qquad
\bar\psi = \left(\bar\lambda_1\ \ \bar\lambda_2\right)
\eeq
This choice is unique up to uninteresting permutations.
\begin{table}[t]
\centerline{
\hbox{\begin{tabular}{|l||r|r|}
\hline
bosons &$ r_1 $&$ r_2 $\\ \hline
$z_1 = \frac{v_0-iv_3}{\sqrt{2}}$&$ 1 $&$ 0 $\\ \hline
$\bar z_1 = \frac{v_0+iv_3}{\sqrt{2}}$&$ -1 $&$ 0 $\\ \hline
$z_2 = -i\frac{v_1-iv_2}{\sqrt{2}}$&$ 0 $&$ 1 $\\ \hline
$\bar z_2 = i\frac{v_1+iv_2}{\sqrt{2}}$&$ 0 $&$ -1$ \\ \hline
\end{tabular}
\quad
\begin{tabular}{|l||r|r|}
\hline
 fermions &$ r_1 $&$ r_2 $\\ \hline
$\phantom{\frac{a}{\sqrt{b}}}\lambda_1$&$0$&$0$\\ \hline
$\phantom{\frac{a}{\sqrt{b}}}\lambda_2$&$-1$&$-1$\\ \hline
$\phantom{\frac{a}{\sqrt{b}}}\bar\lambda_1$&$1$&$0$\\ \hline
$\phantom{\frac{a}{\sqrt{b}}}\bar\lambda_2$&$0$&$1$\\ \hline
\end{tabular}}}
\caption{\sl Assignment of the $Z_N\times Z_N$ charges for the
  variables of the mother theory \eq{momiv}; see \citep{Cohen:2003xe}
  for details.}
\label{table:1}
\end{table}
We can then use these $r$-charges to define a $Z_N\times Z_N$
projection which creates the lattice  shown in
Fig.~\ref{fig:2dlat}.  Note that the placement of each degree of
freedom in this figure is simply determined by the corresponding $\vec
r$ charge appearing in Table~\ref{table:1}.

We won't give any of the details here, but it is not too difficult to
construct the lattice action by substituting the orbifold projected
matrices back into the action of the mother theory, \eq{mom}.  One
then follows the path of deconstruction, expanding the boson fields as
\beq
z_i = \frac{1}{a\sqrt{2}}\,{\mathbf 1}_k + \frac{s_i + i
  v_i}{\sqrt{2}}
\eeq
and taking the continuum limit $a\to 0$ with $g^2 a^2 =g_2^2$ kept
fixed.  Amazingly enough, one finds the target theory \eq{targ2} in
this limit, with the identification
\beq
s = \frac{s_1+i s_2}{\sqrt{2}}\ ,\qquad \psi = \begin{pmatrix}
  \lambda_1\cr \lambda_2\end{pmatrix}\ ,\qquad
\bar\psi = \left(\bar\lambda_1 \ \ \bar\lambda_2\right)\ ,\qquad
v_m = ( v_1\ \ v_2)\ .
\eeq
So what about the list of obstructions mentioned  at the end
of \S4?  How does this theory get around them?  Well, for one thing,
the conundrum of satisfying $\{Q,\bar Q\}=\gamma .P$ when there is no $P$
operator on the lattice is circumvented by the fact that we have a $Q$
charge on our lattice, but no corresponding $\bar Q$!  This follows
because our construction leads to a single
site fermion $\lambda_1$ (which has $\vec r= \{0,0\}$) but no
corresponding site variable $\bar 
\lambda_1$. This is one of
the funny things about supersymmetry
in Euclidean spacetime:  it is possible to have a theory respecting
a single supercharge, which is impossible in Minkowski space. This feature
is related to the strange property of fermions continued to
Euclidean space, that $\bar\psi$ is not related to the hermitean
conjugate of $\psi$, i.e,  $\bar\psi \neq \psi^{\dagger} \gamma_{0}$. 
  
Other questions raised in \S4 remain to be answered:
for example, the
target theory has a chiral $U(1)$ $R$-symmetry which is exact up to
anomalies; how does this symmetry
arise in the lattice theory?  Did we invent a new type of lattice
chiral fermion?  Also, we are
claiming that the scalar $s$ in the target theory is represented on
the lattice by $s_1$ and $s_2$ (the real parts of $z_{1,2}$) which are
link variables;  this means 
that even though $s_1$ and $s_2$ transform into each other
non trivially under lattice rotations, they must be invariant under
rotations in the continuum!  Isn't this absurd, since the continuum
rotations contain lattice rotations as a subgroup, and an object
transforming non trivially under the latter must transform non trivially
under the former?

To understand what is going on, let us first  focus on the quadratic
part of the  boson action, which looks 
like:
\begin{align}
  \begin{split}
   \frac{1}{2 g^2 a^2}\, \sum_{\mathbf n}
    \Tr&\left[ \left(
        s_{1,\mathbf{n}-\xh} - s_{1,\mathbf n} +
    s_{2,\mathbf{n}-\yh} -  s_{2,\mathbf n} \right)^2
    \right. \\\notag
    &+\left.\Bigl\vert \left(s_{1,\mathbf{n} + \yh} - s_{1,\mathbf n} +
        s_{2,\mathbf n} - s_{2,\mathbf{n}+\xh}\right)
    -i\left(v_{1,\mathbf{n} + \yh} - v_{1,\mathbf n} -
    v_{2,\mathbf{n}+\xh}+  
        v_{2,\mathbf n}\right)\Bigr\vert^2 \right]
    \end{split}\\
    =\frac{1}{2 g^2}\, \sum_{\mathbf n}\Tr&\left[
     \sum_{\boldsymbol{\hat{\mu}}} \sum_{i=1,2}\left(\frac{s_{i,\mathbf n}-s_{i,\mathbf
            n-\boldsymbol{\hat{\mu}}}}{a}\right)^2 +
      \left(\frac{v_{1,\mathbf{n} + \yh} - v_{1,\mathbf n}}{a} -
        \frac{v_{2,\mathbf{n}+\xh} - v_{2,\mathbf
            n}}{a}\right)^2\right]\ , \eqn{bosii}
\end{align}
When we take the continuum limit, we get
\beq
\frac{1}{g_2^2} \int d^2x\, \half \Tr \left[(\partial_1 s_1 +
  \partial_2 s_2)^2 + (\partial_2 s_1 - \partial_1 s_2)^2 +
  (\partial_2 v_1 - \partial_1 v_2)^2\right]\ .
\eeq
Note that the first two terms make $(s_1,s_2)$ look like a vector (as
you would expect from link variables!) rather than components of
scalar:  the first term looks like $(\vec\nabla \cdot\vec s)^2$, while
the second term looks like $(\vec\nabla\times\vec s)^2$; neither term
looks like the scalar kinetic term $\left[(\partial_m s_1 )^2 +
  (\partial_m s_2)^2 \right]$...yet amazingly enough, when you add
the two terms and integrate by parts, that is $exactly$ what you get! Not only
do we get the correct $SO(2)$ Euclidean ``Lorentz'' invariance with
$s_i$ being invariant, but we get an $independent$ internal $SO(2)$
symmetry where the $s_i$ rotate into each other while the derivatives
$\partial_m$ remain unchanged.  The latter $SO(2)=U(1)$ is just the
$R$-symmetry's action on the scalar $s$!

\begin{figure}[t]
\centerline{\resizebox{10.0cm}{!}{%
 \includegraphics{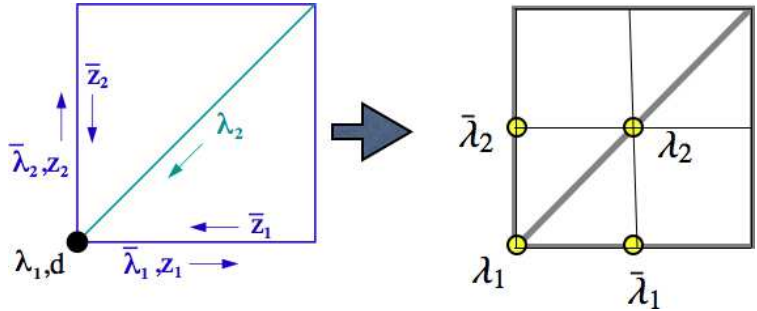} }}
\caption{ The fermions mapped onto a lattice with half the spacing
  can be recognized as reduced staggered fermions.}
\label{fig:stag}
\end{figure}

If we turn to the quadratic part of the fermion action, we find
something more familiar.  If one takes our rather
unconventional lattice, and superimpose upon it a lattice with
spacing $a/2$, the fermions can all be mapped onto sites of this finer
lattice, as shown in
Fig.~\ref{fig:stag}.  Examining the lattice action for these fermions
in the coordinates of this sublattice, one discovers that the fermions
are none other than ``reduced staggered fermions'' as discussed in
\citep{Smit:2002ug}.  Again, you might wonder how a collection of
fermions scattered over different parts of the lattice could
reassemble themselves into a continuum spinor; it seems as mysterious
as how our 
link bosons became a complex scalar.   Understanding these
features goes a long way toward explaining how the obstacles facing
lattice supersymmetry have been circumvented by this orbifold projection
technique.  

%

\subsection{Fine tuning}
\label{sec:9e}

We will finish up this section with a brief discussion about quantum
corrections in our lattice theory for $(2,2)$ SYM. Recall that the
goal of a supersymmetric lattice action was to prevent unwanted
relevant or marginal operators from being radiatively generated which
could spoil supersymmetry in the the continuum limit. 
 The single exact lattice supercharge is enough to protect
the lattice theory from unwanted radiatively induced operators which
could spoil the supersymmetric continuum limit of the lattice theory,
just as we hoped. To show this we can construct the {\it Symanzik
  action} for the theory: we expand the $z$ variables about the flat
direction 
$\vev{z}= {\mathbf 1}_k/a\sqrt{2}$, expand the action for smooth
fields in powers of 
$1/a$,  include all operators allowed by the exact symmetries of the
lattice, and then consider radiative corrections to the coefficients
of these operators, paying special attention to relevant and marginal
operators which violate the full $\CQ=4$ supersymmetry of the target
theory, and whose coefficients by definition do not vanish in the
$a\to 0$ limit. The key is to identify all the operators allowed by
the exact lattice symmetries, which include the single supersymmetry.
This is most easily done by constructing superfields: we introduce a
Grassmann coordinate $\theta$, which has mass dimension $1/2$ (where
spacetime coordinates $x$ have mass dimension $-1$), and define the
exact lattice supercharge to be $Q=\partial_\theta$. With this
definition of $Q$, and knowing the action of $Q$ on the lattice
variables, it is possible to construct superfields as is done in the
more familiar $d=4$, $\CN=1$ supersymmetry \citep{Wess:1992cp}.  One
finds the following superfields on the unit cell at site $\bfn$:
\begin{eqnarray}
Z_1(\bfn) &=& z_1(\bfn) + \sqrt{2}\theta\bar\lambda_1(\bfn)\ ,\cr
Z_2(\bfn) &=& z_2(\bfn) + \sqrt{2}\theta\bar\lambda_2(\bfn)\ ,\cr
\Xi(\bfn) &=& \lambda_2(\bfn) + 2\left[\bar z_1(\bfn+\yh)\bar z_2(\bfn)
  - \bar z_2(\bfn + \xh) \bar z_1(\bfn)\right]\theta\ ,\cr
\Lambda(\bfn)&=&\lambda_1(\bfn) -\left[\bar z_1(\bfn-\xh)z_1(\bfn-\xh)
  - z_1(\bfn)\bar z_1(\bfn)\right.\cr &&\qquad\qquad \left. + \bar
  z_2(\bfn-\yh) z_2(\bfn-\yh) - 
  z_2(\bfn) \bar z_2(\bfn) + i d(\bfn)\right]\theta\ .
\end{eqnarray}
Since
$Q= \partial_\theta$, the most general supersymmetric
action  can be written as
\beq
\frac{1}{g_2^2}\int d\theta \int d^2x \, \sum_\CO C_\CO\CO(x,\theta)
\eeq
where the  $\CO$ are local Grassmann operators.  This expression is
obviously annihilated by $Q$ since it doesn't depend on
$\theta$. Since the action has 
to be dimensionless, if $\CO$ has 
mass dimension $p$, it is easy to check that the operator coefficient
$C_O$ must have dimension $(7/2-p)$. Now, since the action has a
$1/g_2^2$ out front (where $g_2$ has mass dimension 1), radiative
corrections to $C_\CO$ at $\ell$ loops will be of the form
\beq
\delta C_\CO \sim c_\ell a^{(p-7/2)} (g_2^2 a^2)^\ell\ ,
\eeq
where the $c_\ell$ are dimensionless coefficients and can only depend
on $a$ logarithmically.
Since we only care about operator coefficients which do not vanish as
$a\to 0$, we need only consider operators and loops satisfying $p\le
(7/2-2\ell)$.  At $\ell=0$ (tree level) we claim our lattice action
gives the correct target theory in the continuum limit.  At $\ell=1$
we need to consider $p\le 3/2$; 
 it turns out we cannot construct operators with $p\le 1/2$ so
that's it!  It is then a quick job to convince oneself that there are
no bad operators $\CO$ with $p=3/2$ which one can construct.
Therefore,  we can prove that the supersymmetric lattice does what
it was supposed to do: it allows one to realize the supersymmetric target
theory without fine-tuning.  The exact supersymmetry of the lattice
was crucial for this to be possible.  For example, in a  non-supersymmetric lattice formulation, 
scalar mass terms are permitted and needs to be fine-tuned, which is forbidden in a lattice theory with exact supersymmetry. 
 We refer interested readers to
ref.~\citep{Cohen:2003xe} for details of the argument. The analysis for
this theory was simplified by the fact that it is
``super-renormalizable'', namely that each loop correction introduced
positive powers of $a$.   In \S \ref{sec:10d}, we briefly discuss  renormalization of a $d=4$ theory, in which divergent contributions may arise at arbitrary loop order. 

\subsection{Other supersymmetric lattices}
\label{sec:9f}

\begin{figure}[t]
\centerline{\resizebox{10.0cm}{!}{%
 \includegraphics{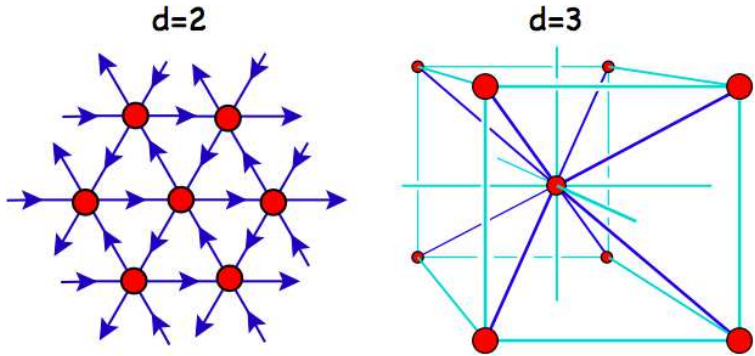} }}
\caption{The lattices for $\CQ=16$ supersymmetry in $d=2$ and $d=3$
  dimensions \citep{Kaplan:2005ta}. In $d=4$ the lattice for $\CN=4$ SYM
has the $A_4^{*}$ lattice structure.}
\label{fig:sixteen}
\end{figure}

So what are the supersymmetric lattices we have constructed to date?
SYM theories exist in $d\ge 2$ with $\CQ=2,4,8,16$.
Since each dimension requires projecting out a $Z_N$ factor, and each
projection costs one half of the remaining supersymmetries of the
mother theory, and we want at least one unbroken supercharge on the
lattice, we can only consider SYM theories with $\CQ\ge 2^{d}$.
That constrains us to 
\begin{eqnarray}
\CQ=4:\quad d&=&2 \cr
\CQ=8:\quad d&=&2,3\cr
\CQ=16:\quad d&=&2,3,4\ ,
\end{eqnarray}
and all of these lattice have been constructed.  As we have
discussed the $\CQ=16$ theories
are especially interesting and have especially symmetric lattices,
shown in Fig.~\ref{fig:sixteen}.

%

The $d=1$ lattices for $\CQ=16$ SYM give an alternative
to the naive lattice actions we introduced before
for simulating  $\CQ=16$ quantum mechanics.  
In addition to pure SYM theories, a lattice for $(2,2)$ SYM has also been
constructed with certain classes of matter fields
\citep{Endres:2006ic}, which we discuss next. 

\subsection{Addition of matter to $(2,2)$ theories}
\label{sec:9g}
The supersymmetric lattice construction of the last section can be generalized to include
charged matter fields 
interacting via a superpotential \citep{Endres:2006ic}.  
In \citep{Giedt:2006dd} Giedt extended this construction to $(4,4)$ theories 
with matter.  Such theories are lower dimensional counterparts of 
super QCD in $d=4$ dimensions. 
 More recently orbifold/twisted constructions have been
obtained with matter in the fundamental representation \citep{Matsuura:2008cf,Sugino:2008yp}. These
are very interesting as they open up the possibility of defining phenomenologically
more realistic models with exact supersymmetry.

\section{${\cal N}=4$ SYM in four dimensions}
\label{sec:10}

In this section we discuss perhaps one of the most interesting applications
of these ideas -- the construction of a lattice model invariant under
a single exact supersymmetry whose naive continuum limit is ${\cal N}=4$
SYM theory in four dimensions. This gauge theory  is thought to
be dual to type IIB string theory in $AdS^5\times S^5$ space. In the large 't Hooft coupling limit, 
it is conjectured to be describe the supergravity limit of 
that string theory.   The lattice theory  
constitutes the only known example of a supersymmetric lattice model
in four dimensions\footnote{Recently 
a paper has appeared which shows how a non-supersymmetric 
formulation of $\CN=4$ SYM can be constructed with 
domain wall fermions which has 
significantly reduced  fine-tuning compared to what one might expect, 
and which may be numerically tractable 
\citep{Elliott:2008jp}; see also \citep{Giedt:2009yd}.
A alternative regularization for ${\cal N}=4$ SYM in the planar
limit was given in \citep{Ishiki:2008te}}. We first summarize the construction of the
supersymmetric orbifold action for this model, then go on to re-derive it
by discretization of an appropriate twist of the continuum
theory.

\subsection{Orbifold action}
\label{sec:10a}

The strategy to obtain the supersymmetric lattice action for the $\CN=4$ target theory is 
four-dimensional generalization of  the one given for  ${\cal N}=(2,2)$ model in two dimensions. 
As discussed in  \S \ref{sec:9b}, 
to build the ${\cal N}=(2,2)$ model we orbifolded the 
${\cal Q}=4$ mother matrix 
theory.
This mother theory  possesses an $ SO(4) \times U(1)$ R-symmetry group with a maximal abelian  $U(1)^3$ subgroup which allowed us to
build a lattice theory in two dimensions.

To obtain the target  $\CN=4$ target theory in four
dimensions, we start with  $\CQ=16$ matrix model. The matrix model may be obtained by dimensionally reducing the $d=10$ dimensional $\CN=1$ SYM theory down to $d=0$ dimensions.  The reduced model possesses  $SO(10)$ $R$-symmetry inherited 
from the Lorentz symmetry of the $d=10$ dimensional theory prior to reduction.    The field content of the mother theory is  ten bosonic and sixteen Grassmann odd fermionic matrices transforming as  ${\bf 10}$ and   ${\bf 16}$ representation of the $R$-symmetry, and 
in the adjoint representation of the gauge group.

To proceed, it is more convenient to decompose  the variables of the mother theory under the $SU(5) \times U(1)$ subgroup of $SO(10)$.  
 \begin{eqnarray}
&&{\rm bosons:} \;  {\bf 10}\to {\bf 5} \oplus\mybar {\bf 5} = z^m   \oplus \mybar z_m  \cr
&& {\rm fermions:} \;  {\bf 16} \to {\bf 1} \oplus {\bf 5} \oplus  \mybar {\bf 10} = \lambda \oplus \psi^m \oplus \xi_{mn}. 
\end{eqnarray}
Written in terms of this $SU(5)\times U(1)$ decomposition, the action
of the type IIB matrix theory becomes 
\begin{eqnarray}
S &=& \frac{1}{g^2} \Tr \Bigl[ \sum_{m,n} \left( \frac{1}{2}[\mybar z_m, z^m][\mybar z_n, z^n] + 
[z^m, z^n][\mybar z_n,\mybar z_m]\right)\Bigr.\cr
&&\qquad\quad\Bigl. +  
{\sqrt 2}\left( \lambda [\mybar z_m, \psi^m] - 
\xi_{mn}[z^m, \psi^n] + \frac{1}{8} \epsilon^{mnpqr}
\xi_{mn}[\mybar z_p, \xi_{qr}] \right) \Bigr]\ .
\eqn{act2}
\end{eqnarray}
 Below, we employ the  $U(1)^5$ abelian subgroup of the $R$-symmetry group to generate the four dimensional  lattice with one exact supersymmetry. As usual, the starting point is the mother theory with  $U(N^d k)$  gauge group with  $d=4$.  An  orbifold projection by   $(Z_N)^4$ symmetry generates the four dimensional lattice.  
The ten bosonic and sixteen fermionic lattice fields, their charges under $U(1)^5$ and 
their   associated ${\bf r}$-charges (which determines the position and orientation  of each lattice field on the unit cell)  are given  in Table~\ref{tab5}.  In Table~\ref{tab5}, we also define five  
${\bfmu}_m $ vectors which will be used to specify the ${\bf r}$-charges directly in terms of 
$SU(5)$ tensor indices.  
 For the further details of this procedure, 
we refer to \citep{Kaplan:2005ta} for details. 

\begin{table}
\begin{tabular}[t]
{|r||r||r|r|r|r|r||rcl|}
\hline
& $2Q_0$\ & $2q_1\ $ & $2q_2\ $ & $2q_3\ $ & $2q_4\ $ & $2q_5\ $ &
&\bfr&
\\ \hline
$z_1$ &  $\, \ 2$ &$\,\ 2$ & $\, \ 0$ & $\, \ 0$ & $\, \ 0$ & $\, \ 0$ & 
$\{1,0,0,0\}$&=&$\ \ \bfmu_1$\\ 
$ z_2$&  $\, \ 2$ &$\, \ 0$ & $\,\ 2$ & $\, \ 0$ & $\, \ 0$ &$\, \ 0$ & 
$\{0,1,0,0\}$&=&$\ \ \bfmu_2$\\
 $z_3$ &$\, \ 2$  & $\, \ 0$ & $\, \ 0$ & $\,\ 2$ & $\, \ 0$ & $\, \
 0$ & %
$\{0,0,1,0\}$&=&$\ \ \bfmu_3$\\
$z_4$ & $\, \ 2$  & $\, \ 0$ & $\, \ 0$ & $\, \ 0$ & $\,\ 2$ & $\, \ 0$&
$\{0,0,0,1\}$&=&$\ \ \bfmu_4 $\\
$z_5$ &  $\,\ 2$ & $\, \ 0$ & $\, \ 0$ & $\, \ 0$ & $\, \ 0$ &$\, \ 2$ &
$\{-1,-1,-1,-1\}$&=&$\ \ \bfmu_5$\\ \hline
$\mybar z_1 $ & $-2$ &  $\,\ -2$  & $\, \ 0$ & $\, \ 0$ & $\, \ 0$ & $\, \ 0$ &
$\{-1,0,0,0\}$&=&$-\bfmu_1$\\ 
$\mybar z_2$& $-2$  & $\, \ 0$ & $\, \ -2$ & $\, \ 0$ & $\, \ 0$ & $\, \ 0$&
$\{0,-1,0,0\}$&=&$-\bfmu_2$ \\
$\mybar z_3$ &  $-2$ &$\, \ 0$ & $\, \ 0$ &  $\,\ -2$ & $\, \ 0$ & $\, \ 0$ &
$\{0,0,-1,0\}$&=&$-\bfmu_3$ \\
$\mybar z_4$ &  $-2$ &$\, \ 0$ & $\, \ 0$ & $\, \ 0$ & $-2$ & $\, \ 0$ &
$\{0,0,0,-1\}$&=&$-\bfmu_4$ \\
$ \mybar z_5$ &  $-2$ &$\, \ 0$ & $\, \ 0$ & $\, \ 0$ & $\, \ 0$ & $\,\ -2$ &
$\{1,1,1,1\}$&=&$-\bfmu_5$\\   \hline   \hline 
$\lambda$ & \, \ $5$ & $\ 1$ & $\ 1$ & $\ 1$ &
$\ 1$ & $\ 1$ &
$\{0,0,0,0\}$&=&\,\ \ {\bf 0}\\ \hline 
$\psi_1$ &  $-3$ & $\ 1$ & $-1$ & $-1$ & $-1$ & $-1$ &
$\{1,0,0,0\}$&=&$\ \ \bfmu_1$
\\
$\psi_2$  & $-3$ & $-1$ & $\ 1$ & $-1$ & $-1$ & $-1$ &
$\{0,1,0,0\}$&=&$\ \ \bfmu_2$ \\ 
$\psi_3$& $-3$ & $-1$ & $-1$ & $\ 1$ & $-1$ &
$-1$ &
$\{0,0,1,0\}$&=&$\ \ \bfmu_3$\\
$\psi_4$ & $-3$ & $-1$ & $-1$ & $-1$ & $\ 1$ & $-1$ &
$\{0,0,0,1\}$&=&$\ \ \bfmu_4 $\\
$\psi_5$ & $-3$ & $-1$ & $-1$ & $-1$ & $-1$ & $\ 1$ &
$\{-1,-1,-1,-1\}$&=&$\ \ \bfmu_5$\\  \hline 
$\xi_{12}$& \, \ $1$ & $-1$ & $-1$ & $\ 1$ & $\ 1$ & $\ 1$ &
$\{-1,-1,0,0\}$&=&$-\bfmu_1 - \bfmu_2$\\
$\xi_{13}$ & \, \ $1$ & $-1$ & $\ 1$ & $-1$ & $\ 1$ & $\ 1$ &
$\{-1,0,-1,0\}$&=&$-\bfmu_1-\bfmu_3$\\
$\xi_{14}$& \, \ $1$ & $-1$ & $\ 1$ & $\ 1$ &
$-1$ & $\ 1$ &
$\{-1,0,0,-1\}$&=&$-\bfmu_1-\bfmu_4$ \\
$\xi_{23}$ & \, \ $1$ & $\ 1$ & $-1$ & $-1$ &
$\ 1$ & $\ 1$ &
$\{0,-1,-1,0\}$&=&$-\bfmu_2-\bfmu_3$\\
$\xi_{24}$& \, \ $1$ & $\ 1$ & $-1$ & $\ 1$ & $-1$ & $\ 1$ &
$\{0,-1,0,-1\}$&=&$-\bfmu_2-\bfmu_4$  \\
$\xi_{34}$ & \, \ $1$ & $\ 1$ & $\ 1$ & $-1$ & $-1$ & $\ 1$ & 
$\{0,0,-1,-1\}$&=&$-\bfmu_3-\bfmu_4$ \\
$\xi_{15}$& \, \ $1$ & $-1$ & $\ 1$ & $\ 1$ &
$\ 1$ & $-1$ &
$\{0,1,1,1\}$&=&$-\bfmu_1-\bfmu_5$ \\
$\xi_{25}$& \, \ $1$ & $\ 1$ & $-1$ & $\ 1$ &
$\ 1$ & $-1$ &
$\{1,0,1,1\}$&=&$-\bfmu_2 - \bfmu_5$ \\ 
$\xi_{35}$ & \, \ $1$ & $\ 1$ & $\ 1$ & $-1$ & $\ 1$ & $-1$ &
$\{1,1,0,1\}$&=&$-\bfmu_3-\bfmu_5$\\
$\xi_{45}$ & \, \ $1$ & $\ 1$ & $\ 1$ & $\ 1$ &
$-1$ & $-1$ &
$\{1,1,1,0\}$&=&$-\bfmu_4-\bfmu_5$\\  \hline
 \end{tabular} 
\caption{ The $Q_0$, ${q_m}$ and $r_\mu = (q_\mu-q_5)$ charges of the
  bosonic variables $v$
  and fermionic variables $\omega$  of the $\CQ=16$ mother  theory under the
  $ SO(10)\supset SU(5)$ decomposition 
$v=10\to 5\oplus\mybar 5 = z^m   \oplus \mybar z_m $, and 
$\omega= 16 \to 1\oplus 5\oplus  \mybar {10} = \lambda \oplus \psi^m \oplus \xi_{mn}$. 
 }
\label{tab5}
\end{table}
The action of the lattice gauge theory that results from the orbifold projection may be written in component form as    \citep{Kaplan:2005ta}  
\begin{equation}
\begin{aligned} 
S =
  \frac{1}{g^2}\sum_{\bfn} \Tr \biggl[ & \half\Bigl(\sum_{m=1}^5 \left(\mybar z_m(\bfn-\bfmu_m)
    z^m(\bfn-\bfmu_m) - z^m(\bfn)\mybar z_m(\bfn)\right)
\Bigr)^2\\&
  +\sum_{m,n=1}^5
\Bigl\vert\, z^m(\bfn) z^n(\bfn + \bfmu_m)-  z^n(\bfn) z^m(\bfn + \bfmu_n)
\Bigr\vert^2
  \\ & 
-\sqrt{2}\Bigl( \Delta_\bfn(\lambda,\mybar z_m, \psi^m)+
  \Delta_\bfn(\xi_{mn}, z^m,\psi^n)
  +\frac{1}{8}\epsilon^{mnpqr}\Delta_\bfn(\xi_{mn},\mybar z_p,\xi_{qr})
\Bigr) 
\biggr] 
\\ &
\eqn{d4lat}
\end{aligned} 
\end{equation} 
 We have introduced the labeling convention that
$z^m(\bfn)$, $\psi^m(\bfn)$ and $\mybar z_m(\bfn)$ live on the same link, running
between site $\bfn$ and site $(\bfn+\bfmu_m)$; similarly
$\xi_{mn}(\bfn)$ lives on the link between sites $\bfn$ and
$(\bfn+\bfmu_m + \bfmu_n)$, while $\lambda(\bfn)$ resides at the site
$\bfn$.  The site vector $\bfn$, a four-vector with integer-valued
components,  should be distinguished from $SU(5)$
indices $n$. 

We have introduced the triangular plaquette function $\Delta_\bfn$
defined as:
\begin{equation}
\begin{aligned} 
 \Delta_\bfn(\lambda,\mybar z_m, \psi^m)=&-
\lambda(\bfn) \Bigl(\mybar z_m(\bfn-\bfmu_m)  \psi^m(\bfn-\bfmu_m) -
\psi^m(\bfn)\mybar z_m(\bfn)\Bigr)\ ,\\ 
 \Delta_\bfn(\xi_{mn}, z^m,\psi^n)=&   \xi_{mn}(\bfn)\Bigl(
   z^m(\bfn) \psi^n(\bfn+\bfmu_m) - \psi^n(\bfn)
   z^m(\bfn+\bfmu_n)\Bigr)\ ,\\
\Delta_\bfn(\xi_{mn},\mybar z_p,\xi_{qr})=&-\xi_{mn}(\bfn)\Bigl(\mybar
z_p(\bfn-\bfmu_p) \xi_{qr}(\bfn+\bfmu_m+\bfmu_n) \\&\qquad\qquad- 
\xi_{qr}(\bfn-\bfmu_q-\bfmu_r)\mybar z_p(\bfn+\bfmu_m+\bfmu_n)\Bigr)
\end{aligned}
\eqn{Deltadef}
\end{equation}
Note that $\Delta$ corresponds to the signed sum of two terms, each of
which is a string of three
variables along a closed and oriented path on the lattice, with the
sign determined by the orientation of the path.
 As discussed in \S~\ref{sec:2}, there is a $U(k)$ gauge
symmetry associated with each site, with $\lambda(\bfn)$ transforming
as an adjoint, while the oriented link variables transform as bifundamentals
under the two $U(k)$ groups associated with the originating and
destination sites of the link. A  string of variables along any closed
path on the lattice, such as we see in the definition of $\Delta$, is
gauge invariant. In the continuum limit, the $\Delta$ terms will form
the gaugino hopping terms and Yukawa couplings of the $\CQ=16$ SYM theory.

It is now simple to write down the action for the lattice theory that
results from the orbifold projection, in a form which is manifestly
$\CQ=1$ supersymmetric.

    After orbifold projection, there are
superfields associated with each lattice site $\bfn$, where $\bfn$ is a four component
vector of integers, each component ranging from 1 to $N$:
\begin{eqnarray}
{\bf Z}^m({\bf n})  &=&  z^m(\bfn) + \sqrt 2  \theta \psi^m(\bfn) 
\cr
{\bf \Lambda}(\bfn) &=& \lambda(\bfn)  -\theta  id(\bfn) \cr
{\bf \Xi}_{mn}(\bfn) &=& \xi_{mn}(\bfn) -2 \theta  
\left[\, \mybar z_m(\bfn+ \bfmu_n) \mybar z_n(\bfn) - \mybar z_n(\bfn + \bfmu_m ) 
\mybar z_m(\bfn)\right]
\end{eqnarray}
In addition there is the singlet field $\mybar z_m(\bfn)$. 
In the above expressions, subscripts and superscripts $m,n=1,\ldots,5$ and repeated indices are
summed over.  Note that the superfields are not
entirely local, and that in the continuum they will depend on
derivatives of fields as well as the fields themselves.

The lattice action we obtained may be written in manifestly  $\CQ=1$ 
supersymmetric form as 
\begin{eqnarray}
S &=& \frac{1}{g^2}\Tr \sum_{\bfn}  \int \;  d \theta \left( 
-\frac{1}{2} {\bf \Lambda}(\bfn) {\partial}_{\theta} {\bf \Lambda}(\bfn)  
- {\bf\Lambda}(\bfn)\Bigl[
\mybar z_m(\bfn - \bfmu_m) 
{\bf Z}^m(\bfn - \bfmu_m)
 - {\bf Z}^m(\bfn) \mybar z_m(\bfn) \Bigr]\right. \cr &&\qquad\qquad\qquad\qquad
 + \left.
\frac{ 1}{2}
{\bf \Xi}_{mn}(\bfn)\Bigl[{\bf Z}^m(\bfn) {\bf Z}^n (\bfn + \bfmu_m)- {\bf
 Z}^n(\bfn) 
 {\bf Z}^m(\bfn + \bfmu_n)\Bigr]
\right) \cr
&&
 +
  \frac{\sqrt 2 }{8}  \epsilon^{mnpqr}{\bf \Xi}_{mn} (\bfn)
 \Bigl[\mybar z_p (\bfn-  \bfmu_p)  {\bf \Xi}_{qr}( \bfn +  \bfmu_m + 
\bfmu_n )
-{\bf \Xi}_{qr}(\bfn -  \bfmu_q - \bfmu_r) \mybar z_p(\bfn +  
\bfmu_m + \bfmu_n) \Bigr]
\cr&&
\eqn{d4latss}
\end{eqnarray}
The auxiliary field $d(\bfn)$ has no hopping term, and after
eliminating it by the equations of motion on can show that the above
action in terms of superfields is equivalent to the lattice action
given in component form in \eq{d4lat}.
Formulating the action in this supersymmetric 
facilitates the analysis of allowed operators and the continuum limit of
the lattice theory.

The lattice defined by the orbifold projection cannot be directly
considered to be a spacetime lattice, as all terms in the lattice
action are trilinear and  conventional
hopping terms are absent. To generate a spacetime lattice and take the continuum
limit one must follow the example of deconstruction
\citep{ArkaniHamed:2001ca} and follow a particular trajectory  out to
infinity in the
moduli space of the theory, interpreting the distance
from the origin of moduli space as the inverse lattice spacing.

As can be seen in \eq{d4lat}, the moduli space in the present theory  corresponds to all values for
the bosonic $z$ variables such that
\begin{equation}
\begin{aligned} 
0=\sum_{\bfn} \Tr \biggl[ & \half\Bigl(\sum_m \left(\mybar z_m(\bfn-\bfmu_m)
    z^m(\bfn-\bfmu_m) - z^m(\bfn)\mybar z_m(\bfn)\right)
\Bigr)^2\\&
  +
\sum_{m,n}
\Bigl\vert\, z^m(\bfn) z^n(\bfn + \bfmu_m)-  z^n(\bfn) z^m(\bfn + \bfmu_n)
\Bigr\vert^2\biggr] .
\end{aligned}
\end{equation}

\subsubsection{A hypercubic lattice}
\label{sec:10b.1}

There are clearly a large class of solutions to these equations.
One possibility is
\begin{eqnarray}
z^m(\bfn) &=& \mybar z_m(\bfn) = \frac{1}{a\sqrt{2} }   {\bf 1}_k
,\qquad  m= 1, \ldots, 4,\cr
z^5(\bfn) &=& \mybar z_5(\bfn) = 0\ ,
\end{eqnarray}
where $a$ is the length scale associated with the lattice spacing,
interpreted as the physical length (up to a factor of $4/5$) of the links on which $z_m$ and
$\mybar z^m$ variables reside, for $m=1,\ldots,4$. Such a  lattice can be interpreted as a hypercubic lattice
of length $a$ on an edge, since the $\bfr$ charges for these variables
correspond to Cartesian unit vectors,
as seen  in Table~\ref{tab5}.  In this case, the physical location of
site $\bfn$ is simply the four-vector  $\bfR = a \bfn$.
Various fields of the $SU(5)$ multiplets distribute to the  hypercubic lattice as follows: 
\begin{eqnarray} 
&&\lambda \rightarrow \lambda, \qquad   0-{\rm cell} \cr
&& \psi^m  \rightarrow (\psi^\mu, \psi^5)  =  (\psi^\mu, \frac{1}{4!}\epsilon^{\mu \nu \rho \sigma} \psi_{\mu \nu \rho \sigma}), \qquad    ( 0-{\rm cell},    4-{\rm cell})  \cr 
&& \xi_{mn}  \rightarrow (  \xi_{\mu \nu}, \xi_{\mu 5} )  = 
 (  \xi_{\mu \nu}, \frac{1}{3!}\epsilon_{\mu \nu \rho \sigma} \xi^{\nu \rho \sigma}), \qquad    ( 2-{\rm cell},    3-{\rm cell})  
 \eqn{split3}
\end{eqnarray}
In other words, the fermions are totally anti-symmetric $p$-cell variables, which one would naturally associate with the
$p$-form representation of $SO(4)$ of the continuum.
Thus the fermionic content of the hypercubic lattice construction
provides
an explicit realization of \KD fermions.   
The distributions of bosons  such as 
$z^m \rightarrow  (z^{\mu}, z^5) =   (z^{\mu},z^5) = (z^\mu, \frac{1}{4!}\epsilon^{\mu \nu \rho \sigma} z_{\mu \nu \rho \sigma}) $ 
as well as their orientations are dictated by the fermions because of exact supersymmetry. The  symmetry of the hypercubic lattice action is  
 $S_4$,  much smaller than the hypercubic group, since the fields are oriented.

\subsubsection{The $A_4^*$ lattice and point group symmetry}
\label{sec:10b.2}

Instead of the above trajectory, we can examine the most 
symmetric solution, in the hopes
that the greater the symmetry of the spacetime lattice, the fewer
relevant or marginal operators will exist.  A solution which
treats all five $z^m$ symmetrically (and thus preserves an
$S_5$ permutation symmetry) is to have the five links on which they
reside correspond to the vectors connecting the center of a 4-simplex to
its corners. The lattice generated by such vectors is 
known to mathematicians as $A_4^*$\footnote{The $A_4$
  lattice is generated by the 
simple roots of $SU(5)=A_4$; then $A_4^*$ is the dual lattice,
generated by the fundamental weights of $SU(5)$, or equivalently, by
the weights of the defining representation of $SU(5)$. Lower dimension
analogues are  $A_2^*$, the triangular lattice, and $A_3^*$, the
body-centered cubic lattice.}.
We thus expand about the 
symmetric point:
\begin{eqnarray}
z^m(\bfn) &= \mybar z_m(\bfn) = \frac{1}{a\sqrt{2}}    {\bf 1}_k
,\qquad  m= 1, \ldots, 5\ .
\eqn{s5}\end{eqnarray}
Once again  $a$ is interpreted as the spacetime length of the link
that each $z^m$ resides upon.    

The physical point group symmetry   of the lattice is isomorphic to permutation
group $S_5$,  the Weyl group of $SU(5)$, corresponding to the permutations of the 
group indices of $SU(5)$. 
The character table and conjugacy classes of $S_5$ are 
given in  Table \ref{tab:S5}. 
\begin{table}[t]
\centerline{
\begin{tabular}
{|c|c|c|c|c|c|c|c|} \hline
classes: & (1)  & (12) & (123) & (1234)&(12345) & (12)(34) & (12)(345)\\ \hline
sizes:   & 1  & 10   &  20   & 30    &  24    &  15      & 20       \\ \hline
$\chi_1$ & 1  &   1  &   1   &  1    &   1    &    1     &   1      \\ 
$\chi_2$ & 1  &  -1  &   1   & -1    &   1    &    1     &  -1      \\ 
$\chi_3$ & 4  &   2  &   1   &  0    &  -1    &    0     &  -1      \\ 
$\chi_4$ & 4  &  -2  &   1   &  0    &  -1    &    0     &   1      \\ 
$\chi_5$ & 5  &  -1  &  -1   &  1    &   0    &    1     &  -1      \\ 
$\chi_6$ & 5  &   1  &  -1   & -1    &   0    &    1     &   1      \\ 
$\chi_7$ & 6  &   0  &   0   &  0    &   1    &    -2     &   0     \\ \hline
\end{tabular} }
\caption{\sl The character table of $S_5$, the point symmetry group of
  $A_4^*$ lattice. The even permutations are spacetime rotations, the odd 
permutations involves parity operations and hence improper rotations.}
\label{tab:S5}
\end{table}
The group has $5!=120$ elements and seven conjugacy classes
shown in Table~\ref{tab:S5}.  The symmetry of the lattice action is 
composed of the elements of $S_5$. It is easy to show that {\bf even} permutations
with determinant  one (which can be read off from the  $\chi_2$(or sign)  representation) 
are pure rotational symmetries of the action.  
We see from Table~\ref{tab:S5} that the odd 
permutations have determinant 
minus one, and are not proper rotations. 
In fact, the odd permutations accompanied by  
\begin{equation}
\lambda \rightarrow \lambda, \;  (\psi^m, z^m, \mybar z_m) \rightarrow -(\psi^m, z^m, \mybar z_m),  \;   \xi_{mn} \rightarrow \xi_{mn}
\eqn{oddextra}
\end{equation}
generate additional symmetries of the action. 
Notice that the symmetry of the action is not the full
symmetry of the $A_4^*$ lattice, as reflection symmetries which
exchange $z_m$ and $\mybar z_m$ are not symmetries of the action.

The point group symmetry combined with gauge invariance of the lattice action and exact supersymmetry is very powerful in constraining  
the possible fine tuning required in the continuum limit.  
Representation theory of $S_5$ is  also 
useful in identifying the precise relation between the Marcus's twist and $A_4^*$ lattice formulation of $\CN=4$ SYM. In particular, we will show that the $S_5$ symmetry 
of the $A_4^*$ lattice lives in the twisted Lorentz group, the diagonal sum of the  $R$-symmetry and Lorentz symmetry of the original theory as shown in 
Fig.\ref{fig:sym}. 

To relate the lattice site $\bfn$ with a physical location in
spacetime, we introduce a specific basis, in the form of  five,
four-dimensional lattice vectors 
\begin{eqnarray}
    {\bf e_1} &=& (\frac{1}{\sqrt 2},  \frac{1}{\sqrt 6}, 
\frac{1}{\sqrt{12}}, \frac{1}{\sqrt{20}})  \cr
{\bf e_2} &=& (-\frac{1}{\sqrt 2},  \frac{1}{\sqrt 6},
\frac{1}{\sqrt{12}}, \frac{1}{\sqrt{20}})  \cr
{\bf e_3} &=& (0,  -\frac{2}{\sqrt 6},
\frac{1}{\sqrt{12}}, \frac{1}{\sqrt{20}})  \cr
{\bf e_4} &=& (0, 0,
-\frac{3}{\sqrt{12}}, \frac{1}{\sqrt{20}})  \cr
{\bf e_5} &=& (0, 0,
0, -\frac{4}{\sqrt{20}}).  
\eqn{latvec}\end{eqnarray}
These vectors satisfy the relations
\beq
\sum_{m=1}^5 \bfe_m = 0\ ,\qquad  \bfe_m\cdot \bfe_n =
\left(\delta_{mn}-\frac{1}{5}\right)\ ,\qquad \sum_{m=1}^5
(\bfe_m)_\mu (\bfe_m)_\nu = \delta_{\mu\nu}\ .
\eqn{eprop}
\eeq
The lattice vectors \eq{latvec} are simply related to the $SU(5)$
weights of the {\bf 5} representation, and  the  $5\times 5$ matrix
$\bfe_m\cdot \bfe_n$ can be recognized as the 
Gram matrix for $A_4^*$ \citep{Conway:1991}. 
The site $\bfn$ on our
lattice is then defined to be at the spacetime location 
\beq
\bfR = a \sum_{\nu=1}^4 (\bfmu_\nu\cdot \bfn)\, \bfe_\nu = a
\sum_{\nu=1}^4 \, n_\nu \,\bfe_\nu \ ,
\eqn{ra4}\eeq
where $a$ is the lattice spacing introduced in \eq{s5}, and the
vectors $\bfmu_\nu$ (which have integer components) were defined in
Table~\ref{tab5}.  By making use of the fact that $\sum_m\bfe_m=0$, it is
easy to show that a small lattice displacement of the form $d\bfn =
\bfmu_m$ corresponds to a spacetime translation by $(a\,\bfe_m)$:
\beq
d\bfR =  a \sum_{\nu=1}^4 (\bfmu_\nu\cdot d\bfn)\, \bfe_\nu =  a
\sum_{\nu=1}^4 (\bfmu_\nu\cdot \bfmu_m)\, \bfe_\nu =a\,\bfe_m\ .
\eqn{dra4}\eeq
Thus from the last column in Table~\ref{tab5} one can read
off the physical location of each of the variables.  For example, at
the site $\bfn={\bf 0}$, $z^1({\bf 0})$ lies on the link directed from $\bfR={\bf 0}$ to
$\bfR = a \,\bfe_1$, while $\xi_{45}({\bf 0})$  lies on the link
directed from the site
$\bfR= a \,(\bfe_4 + \bfe_5)$ to the site $\bfR={\bf 0}$. From the
relation \eq{dra4} we see that each of the five links occupied by the
five $z^m$ variables has length $|a\,\bfe_m|=\sqrt{\frac{4}{5}}\, a$, unlike
the case of the hypercubic lattice mentioned above, where $z^5$
resided on a  link twice as long as the links occupied by the other four $z^m$ variables.

\subsection{Twisted construction}
\label{sec:10c}

\subsubsection{Continuum theory -- Marcus twist}
\label{sec:10c.1}

There are three  inequivalent twists of the $\CN=4$ SYM theory in 
four dimensions \citep{Vafa:1994tf, Marcus}.  Two of those do not emerge from the lattice construction due to reasons to be explained later. 
The one we will consider and which corresponds to the orbifold lattice 
construction is due to Marcus.  In addition to its
application in 
lattice supersymmetry \citep{Catterall:2007kn,Unsal:2006qp}, it
also plays an important role
in the geometric Langlands program \citep{Kapustin:2006pk}.  Here, we briefly outline this  twist.

The $\CN=4$ SYM theory in $d=4$ dimensions possesses  a global Euclidean 
Lorentz symmetry 
$SO(4)_E \sim   SU(2) \times SU(2)$ and  a global $R$-symmetry group  
$SO(6) \sim SU(4)$. The $R$-symmetry contains  a subgroup $SO(4)_R \times U(1)$.
To construct the twisted   theory, we take  the diagonal
sum of $SO(4)_E \times SO(4)_R $ and declare it  the new rotation
group. Since the $U(1)$ part of the symmetry group is undisturbed, 
it remains as  a  global $R$-symmetry of the twisted theory.    
Under the global 
 $ G= \big( SU(2) \times SU(2) \big)_E \times 
\big( SU(2) \times SU(2) \big)_R $  symmetry, the fermions transform as 
$(2,1,2,1) \oplus  (2,1,1,2) \oplus 
  (1,2,1,2) \oplus    (1,2,2,1) $.
The same fields, under $G'=SU(2)' \times SU(2)' \times U(1)$ (or  under 
$SO(4)' \times U(1)$) transform 
as\footnote{Twice of the $U(1)$  charge is usually called the ghost number in 
the topological counterpart of this  theory.}    
\begin{eqnarray} 
{\rm fermions} \; && \rightarrow (1,1)_{\frac{1}{2}} \oplus 
(2,2)_{-\frac{1}{2}} 
\oplus [(3,1) \oplus (1,3)]_ {\frac{1}{2}}
\oplus  (2,2)_{-\frac{1}{2}}  \oplus (1,1)_{\frac{1}{2}}  \cr
&&  \rightarrow 1_{\frac{1}{2}} \oplus 
4_{-\frac{1}{2}} 
\oplus 6_ {\frac{1}{2}}
\oplus  4_{-\frac{1}{2}}  \oplus 1_{\frac{1}{2}} \; .
\eqn{twistM}
\end{eqnarray}
 The magic of this particular embedding 
is clear. There are now
two spin zero fermions, while the remaining fermions are now 
in integer spin representations  of the 
twisted Lorentz symmetry  $SO(4)'$.  
They transform as  scalars, vectors, and 
higher rank $p$-form tensors.   We parameterize these 
Grassmann valued tensors, accordingly, 
 $( \lambda,  \psi^{\mu},  \xi_{\mu \nu}, 
\xi^{\mu \nu \rho},  \psi_{\mu \nu \rho \sigma})$. 

The gauge boson $V_{\mu}$ which 
transforms as $(2,2,1,1)$ 
under the  group $G$  becomes $(2,2)$ under $G'$. 
Similarly, four of the  scalars $S_{\mu}$ which originally
transformed as $(1,1, 2 ,2)$ 
are now elevated to the same footing as the gauge boson and transform as  
$(2,2)$  under the twisted rotation group. The resulting theory is
most compactly described using a complex vector field
\footnote{The indices   $\mu, \nu, \rho, \sigma
\ldots  $   are $SO(4)'$ or $4$-dimensional 
hypercubic  indices  and summed over $1, \ldots 4$. The indices $ m, n,
\ldots$ are indices for permutation group $S_{5}$  (for $A_4^*$ lattices)  
and   are summed over  $1,\ldots ,5.$ }
\begin{eqnarray}
z^{\mu}= (S^{\mu} + i V^{\mu})/ \sqrt 2,  \qquad 
 \mybar z_{\mu}= (S_{\mu} - i V_{\mu})/ \sqrt 2  \qquad \mu=1, \dots, 4
\eqn{vector}
\end{eqnarray} 
Since there are two types of vector  fields, there are indeed two types of 
complexified gauge covariant derivative appearing in the formulation. 
These are holomorphic and antiholomorphic in character   
\begin{eqnarray}
 \CD^{\mu}\, \cdot  = \partial^{\mu} \cdot + \sqrt 2 [ z^{\mu}, 
  \, \cdot \,
], 
\qquad 
 {\mybar \CD}_{\mu} \, \cdot  = -\partial_{\mu} \cdot  +  \sqrt 2 
[ \mybar z_{\mu}
 , \, \cdot \, ] \, , 
\eqn{cov}
\end{eqnarray}
In fact only three combinations of the covariant derivatives (similar to the  $F$-term
and  $D$-term in $\CN=1$ gauge theories)  appear in the
formulation. These are 
\begin{eqnarray}
\cF^{\mu \nu}&&= -i [\CD^\mu, \CD^\nu] 
= F_{\mu\nu} -i [S_{\mu}, S_{\nu}] 
-i (D_{\mu}S_{\nu}- D_{\nu}S_{\mu}) \cr  
 {\mybar \cF}_{\mu \nu}&&= 
-i [{\mybar \CD}_\mu, {\mybar \CD}_\nu] =  F_{\mu\nu} -i [S_{\mu}, S_{\nu}] 
+i (D_{\mu}S_{\nu}- D_{\nu}S_{\mu}) \cr
(-id) &&= \half  [\mybar {\cal D}_{\mu}, {\cal D}^{\mu}] + \cdots   = 
-D_{\mu}S_{\mu} + \cdots
\eqn{strength}
\end{eqnarray}
where $D_{\mu}\, \cdot  = \partial_{\mu} \cdot  + i [ V_{\mu}, \, \cdot \,]$
is the usual covariant derivative and   $F_{\mu\nu} = -i [D_\mu, D_\nu]$
is the nonabelian  field strength.  
The field strength $\cF^{\mu \nu}(x) $ is holomorphic;
depending only on the complexified vector field $z^{\mu}$   and not on $\mybar
z_{\mu}$. Likewise,   $\mybar \cF_{\mu \nu}$  is anti-holomorphic. 
The $(-id)$ will come out of the solutions of equations of motion for
auxiliary field $d$ and 
ellipses  stand for possible scalar  contributions. 
These combinations arise naturally  from {\bf all} of the  orbifold lattice constructions in any dimensions and is one of the reasons for considering this class  of twist (we saw this already in our discussion of the
self-dual twist of the $(2,2)$ YM theory in two dimensions). 

Finally, the two other scalars remains as scalars under the twisted 
rotation  group. Since one of the scalars is the  superpartner 
(as will be seen below)  
of  the four form fermion, we label them as  
$(z_{\mu \nu \rho \sigma}, \mybar z^{\mu \nu \rho \sigma} )$. 
To summarize,  the bosons transform under $G'$ as   
\begin{eqnarray} 
{\rm bosons} \rightarrow  
z_{\mu \nu \rho \sigma} \oplus  z^{\mu} \oplus  \mybar z_{\mu} \oplus 
\mybar z^{\mu \nu \rho \sigma}   \; \rightarrow  
[ (1,1)_1 \oplus (2,2)_0 +  (2,2)_0 + (1,1)_{-1} ]
\eqn{bosons}
\end{eqnarray}
As can be seen easily from the decomposition of the fermions, there are two
Lorentz singlet supercharges $(1,1)$ under the twisted Lorentz group and 
either of these (or their linear  combinations) can be used to write down 
the Lagrangian  of the  four dimensional theory in ``topological'' form.  
Below, we use the scalar supercharge associated with  $\lambda$. 
This produces the transformations given by  
\citep{Marcus}. 

The  continuum off-shell  supersymmetry transformations are given by  
\begin{eqnarray}
&&Q \lambda =  -id,   \qquad    Q d = 0 \cr 
&& Q z^{\mu} =   \sqrt 2 \, \, \psi^{\mu}, \qquad  Q \psi^{\mu}=0 \cr
&& Q \mybar z_{\mu} = 0  \cr
&&Q \xi_{\mu\nu} = -i \mybar \cF_{\mu \nu} \cr
&&Q \xi^{\nu \rho \sigma} = \sqrt 2 \, {\mybar \CD}_{\mu}  
\mybar z^{\mu \nu \rho \sigma}  \cr
&&Q z_{\mu \nu \rho \sigma} = \sqrt{2} \psi_{\mu \nu \rho \sigma}, \qquad 
Q \psi_{\mu \nu \rho \sigma} = 0    \cr 
&&Q \mybar z^{\mu \nu \rho \sigma} =0  
\eqn{QAoffshell}
\end{eqnarray}
where $d$ is an auxiliary field introduced for the  off-shell completion of 
the  supersymmetry algebra.  
Clearly, the scalar    supercharge is nilpotent  
\begin{eqnarray} Q^2 \; \cdot  = 0. \end{eqnarray}
owing to the anti-holomorphy of $\mybar \cF_{\mu \nu}$ etc.    
The fact that the subalgebra  ($Q^2=0$) does not 
produce any spacetime translations makes it possible to carry it easily onto
the lattice.
This exact nilpotent property, in contrast to nilpotency
only up to gauge
transformation,  has a technical advantage - it 
admits a superfield formulation of the target supersymmetric field theory. 

The twisted  Lagrangian  may be written as a sum of $Q$-exact and  $Q$-closed 
terms:
\begin{eqnarray}
g^2 \CL = && \CL_{exact} +\CL_{closed} =  \CL_1 +  \CL_2 +   \CL_3 =  
Q {\widetilde \CL_{exact}} +  \CL_{closed},
\eqn{LT41} 
\end{eqnarray} 
where $g$ is coupling constant and 
 $\widetilde \CL_{exact}= {\widetilde \CL}_{e,1} + 
{\widetilde \CL}_{e,2} $ is given by  
\begin{eqnarray}
&& {\widetilde \CL}_{e,1} =    \Tr  \Big( \lambda ( \half id + \half 
[ \mybar {\cal D}_{\mu}, {\cal D}^{\mu} ] +
\textstyle{\frac{1}{24}}[\mybar z^{\mu\nu \rho \sigma}
, z_{\mu\nu \rho \sigma}] ) \Big) \cr
&& {\widetilde \CL}_{e,2}= 
  \Tr  \Big( 
\fourthi \xi_{\mu\nu} \cF^{\mu\nu} +  \textstyle{\frac{1}{12 \sqrt2}}
\xi^{\nu \rho \sigma} {\cal D}^{\mu} z_{\mu \nu \rho \sigma} 
\Big) 
\end{eqnarray}
and   $\CL_{closed}$ is given by 
\begin{eqnarray}
 \CL_{closed}= \CL_3 = \Tr 
\half \xi_{\mu\nu }  \mybar {\cal D}_{\rho} \xi^{\mu \nu \rho} +  
\textstyle{\frac{\sqrt2}{8}} \;  \xi_{\mu\nu } [\mybar 
z^{\mu \nu \rho \sigma}, \xi_{\rho \sigma}] \qquad
\end{eqnarray}
By using the transformation properties  of fields and 
the equation of motion of the auxiliary field $d$  
\begin{eqnarray}
(-id)= \half  [ \mybar {\cal D}_{\mu}, 
{\cal D}^{\mu} ] +
\textstyle{\frac{1}{24}}[\mybar z^{\mu\nu \rho \sigma}
, z_{\mu\nu \rho \sigma}] \; ,
\end{eqnarray}
we obtain the Lagrangian expressed in terms of propagating degrees of 
freedom: \footnote{Notice that the splitting  of the exact terms in 
Lagrangian  into $\CL_1$ and $\CL_2$  is not identical to the one used 
by Marcus.  The reason for the above splitting lies in the symmetries of the
cut-off theory ($A_d^*$ lattice theory)  that will be discussed later.}
\begin{eqnarray} 
 &&  \CL_1 =   \Tr  \Big( \half (  \half [\mybar {\cal D}_{\mu} , 
 {\cal D}^{\mu}] + \textstyle{\frac{1}{24}}[\mybar z^{\mu\nu \rho \sigma}
, z_{\mu\nu \rho \sigma}])^2 +   
\lambda ( \mybar {\cal D}_{\mu} \psi^{\mu} +   
\textstyle{\frac{1}{24}}[\mybar z^{\mu\nu \rho \sigma}
, \psi_{\mu\nu \rho \sigma}]) \Big)  \cr 
&&  \CL_2 =   \Tr  \Big(  \fourth \mybar {\cF}_{\mu\nu} {\cF}^{\mu\nu}    
+   \xi_{\mu\nu}  {\cal D}^{\mu} \psi^{\nu} 
+ {\textstyle \frac{1}{12}} |{\cal D}^{\mu} z_{\mu \nu \rho \sigma}|^2
+  {\textstyle \frac{1}{12}} 
 \xi^{\nu \rho \sigma} {\cal D}^{\mu} \psi_{\mu \nu \rho \sigma} 
+  {\textstyle \frac{1}{6 \sqrt 2 }}
 \xi^{\nu \rho \sigma} [\psi^{\mu}, z_{\mu \nu \rho \sigma}]
\Big)   \cr 
&&  \CL_3 = \Tr \Big(
\half \xi_{\mu\nu }  \mybar {\cal D}_{\rho} \xi^{\mu \nu \rho} +  
\textstyle{\frac{\sqrt2}{8}} \;  \xi_{\mu\nu } [\mybar 
z^{\mu \nu \rho \sigma}, \xi_{\rho \sigma}] 
\Big) \; .
\eqn{LT4}
\end{eqnarray}
The $Q$-invariance of the  $\CL_{exact}$ is obvious and  
 follows from supersymmetry algebra $Q^2=0$.
To show the invariance of $Q$-closed term  
requires the  use of the Bianchi 
(or Jacobi identity for covariant derivatives)  identity   
\begin{eqnarray}
\epsilon^{\sigma \mu\nu \rho } \mybar {\cal D}_{\mu}  \mybar \cF_{\nu \rho}=
\epsilon^{\sigma \mu\nu \rho } [\mybar {\cal D}_{\mu}, [ 
\mybar {\cal D}_{\nu},  \mybar {\cal D}_{\rho} ]] = 0
\eqn{Jacobi}
\end{eqnarray}
and a similar identity involving scalars. 
The action is expressed in terms of the  twisted Lorentz multiplets, and  
the   $SO(4)'\times  U(1)$ symmetry is manifest. 
The Lagrangian  \Eq{LT4} arises in the classical continuum limit  of  the hypercubic lattice and  the $A_4^*$ lattice action. In the former, a lattice $p$-cell field is identified with a continuum  $p$-form under twisted $SO(4)'$. 
In the latter case, the matching of the fields can be deduced 
by the representation theory of $S_5$ as we will see.  

Up to trivial rescalings this is the action (with gauge
parameter $\alpha=1$) of twisted ${\cal N}=4$ Yang-Mills in four
dimensions written down by Marcus \citep{Marcus}.
This twisted action is well known to be fully equivalent to the usual
form of ${\cal N}=4$ in flat space.

\subsubsection{A shortcut derivation of the Marcus twist}
There is a slick way to obtain the twisted 
theory
described in the previous section.  The idea is 
to amalgamate the four complexified gauge fields  \Eq{vector} 
and the extra scalar 
into a single five-component ``gauge connection".
\beq
\left( z^{\mu}= \frac{S_\mu + i V_{\mu}}{\sqrt 2}, \;  z^5=  \frac{S_5 + i S_6} {\sqrt 2}  \right)\rightarrow 
z^m, \qquad m=1, \ldots 5
\eqn{amalgamate}
\eeq 
The theory may then
be realized as a five dimensional $\CQ=16$ theory dimensionally reduced to $d=4$ dimensions. In five 
dimensions,   $z^5=  \frac{S_5 + i V_5} {\sqrt 2}$ and we may identify $S_6$ with $V_5$ upon dimensional reduction. 
Paralleling the four supercharge theory we introduce an additional auxiliary bosonic scalar field $d$ and a set of five dimensional antisymmetric tensor fields  to represent the fermions $\Psi=(\lambda, \psi^m,\xi_{mn})$. 
This latter field content corresponds to considering just one of the two \KD fields used to represent the 32 fields of the five dimensional theory. 
Again, a nilpotent symmetry relates these fields
\begin{eqnarray}
\cQ\; z^m&=& \sqrt 2 \psi^m \nonumber\\
\cQ\; \psi^m&=&0\nonumber\\
\cQ\; \mybar z_m&=&0\nonumber\\
\cQ\; \xi_{mn}&=&- i \cFb_{mn}\nonumber\\
\cQ\; \lambda&=&id\nonumber\\
\cQ\; d&=&0
\end{eqnarray}
and remarkably 
we may extract the Marcus
theory from the same $\cQ$-exact action that was employed in
\S.\ref{sec:8c} for $(2,2)$ Yang-Mills in
two dimensions $S=\beta\cQ \Lambda$ with
\beq
\Lambda=\int
\Tr \Big ( 
 \lambda ( \half id + \half 
[ \mybar {\cal D}_{m}, {\cal D}^{m} ] )  + \fourthi \xi_{mn} \cF^{mn} 
 \Big)
\eeq
where we have again employed complexified covariant derivatives.
Carrying out the $\cQ$-variation and subsequently integrating out the
auxiliary field as for the $\cQ=4$ theory leads to the action
\beq
S_{\rm exact}=\int\Tr \left(  
\fourth \mybar {\cF}_{mn} {\cF}^{mn}  +  \eight ([\mybar {\cal D}_{m} , 
 {\cal D}^{m}])^2 +
\lambda  \mybar {\cal D}_{m} \psi^{m}     
+   \xi_{mn}  {\cal D}^{m} \psi^{n} 
\right)\eeq
Actually in this theory there is another fermionic term one can write down which is also
invariant under this supersymmetry:
\beq
S_{\rm closed}=  
\frac{1}{8}\epsilon^{mnpqr} 
\xi_{mn}\mybar {\cal D}_p \xi_{qr}
\eeq
The invariance of this term is just a result of the Bianchi
identity $\epsilon_{mnpqr}\cD_p \cFb_{qr}=0$. The final action we will
employ is the sum of the $\cQ$-exact piece and this $\cQ$-closed term
and reproduces, after dimensional reduction, the four dimensional
$\cQ$-closed term we have already discussed. 
The coefficient in front of this term is determined by the requirement
that the theory reproduce the Marcus twist of ${\cal N}=4$ Yang-Mills.

Splitting  $\cF^{mn} \to \cF^{\mu\nu}\oplus\cD_\mu z^5, \;\;
\left[\cDb_m,\cD^m\right] \to\left[\cDb_\mu,\cD^\mu\right]\oplus
\left[\mybar z_5, z^5\right] $ and using \Eq{strength} and 
\Eq{split3} gives the $\CN=4$ SYM action in the twisted form shown in 
\Eq{LT4}.

\subsubsection{Lattice theory}
\label{sec:10c.2}

The discretization scheme that is
employed is precisely the same as the $\CN=(2,2)$
target theory in $d=2$ dimensions as described in \S.\ref{sec:8d}. 
Specifically the continuum gauge field is exponentiated into
a non-unitary  
link field with  
\begin{eqnarray}
\cU^{\mu}  = \frac{1}{\sqrt 2 a} e^{a(S_{\mu, \bfn} + i V_{\mu, \bfn})}, \;\;
\cU^{5} = \frac{1}{\sqrt 2 a} e^{a(S_{5, \bfn} + i S_{6, \bfn})}
\eqn{defU2}
\end{eqnarray}
as described in the continuum in \Eq{amalgamate} \footnote{The definition of  \Eq{defU2}  is rescaled relative to the discussion in \S.~\ref{sec:8d} by a factor of $\sqrt 2$. With this modification, the small field  expansion of   non-unitary link field is
$\cU^{\mu} =  \frac{1}{\sqrt 2 a}  +  \frac{S_{\mu, \bfn} + i V_{\mu, \bfn}}{\sqrt 2}$, and  same as the one used in deconstruction/orbifold approach as in   \S.~\ref{sec:9} and   \Eq{exp1orb}. }. 
The $\cQ$-supersymmetry is
essentially the same as in the continuum and remains nilpotent
\begin{eqnarray}
\cQ\; \cU^m&=&\sqrt 2 \psi^m\nonumber\\
\cQ\; \cUb_m&=&0\nonumber\\
\cQ\; \psi^m&=&0\nonumber\\
\cQ\; \xi_{mn}&=& -2 \left(\cF^L_{mn}\right)^\dagger\nonumber\\
\cQ\; \lambda&=& id\nonumber\\
\cQ\; d&=&0
\label{latticeQ}
\end{eqnarray}
where the lattice field strength $\cF^L_{ab}$ is given by eq.~(\ref{field})
as before. 

As for the $(2,2)$ twisted SYM model the twisted fermions are to be
placed on $p$-cells in the lattice. However, there is one remaining
wrinkle in this mapping; 
for each $p$-cell  (with $1 \leq p \leq 4$)  field 
associated with hypercubic lattice, we may have two possible orientations. 
This orientation is physical and determines the gauge rotation properties of the fields. We need to give 
a prescription to go from Marcus's twist to the lattice. As we will see, exact supersymmetry also plays an important role here.

For the moment let us base our 
discretization scheme around a hypercubic lattice.  Then the gauge links
$\cU^\mu(\bx) \equiv \cU^m, m=1\ldots 4$ 
should live on   elementary coordinate directions in the unit hypercube,  running 
from  
$\bx \to \bx+\bmu$.  
 We will adopt the notation that these four
basis vectors are labeled $\bmu_a, a=1\ldots 4$.  
This assignment then implies that the
superpartners of the gauge links $\psi^\mu (\bx)$ should also live on
the same links and be oriented identically. Evidently,  $\cUb_m(\bx)$ is oriented oppositely, 
running from $ \bx+\bmu \to  \bx$.  By eq.(~\ref{field}), the complexified field strength  runs 
from $\bx \to \bx+\bmu + \bnu$, hence  $\left(\cF^L_{mn}\right)^\dagger$ and by exact supersymmetry $\xi_{\mu \nu}$ runs oppositely.  The reader may have a feeling that, in this  way, we are essentially re-constructing Table  \ref{tab5}, and indeed, this is true. 

However, a priori,  the assignment of $\psi^5$  is not
immediately obvious -- a naive assignment to a site field
would result in {\it two} fermionic 0-forms which
is not what is expected for a four dimensional \KD field.  \KD decomposition demands a 4-form,  
associated with the chiral matrix of the
four dimensional theory $\Gamma^5=\gamma_5=\gamma_1\gamma_2\gamma_3\gamma_4$.
This motivates assigning the lattice field to the body diagonal of the unit hypercube, a 4-cell. The ability to construct gauge invariant expressions involving the 4-cell 
fields (such as the last term in ${\cal L}_3$ in \Eq{LT4}) demands that $\psi^5$ and $z^5$ fields 
to be oriented along the vector  $\bmu_5=(-1,-1,-1,-1)$. Notice that
this assignment also ensures that $\sum_{m=1}^5\bmu_m=\bzero$ which will be 
seen to be crucial for constructing gauge invariant quantities. 

To  summarize  the  $p$-cell {\it  and}  orientation assignments of lattice fields, we write down 
their lattice gauge transformations:
\begin{eqnarray}
\lambda(\bx)&\to& G(\bx) \lambda(\bx) G^\dagger(\bx)\nonumber\\
\psi^{m}(\bx)&\to& G(\bx)\psi^{m}(\bx) G^{\dagger} (\bx+\bmu_m)\nonumber\\
\xi_{mn}(\bx)&\to&G(\bx + \bmu_m + \bmu_n)\xi_{mn}(\bx)G^\dagger(\bx)
\nonumber\\
\cU^m(\bx)&\to&G(\bx)\cU^m (\bx)G^\dagger(\bx+\bmu_m)\nonumber\\
\cUb_m(x)&\to&G(\bx+\bmu_m)\cUb_m(\bx)G^\dagger(\bx)\nonumber\\
\end{eqnarray}

Notice also that these link choices and orientations
match exactly the ${\bf r}$-charge assignments of the orbifold action for the sixteen
supercharge theory in four dimensions given in  Table \ref{tab5}.
As for two dimensions, successive components of the
resultant fermionic \KD field alternate in orientation which will be the
key to writing down gauge invariant fermion kinetic terms.  Switching back to the four component 
anti-symmetric index notation, the set of four
Majorana fermions required for $\cN=4$ YM are now compactly
expressed in matrix form
 \beq
   (\Psi^{\rm Maj})_{\Upsilon I} = \left(\lambda 1 + \psi^{\mu} \gamma_{\mu} + \xi_{\mu \nu}  \gamma^{[\mu\nu]} + \xi^{\mu \nu \rho}  \gamma_{[\mu\nu \rho]} + \psi_{\mu \nu \rho \sigma}  \gamma^{[\mu\nu \rho \sigma]} 
    \right)_{\Upsilon I}, \Upsilon, I=1, \ldots, 4
     \eqn{Weyls}
 \eeq 
 where upper index means oriented along the unit vectors and lower index means oppositely oriented.  Thus, 1 and 3-form fermions $(\psi^{(1)}, \psi^{(3)})$ are positively oriented and 
 0, 2 and 4-forms  $(\psi^{(0)}, \psi^{(2)},  \psi^{(4)}, )$  are negatively oriented  
This property is crucial (for any  supersymmetric (orbifold) lattice in any dimension) both for gauge invariance and an absence of fermion
doubling in any supersymmetric (orbifold) lattice theory.

Using the prescription of \S \ref{sec:8d}, and eq.~(\ref{derivs}) and
 eq.~(\ref{field}) produces precisely the supersymmetric  
lattice action for the  $\CN=4$ SYM target theory given in  \Eq{d4lat}, 
modulo the replacement $z^m(\bfn)\rightarrow \cU^m(\bfn)$. The $\cQ$-exact part becomes $S_{\rm exact}=\cQ\Lambda$ where
\beq
\Lambda=\sum_x
\Tr\left( -\half \xi_{mn}\cF^L_{mn} - \lambda\cDb^{(-)}_m\cU_m +\half \lambda
(id)\right)
\label{5dgauge}
\eeq
which after $\cQ$-variation and elimination of the auxiliary $d$ yields
\beq
S=\sum_{\bx}\Tr \left( \cF^{L\dagger}_{mn}\cF^{L}_{mn}+
\half \left(\cDb^{(-)}_m \cU_m\right)^2-\sqrt{2}\left(\lambda \cDb^{(-)}_m\psi_m +
\xi_{mn}\cD^{(+)}_{\left[m\right.}\psi_{\left.n\right]}
\right)\right)\eeq
where the lattice field strength is given by the same expression as
in \S\ref{sec:8d}. The third triangular fermion plaquette term arising in
the orbifold action is now seen to be
a discretized version of the $\cQ$-closed term
\beq
S_{\rm closed}=-\frac{\sqrt 2}{8}\sum_{\bx}\Tr 
\epsilon_{mnpqr}\xi_{qr}(\bx+\bmu_m+\bmu_n+\bmu_p)
\cDb^{(-)}_p\xi_{mn}(\bx+\bmu_p)\label{closed}\eeq
where
\beq
\cDb^{(-)}_p\xi_{mn}(\bx)=\xi_{mn}(\bx)\cUb_p(\bx-\bmu_p)-
\cUb_p(\bx+\bmu_m+\bmu_n-\bmu_p)\xi_{mn}(\bx-\bmu_p)
\eeq
Notice that the $\epsilon$-tensor forces all indices to be distinct
and the gauge invariance of this result follows from
the fact that $\sum_{m=1}^5\bmu_m=\bzero$.
In the continuum the invariance of this term under $\cQ$-transformations
requires use of the
Bianchi identity.
Remarkably, the lattice difference operator
satisfies a similar identity (see \citep{Aratyn} for the four dimensional
result)
\beq
\epsilon_{mnpqr}D^{(+)}_p\cF^{L}_{qr}=0\eeq
Furthermore, since the bosonic and fermionic link fields 
entering each lattice site in a hypercubic lattice construction are
the same as the  $A^*_4$ lattice, 
we can
obtain both hypercubic and $A^*_4$ lattices from the twisted 
construction as well. 

\begin{figure}[t]
\begin{center}
\includegraphics[height=60mm]{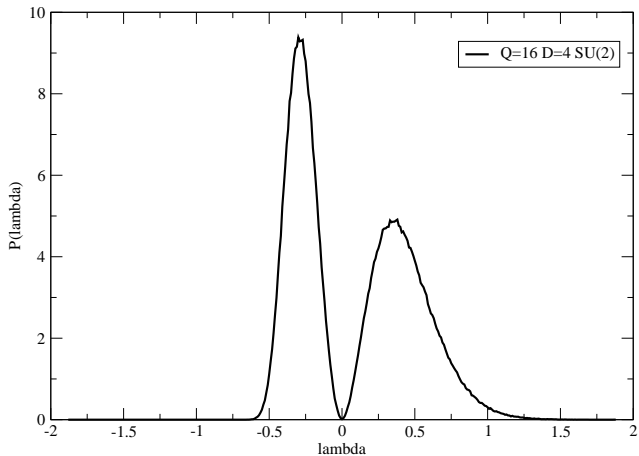}
\caption{Probability distribution of scalar
eigenvalues for $\cQ=16$, $D=4$, $L=2$ and $SU(2)$}
\label{scalar4}
\end{center}
\end{figure}

Preliminary simulations of this model have already been
performed with encouraging results
\citep{Catterall:2008dv}. Table~\ref{table8} shows the mean
bosonic action for lattices with volume
$2^4,3^4$ at fixed
't Hooft coupling $\lambda=0.5$ (the data corresponds to $6000$ and
$1000$ configurations for linear size $L=2$ and $L=3$ respectively). As for
the $\CN=(2,2)$ model this observable can be computed exactly using
a $\cQ$-Ward identity yielding the exact result quoted in the last
column. 
\begin{table}
\begin{center}
\begin{tabular}{||c|c|c||}\hline
$L$ & $<S_B>$ & $S_B^{\rm exact}$\\\hline
2 &211.2(2)& 216.0 \\\hline
3 &1075.0(35)& 1093.5 \\\hline
\end{tabular}
\caption{Observables for $SU(2)$ $\cQ=16$ model in $D=4$ at $\lambda=0.5$}
\label{table8}
\end{center}
\end{table}
For comparison, table~\ref{table9} shows the same quantities for
't Hooft coupling $\lambda=0.25$.
\begin{table}
\begin{center}
\begin{tabular}{||c|c|c||}\hline
$L$ & $<S_B>$ & $S_B^{\rm exact}$\\\hline
2 &211.5(5)& 216.0\\\hline
3 &1080.5(45)& 1093.5\\\hline
\end{tabular}
\caption{Observables for $SU(2)$ $\cQ=16$ model in $D=4$ at $\lambda=0.25$}
\label{table9}
\end{center}
\end{table}
Notice that as we approach weak coupling and smaller lattice spacings
the bosonic action moves towards its exact supersymmetric value as
expected.\footnote{The small breaking of susy seen in this data is associated with the
truncation $U(2)\to SU(2)$ employed in the simulations. This was necessary
to avoid a vacuum instability problem. For further details on
this and the issue of the Pfaffian phase we refer
the reader to \citep{Catterall:2008dv}.} 
Finally the scalar eigenvalue distribution is shown in
figure~\ref{scalar4} and looks qualitatively similar to
what was seen for $(2,2)$ YM with the important caveat that the tail
of the distribution is much more rapidly damped in the $\cQ=16$ supercharge
case. This is similar to what had been observed before in
simulations of the corresponding matrix models \citep{Krauth:1999qw}.

To conclude, we have shown how to derive the supersymmetric orbifold
lattice corresponding to $\cN=4$ SYM
\citep{Kaplan:2005ta} by geometrical discretization of the continuum
twisted SYM theory. This connection is not
unexpected -- it was shown earlier in \citep{Unsal:2006qp}
that the naive continuum limit of the $\cQ=16$
orbifold theory in four dimensions corresponded to the Marcus twist of
${\cal N}=4$ and more recent work by Damgaard et al.
\citep{Damgaard:2007xi} and Takimi \citep{Takimi:2007nn} have exhibited
the strong
connections between discretizations of the twisted theory and orbifold
theories. In this section we 
have completed this
connection -- the two approaches
are in fact identical provided one chooses the exact lattice supersymmetry
carefully - we must use the {\it self-dual twist} introduced earlier
and employ the geometric discretization proposed in
\citep{Catterall_n=2}. Additionally, as was pointed out by
Damgaard et al.  \citep{Damgaard:2007eh}
this lattice theory is essentially equivalent to the one proposed
by d'Adda and collaborators
\citep{D'Adda_2d} {\it provided} that the fermionic shift parameter
employed in that model is chosen to be zero and we restrict 
our
attention solely to the corresponding scalar supercharge.

This connection between the 
twisting and orbifolding methods
is most clearly exhibited by recasting
the usual Marcus twist of $\cN=4$ Yang-Mills
as the dimensional reduction of a very
simple five dimensional theory. The $\cQ$-exact part of the action is
then
essentially identical to the 
two dimensional theory with $(2,2)$ supersymmetry
with the primary difference
between the two theories arising because of the appearance of a new
$\cQ$-closed term which was not possible in two dimensions. Nevertheless
discretization proceeds along the same lines, the one subtlety being the
lattice link assignment of the fifth component of the complex gauge field
after dimensional reduction. The key requirement governing
discretization is that successive components
of the \KD field representing the fermions have opposite orientations.
This allows the fermionic action to be gauge invariant without any
additional doubling of degrees of freedom. It seems likely that all
the orbifold actions in various dimensions can be obtained in this manner.
\subsubsection{Absence of fermion doubling}
\label{sec:10c.3}
There are two independent ways to demonstrate the absence of unwanted doublers in our formulation.  One is, to calculate the spectrum of scalars (which is technically simpler) and show that the bosonic action is  doubler-free. Then, by exact supersymmetry,  the fermionic spectrum is as well doubler free.  This was the point of view taken in Appendix B of \citep{Kaplan:2005ta}.

There is also an elegant way, which is made manifest by the geometric approach 
and  which makes  it easier to understand why this lattice theory does not suffer from doubling problems. This argument does not rely on supersymmetry, hence it is also useful for doubler-free 
formulations of four-flavor non-supersymmetric theories. 

We will analyze this question in the context of the
hypercubic lattice discretization. Clearly most of
the fermionic kinetic terms manifestly satisfy the double free discretization 
prescription given by Rabin \citep{Rabin:1981qj}. This prescription is;
use forward lattice difference  $(\cD^{\mu})^{(+)}$ whenever the continuum derivative acts as a gauged
exterior derivative and use the backward difference $\cDb^{(-)}_\mu$ whenever the 
continuum derivative appears as an adjoint of the
exterior derivative. Most of the terms appearing in this action
manifestly satisfy the requirements for this theorem. 

The only subtleties arise
when one or more tensor indices of the fields equal $m=5$. Expressions involving
these fields are not located wholly in the positively oriented unit
hypercube and must be translated into the hypercube before they
can examined from the perspective of this prescription. This has
the effect of changing a forward difference to a backward difference
operator after which it is easily seen that the term satisfies the
requirements of this theorem. For more details, see \citep{Catterall:2007kn}.

\subsection{$A_4^{*}$ lattice and  \KD fermions}
\label{sec:irreps}
We have seen that the $\CQ=1$ hypercubic supersymmetric lattice 
provides a 
realization of \KD fermions and a natural latticization of Marcus's twist. 
This discussion also makes it clear that the hypercubic lattice resides in 
the diagonal sub-space of the $R$-symmetry and original Lorentz symmetry. 
In this section, we wish to identify the relation between 
\KD fermions and the fermions of the 
$A_4^{*}$ lattice. Recall that 
the $A_4^{*}$ lattice is the 
maximally symmetric lattice realization of $\CN=4$ SYM 
theory in four dimensions. In $A_4^{*}$, the fermions are distributed as 
single site fermion $\lambda$, five link fermions $\psi^m$ and an
additional ten face fermions $\xi_{mn}$. 
The symmetry of $A_4^*$ makes it clear that all 5 link fermions are on equal  footing and all 10 face fermions are on equal standing as well.  However, we know that in the continuum, 
under the twisted rotation group $SO(4)'$, the fermions must fill in antisymmetric tensor representations as shown in \Eq{twistM}.
In particular, it is evident that the 5 link and 10 face fermions of the $A_4^{*}$ lattice 
must be reducible.   In order to see this explicitly, we need to decompose the lattice fields in terms of 
irreducible representation of $S_5$, as shown in Table \ref{tab:S5}.

The symmetry operations and characters of the $S_5$ point group symmetry are given in 
Table \ref{tab:S5}. 
By choosing a representative  from each symmetry conjugacy  class,  
we can calculate the character of the corresponding group element.  
In Table~\ref{tab:op}, we show how a particular representative 
 from each  conjugacy class acts on the fermion link fields and calculate 
\begin{table}[h]
\centerline{
\begin{tabular}
{|l|c|c|} \hline
Operation   & $ (\psi^1, \psi^2, \psi^3, \psi^4, \psi^5) $ & $\chi(g_{(rep)})$ 
\\ \hline
(1)    & $  (\psi^1, \psi^2, \psi^3, \psi^4, \psi^5) $ & 5   
\\ \hline
(12)    & $  -(\psi^2, \psi^1, \psi^3, \psi^4, \psi^5) $ & -3   
\\ \hline
(123)  & $ (\psi^3, \psi^1, \psi^2, \psi^4, \psi^5) $  & 2 
\\ \hline 
(1234)  & $ - (\psi^4, \psi^1, \psi^2, \psi^3, \psi^5) $  & -1 
\\ \hline 
(12345)    & $(\psi^5, \psi^1, \psi^2, \psi^3, \psi^4)  $  & 0
 \\ \hline
(12)(34)    & $ (\psi^2, \psi^1, \psi^4, \psi^3, \psi^5 ) $  & 1
 \\ \hline
(12)(345)   & $ - ( \psi^2, \psi^1, \psi^5, \psi^3, \psi^4) $  &0
\\ \hline
\end{tabular}}
\caption{\sl A representative of each conjugacy class and their action  
on the site and link fields are shown in the table. The five link fermions 
$\psi^m$ transform in the same way with $z^m$. The transformation of 
 ten fermions $\xi_{mn}$ can be deduced from the antisymmetric product
 representation of $\mybar z_m$ with itself. }
 \label{tab:op}
\end{table}
the character $\chi(g)= \Tr( O(g))$, where $g$ is a 
representative of each class and $O$ is a 
matrix representation of the operation.  
Since the  character is a class function, it is independent of 
representative.  For the fermion fields, we obtain 
\beq
 \chi(\psi^m )= (5,-3,2,-1,0,1,0)   
\eeq
Note that the odd permutations are accompanied by the transformation \Eq{oddextra}, 
since the combined operation is a symmetry of the action. 
Inspecting the character table of $S_5$, we see that this is not an irreducible representation. It is a linear combination of two irreducible representations, 
\begin{equation}
\chi(\psi^m)  =  \chi_4 \oplus \chi_2, 
  \end{equation}
 a four-dimensional pseudo-vector and a singlet pseudo-scalar.   We can also relate this representation theory argument to the detailed calculation given
 in \citep{Kaplan:2005ta}.  
 Recall that under a group operation (see 
Table~\ref{tab:op}), $\psi^m \rightarrow O^{mn}(g)\psi^n$. 
 The fact that the group action on the link field is
reducible means there is a similarity transformation which takes all of the 
$O(g)$ into a block diagonal form. In this case, two blocks have  sizes 
$1\times 1$ and  $4\times 4$.  Now, let us introduce the 
orthogonal  matrix $\CE$ that block-diagonalizes $O(g)$ for all $g\in S_5$. It is, not surprisingly, 
related to the  basis  vectors $\bfe_m$  of  the $A_4^{*}$ lattice. 
\beq
\CE_{m\mu}= (\bfe_m)_{\mu},  \qquad \CE_{m5}= \frac{1}{\sqrt5} \, .
\eqn{connector}
\eeq
 The matrix $\CE_{m n}$ then forms a bridge between the irreducible 
representation of  $S_5$ and the  representations
 of the twisted Lorentz  group $SO(4)'$. 
Thus, we obtain the following relations dictated by  symmetry arguments:
\beq
\psi^{\mu} =
 \CE_{m\mu } \psi^m,   \qquad 
  \frac{1}{4!}\epsilon_{\mu \nu \rho \sigma} \psi^{\mu\nu \rho \sigma} = 
\CE_{ m5} \psi^{m} = \frac{1}{\sqrt 5} \sum_{m=1}^{5} \psi^m 
\eeq
Obviously, we could have easily guessed the form of the singlet.

Performing the same exercise  for all lattice fields, we obtain 
\begin{eqnarray}
&&\chi(\lambda)=\chi(d)=  (1,1,1,1,1) \sim \chi_1  \cr
&&\chi(z^m)=  \chi(\psi^m )= \chi(\mybar z_m)= (5,-3,2,-1,0,1,0)   
\sim  \chi_4 \oplus \chi_2  \cr
&&\chi(\xi_{mn})=\chi([\mybar z_m, \mybar z_n])=  (10, 2,  1, 0, 0, -2,-1) \sim \chi_7 
\oplus  \chi_3 \, . 
\eqn{S5splitting}
\end{eqnarray}
This means that the sixteen fermions appearing in
 the unit cell of the $A_4^{*}$ lattice branch into 
\begin{eqnarray}
&& {\rm  fermions}
\rightarrow  \chi_1 \oplus  \chi_3 \oplus  
\chi_7 \oplus \chi_4 \oplus  \chi_2
\eqn{split}
\end{eqnarray}
irreducible representations of $S_5$, which is nothing but \Eq{twistM}.  
We may also write expressions relating the 
$A_4^{*}$ lattice fields to the continuum twisted   \KD fermions, by using the \Eq{connector}. 
It is 
\begin{eqnarray}
 \xi_{\mu\nu} = \xi_{mn} \CE_{m\mu}\CE_{n\nu}, \qquad 
 \frac{1}{3!}\epsilon_{\mu
  \nu \rho \sigma}\xi^{\nu \rho \sigma} =\xi_{mn}  \CE_{m\mu}\CE_{n5}\; .
  \end{eqnarray}
This completes 
our discussion of the relation between the $A_4^{*}$ fermions  and fermions in the 
twisted theory \Eq{LT4}.  

Retrospectively, these relations are not surprising.  We already knew that the fermions on the hypercubic lattice are \KD fermions. We may  deform the hypercubic lattice into a 
$A_4^{*}$ lattice while remaining within the moduli space of our orbifolded matrix model 
\Eq{d4latss}.  The number of bosonic and fermionic fields leaving and entering each lattice site is equal in these two lattice constructions. 
Obviously, the bosons  work in a similar manner, which follows from exact 
supersymmetry.

In the continuum, the point group symmetry $S_5$ of the lattice action enhances to the twisted rotation group 
$SO(4)'$. 
\beq 
S_5 \subset SO(4)' 
\eeq
without any fine-tuning, thanks to the microscopic symmetries. 
In the renormalization discussion of \S \ref{sec:10d},  we will show that there are no 
relevant or marginal  twisted  $SO(4)'$ violating operators. 
Hence, in the continuum, we are guaranteed to get a $\CQ=1$, twisted Lorentz symmetry invariant 
gauge  theory without any fine tuning.  (In this sense, the enhancement of $S_5$ into  twisted Lorentz symmetry is analogous to the pure YM theory on lattice, where hypercubic symmetry enhances to Lorentz symmetry.)
Unfortunately, this does {\bf not} imply that we can immediately undo the twist and obtain the $\CQ= 16$ target theory. In particular, there are a
few relevant or marginal operators 
which respect  $SO(4)'$, gauge symmetry and $\CQ=1$, but not $SO(4)_E$. This means, 
some amount of fine tuning may be 
necessary in order attain the desired    $\CQ= 16$  
target theory in the continuum limit. We examine these issues in
more detail in the next section.

\subsection{Renormalization}
\label{sec:10d}

The immediate question that arises for this discretization of
${\cal N}=4$ super Yang Mills theory is how much residual fine tuning
will be required to ensure the restoration of full supersymmetry
in the continuum limit. Clearly the existence of one exact
supersymmetry improves the situation over any naive discretization
but it is not immediately clear what additional counter terms will be needed
to realize the full supersymmetry of the continuum theory. 

Unlike the case of  $d \leq 3$ dimensions power counting reveals that
the continuum four dimensional theory has an infinite number of superficially
divergent Feynman diagrams occurring at all orders of perturbation
theory. Of course in the continuum target  theory all of these potential
divergences cancel between diagrams to render the
quantum theory finite. However,
since the lattice theory does not possess all the
supersymmetries of the continuum theory, it is not clear
how many of these will continue to cancel in the lattice theory.

As a first step to understanding the structure
of the effective action that
arises in this lattice theory as a result of
radiative corrections one can attempt to write down
the structure of all possible counterterms which are consistent
with the exact lattice symmetries.  In the case of $A_4^{*}$ lattice, these symmetries are
\begin{itemize}
\item[{\bf a)}] Exact  $\cQ=1$ supersymmetry. 
\item [{\bf b)}] Gauge invariance
\item [{\bf c)}] $S_5$ point group symmetry and discrete translations.
\end{itemize}
In fact, other than exact lattice supersymmetry, the $U(k)$ lattice gauge theory also has 
a second fermionic  symmetry,  given by
\beq
\lambda(\bf n) \rightarrow \lambda(\bfn) + \epsilon 1_k, \qquad  \delta (\rm all \; other \; fields) =0 
\eeq
where $\epsilon$ is an infinitesimal Grassmann parameter.  Thus, we extend our list to include 
\begin{itemize}
\item[{\bf d)}] Fermionic shift symmetry
\end{itemize}

In practice we are primarily interested in  relevant or marginal   operators;
that is operators whose mass dimension is less than or equal
to four. We will see
that the set of relevant counterterms in the lattice theory is
rather short -- the lattice symmetries, gauge invariance in particular,
being extremely restrictive in comparison to the equivalent situation
in the continuum. The argument starts by assigning canonical
dimensions to the fields 
$[\cU_a]=[\cUb_a]=1$, $[\Psi]=\frac{3}{2}$
and $[\cQ]=\frac{1}{2}$ where $\Psi$ stands for any of the twisted
fermion fields $(\lambda,\psi^m,\xi_{mn})$. Invariance under
$\cQ$ restricts the possible counterterms to be either of a  $\cQ$-exact form, 
or  of  $\cQ$-closed form. There is only one $\cQ$-closed operator permitted by lattice 
symmetries, and it corresponds to
the continuum term ${\cal L}_3$ in \Eq{LT4}. Thus, 
we need to look to the set of $\cQ$-exact counterterms.    
Any such counterterm must be of the form $\mathcal{O}=\cQ\Tr(\Psi f(\cU,\cUb))$. There are thus no terms permitted by symmetries and with dimension less than two.
In addition gauge invariance tells us that each term must correspond to
the traces of a closed loop on the lattice.  
The smallest dimension gauge invariant
operator is then just $\cQ(\Tr\psi^m \cUb_m)$. But this vanishes identically
since both $\cUb_m$ and $\psi_m$ are singlets under $\cQ$. No
dimension $\frac{7}{2}$ operators can be constructed with this structure
and we are left with just dimension four counterterms. 
Notice, in particular
that lattice symmetries permit no simple fermion bilinear mass terms.  However, 
gauge invariant 
fermion bi-linears with link field insertions are possible and their effect should be accounted for carefully.  

Possible  dimension four operators are, schematically, 
\begin{eqnarray}
&& \cQ \Tr (\xi_{mn}\cU^m \cU^n)  \cr
&& \cQ \Tr (\lambda \cU^m \cUb_m) \cr 
&& \cQ \Tr (\lambda)  \Tr(\cU^m \cUb_m),
\eqn{ops}
\end{eqnarray}
The  first operator  can be simplified on account of the
antisymmetry of $\xi_{mn}$ to simply $\cQ (\xi_{mn}\cF^{mn})$, which is nothing but the continuum term
${\cal L}_2$ in \Eq{LT4}.  

The second term  in \Eq{ops} requires  more care.  There are two operators of this type permitted by lattice symmetries, not including the fermionic shift symmetry. These are 
\begin{eqnarray}
{\CL}_{1, \mp} =   \cQ  \Tr \lambda(\bfn) \Bigl(\cUb_m(\bfn-\bfmu_m)  \cU^m(\bfn-\bfmu_m)   \mp
\cU^m(\bfn) \cUb_m(\bfn)\Bigr)  
\end{eqnarray}
where both anti-commutator  and commutator structure are allowed. Clearly, the operator 
with the relative minus sign is ${\cal L}_1$ in \Eq{LT4}, but the one involving the anti-commutator is a new operator not present in the bare Lagrangian.  
The only operator of the third type is a double-trace operator
 \begin{eqnarray}
 {\CL}_{1, +}^{\rm d.t.} =  \cQ  \Tr \lambda(\bfn)  \Tr  \Bigl(\cUb_m(\bfn-\bfmu_m)  \cU^m(\bfn-\bfmu_m)  + \cU^m(\bfn) \cUb_m(\bfn)\Bigr)  
\end{eqnarray} 

Note that both ${\CL}_{1, +}$  and $ {\CL}_{1, +}^{\rm d.t.} $   transform non-trivially under the fermionic shift symmetry, but a linear combination  of the two 
\beq 
\CL_4 =  \CL_{1, +} - \frac{1}{k} {\CL}_{1, +}^{\rm d.t.}  
\eqn{ren4}
\eeq
 is invariant under the shift symmetry with
  $k$ the rank of the gauge group  $U(k)$.

By these arguments it appears that the only relevant counterterms correspond
to renormalizations of operators already present in the bare action 
together with $\CL_4 $. 
This is quite remarkable. The most general form for the
renormalized lattice Lagrangian is hence 
\begin{equation}
 g^2{\cal L} =   \CL_1 +  \alpha \CL_{2}  + \beta \CL_{3}  + \gamma \CL_{4}
 \eqn{generic}
\end{equation}
where $\alpha,\beta, \gamma $ are dimensionless numbers taking the value $(1,1,0)$
in the classical lattice theory and $g^2$ is a renormalized coupling constant.
Thus it appears that at most three  couplings 
might need to be fine tuned to approach $\cN=4$ Yang-Mills in the
continuum limit.

In order to see the explicit form of the  $\CL_{4}$ operator close to the continuum limit, 
we expand the action around  $\cUb= \frac{1}{\sqrt 2 a}$. The result is 
\beq 
\CL_4 \sim \frac{1}{a}\left[ \Tr \lambda  (\sum_{m=1}^5 \psi^m)   - \frac{1}{k} \Tr \lambda 
\Tr (\sum_{m=1}^5 \psi^m)  + \right] 
\ldots
\eeq
where ellipsis are dictated by supersymmetry.   The reader will immediately realize that 
$ (\sum_{m=1}^5 \psi^m) =\CE_{5m} \psi^m $ is nothing but the $S_5$  (and twisted 
$SO(4)'$) singlet  identified in \S \ref{sec:irreps}. Indeed, it is the only field that could form a fermion mass term by pairing  with $\lambda$.

As we remarked earlier 
twisted Lorentz invariance $SO(4)' = {\rm Diag}(SO(4)_E \times SO(4)_R)$
emerges  in the continuum without any fine tuning 
due to microscopic symmetries of the lattice action, and gauge symmetry.  
Furthermore, the issue of the restoration
of (untwisted) rotational invariance $SO(4)_E$ , non-abelian R-symmetry invariance 
$SO(6)_R$,  and full supersymmetry can now be
formulated in terms of the
magnitudes of the dimensionless
coefficients $\alpha, \beta, \gamma$. 
For example, the theory with 
$(\alpha, \beta, \gamma) = (1,1,0) $ is the Marcus twist of $\cN=4$ with full supersymmetry.
The classical lattice theory is also defined with these initial conditions. However, it is  currently not  known whether the lattice theory flows to the desired  target theory or not as the lattice spacing is sent to
zero.
A one loop calculation of $\alpha, \beta, \gamma$ has yet to be
done but clearly is of the utmost interest in this regard.
The theories for which $\gamma=0$ also enjoy a global $U(1)_R$  symmetry,   
the so-called ghost number symmetry in the topological field theory literature. 
This $U(1)_R$ is the $SO(2)$ subgroup  of the $R$-symmetry prior to twisting 
$SO(6)_R \supset SO(4)_R \times SO(2)_R $.  The charges under $U(1)_R$  are 
given in \Eq{twistM} and \Eq{bosons}, and apparently, the $\CL_4$ operator explicitly violates it.  The physical reason for the  appearance of this mass  operator in the continuum is then the absence of 
a continuous global chiral symmetry in $A_4^*$ lattice formulation. In this sense, this is similar to the appearance of a Wilson mass term, in the continuum limit of a lattice theory 
without exact $U(1)$ chiral symmetry.   

Finally, the class of theories for which $(\alpha, \beta, \gamma)\ne(1,1,0)$ 
correspond 
$\CN=1/4$ deformations of $\CN=4$ SYM theory and their physical interpretation is currently not known.  

\section{General aspects of supersymmetric lattices}
\subsection{Supersymmetric lattices and topological field theory}

While having a non-perturbative definition of a supersymmetric gauge theory is important 
in its own right, it is also expected that lattice supersymmetry may 
lead to new insights and 
understanding in supersymmetric gauge dynamics. It may also offer a new
non-perturbative window into general problems
in quantum gravity and string theory via the AdS/CFT correspondence. 

In the previous sections of this Report we have seen that 
supersymmetric lattices always lead to twisted
supersymmetric theories in the continuum limit.
These twisted theories are not topological,
however, if desired, one can make them topological by declaring the scalar supercharge $Q$ to be a true
BRST operator. In this case the space of
physical states of the theory is
truncated to include only those
$|\Omega\rangle$ annihilated by $Q$ i.e
$Q|\Omega\rangle=0$, modulo those which can be written as  
$|\Omega'\rangle \sim Q|\Omega''\rangle $.
In this sense, there is an intimate connection between topological 
field theories and supersymmetric lattices. 
The utility of topological field theory in the derivation of certain exact dualities of 
$\CN=4$ SYM 
within  the restricted Hilbert spaces of the associated
topological  theory,
as well as in the theory of 4-manifolds is well known \citep{Witten:1988ze, Birmingham:1991ty, Vafa:1994tf, 
Kapustin:2006pk}.  One of our hopes is that the lattice construction of the supersymmetric theories will shed light into the dynamics and 
dualities in these gauge theories  beyond their topological subsectors.

\begin{figure}[t]
\centerline{\resizebox{10.0cm}{!}{%
 \includegraphics{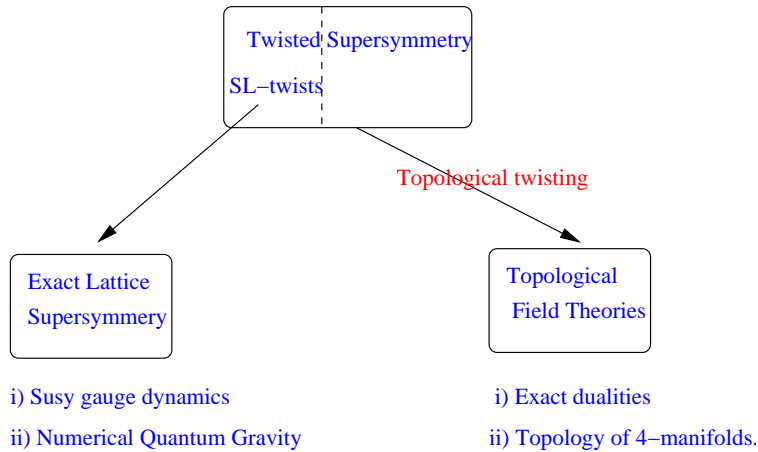} }}
\caption{Classification for twisted supersymmetry, and interrelations between general twists, supersymmetric lattice (SL) twists and topological twists. SL-twists only form a special subset of 
all possible twists. 
Few selected applications to physics and mathematics are also shown.  }
\label{fig:twistclass}
\end{figure}

%

A common thread in both topological field theory and lattice supersymmetry is the existence of a nilpotent scalar supercharge $Q$. However, although all supersymmetric lattices are associated with  
topological  field theories, the converse statement is not true. Given a supersymmetric twist with a scalar supercharge, we are not guaranteed to have a working  supersymmetric  lattice formulation.  Below, we  examine 
this in connection with  the twists of the $\CN=4$ SYM theory in $d=4$.

\subsubsection{Three   twists of $\CN=4$ SYM in $d=4$} 
\label{sec:threetwists}

The  $\CN = 4 $ theory on ${\mathbb R}^4$  has three inequivalent twists -- to 
be described below --  all  of which admit  a nilpotent scalar 
supercharge, with $Q^2\; \cdot=0$. 
All these twists are consistent with not having infinitesimal 
translation generators $P_{\mu}$ on the lattice, and in this sense, provide a 
solution to the problems quoted in \S \ref{sec:3e}.   
However, only one of 
these twists arises naturally in the context of
supersymmetric orbifold lattices, and 
has a natural mapping into a lattice theory
within the twisted/geometric  approach. 
In this section
we will explain what distinguishes these three twists and why
only one admits a supersymmetric lattice construction. This will 
also  shed light on the question
of why the $\CN=2$ theory in $d=4$, which also 
admits twisting and a nilpotent scalar supercharge, cannot be latticized 
in any simple way by the techniques described in this work.   

Recall that the $\CN = 4 $ theory on ${\mathbb R}^4$ can be obtained 
as the dimensional 
reduction of the 
$\CN=1 $ gauge theory on ${\mathbb R}^{10}$ down to ${\mathbb R}^4$. The ten 
 dimensional
 theory possesses an 
$SO(10)$ Euclidean Lorentz rotation group. Upon reduction, 
 the $SO(10)$ group decomposes into 
 \beq
SO(10) \longrightarrow 
\left( \begin{array}{c|c}
     SO(4) & \\
\hline     
 & SO(6)
       \end{array} \right)
\eeq
where $SO(4) \sim SU(2)_L \times SU(2)_R$  is the four dimensional 
Lorentz symmetry action 
on ${\mathbb R}^4$ and $SO(6)_{\cal R} \sim SU(4)_{\cal R}$ is the internal
 ${\cal R}$-symmetry group. 
 The ${\bf 16}$ dimensional positive chirality spinor of $SO(10)$  decomposes as 
\begin{equation}
  Q_{\alpha, I} \oplus  \mybar Q_{\dot \alpha, I} \sim    ({\bf 2,1,4})  
 \oplus  ({\bf 1,2,\bar 4})  \in 
   SU(2)_L \times SU(2)_R \times SU(4)_{\cal R}
   \eqn{super}
  \end{equation}
The twisting procedure corresponds to
a choice of a non-trivial  $[SU(2) \times SU(2)]'$ 
embedding into  $SU(2) \times  SU(2) \times SU(4)_{\cal R}$.

The  
$\CN=4$ SYM theory in $d=4$  has three inequivalent twists, 
i.e, three inequivalent 
embeddings of an $SU(2) \times SU(2) $ into $SU(4)_{\cal R}$ symmetry, 
 each of which results in  one or two scalar supersymmetries for which 
\begin{equation}
Q^2 \; \cdot  = 0  \qquad ({\rm up \;  to\;  gauge \; rotations})
\end{equation}
These twists were first discussed  in \citep{Vafa:1994tf, Marcus} in the 
context of topological  $\CN=4$ SYM theory.

However,
only  a subclass  of these twisted theories 
may be defined on a lattice  consistently \citep{Unsal:2008kx}. 
We may refer to this class as {\it supersymmetric lattice twists} or 
{\it SL-twists} for short.  For example, Marcus's twist is in 
SL-twist 
category, but not the other two.
While the  existence of a  nilpotent scalar supersymmetry  $Q^2 =0$  is  sufficient to formulate 
a  topologically twisted version of a supersymmetric gauge theory on 
curved  space,  it is not sufficient   to allow 
a lattice construction due to the other strictures of the latter.

 The three independent twists of $\CN=4$ SYM  
 are most easily described  by providing  the decomposition of the
 ${\bf 4}$ of 
$SU(4)$ in \Eq{super} under an $SU(2) \times SU(2)$ symmetry
\begin{eqnarray}
&& i) \;\; ({\bf 2,1}) \oplus ({\bf 1, 2}), \qquad  ({\rm SL-twist})    \cr 
&&  ii)\;\; ({\bf 1, 2}) \oplus ({\bf 1, 2})  \cr 
&& iii) \;\;  ({\bf 1, 2})  \oplus 
 ({\bf 1,1}) \oplus ({\bf 1,1}) \; .
 \eqn{embed}
 \end{eqnarray}
Under the twisted rotation group 
\beq
 [SU(2)_L \times SU(2)_R ]'  \times (G_a)  
 \subset  [SU(2)_L \times SU(2)_R ] \times  
 SU(4)_{\cal R}
\eeq
where   $G_a$ $(a=i, ii, iii)$ is the global ${\cal R}$-symmetry of the 
twisted 
theory,  the supercharges (and fermions) transform as 
 \begin{eqnarray}
i) \; \;  {\rm fermions} \; && \rightarrow ({\bf 1,1}) \oplus 
({\bf 2,2})
\oplus [({\bf 3,1}) \oplus ({\bf 1,3})]
\oplus  ({\bf 2,2})  \oplus ({\bf 1,1})   \cr \cr
&&  \rightarrow 
{\bf  1 \oplus  4 \oplus 6 \oplus  4  \oplus 1 }    \qquad  ({\rm SL-twist}) 
 \;  \cr \cr
ii) \; \;  {\rm fermions} \; && \rightarrow 2 \times  \Big[  {\bf (1,1)  \oplus 
(2,2) 
\oplus (3,1) } \Big] \cr \cr
iii) \; \;  {\rm fermions} \; && \rightarrow   \Big[ {\bf (1,1)  \oplus 
(2,2) 
\oplus (3,1)} \Big]  \oplus 2 \times \Big[ ({\bf 2,1}) \oplus ({\bf 1, 2}) \Big]
\end{eqnarray}
Notice that we have dropped the transformation properties under 
$G_a$ which are not important for our  purposes.
The gauge boson, which is a $SU(4)$  singlet,   transforms as 
$(\bf 2, 2)$.   The   scalars are a singlet under the 
Lorentz symmetry and transform in the
 ${\bf 6}={\bf  4 }\wedge {\bf 4} $, anti-symmetric representation  of $SU(4)$. 
 Therefore, \Eq{embed} 
 uniquely fixes  the decomposition  of   the  ${\bf 6}$ under the twisted rotation group, for example,  
\beq
i)\;\;  [   ({\bf 2,1}) \oplus ({\bf 1, 2})  ]  \wedge [   ({\bf 2,1}) \oplus ({\bf 1, 2})  ] 
= ({\bf 2}, {\bf 2}) \oplus 2({\bf 1,1}),   
\eeq
and similarly, 
\begin{eqnarray}
  ii)\;\; 3({\bf 1,1} ) \oplus ({\bf 1, 3})   \qquad 
 iii) \;\; 2 ( {\bf1,  2})  \oplus   2 (1,1)  
 \end{eqnarray}
As we stated above, all three of these 
twists support the existence of  at least one 
nilpotent scalar supercharge $Q \sim  ({\bf 1,1}) $, with $Q^2 =0$, 
modulo gauge rotations. 
 Indeed, the first two twists have
 two such nilpotent charges.
  One would naively expect that, since 
 $Q^2=0$ does not interfere with any translation, it should be
 possible to implement all these twisted theories
on the lattice. 
 This intuition is not completely correct as we shall now argue.
 
First, note that all three twists  have  a copy of the twist of $\CN=2$ SYM 
 theory in 
$d=4$    \citep{Witten:1988ze} where 
eight   supercharges decompose as $   {\bf (1,1)  \oplus  (2,2) 
\oplus (3,1)} 
$. This structure exists in  a $L' \leftrightarrow R'$ symmetric manner in the 
first twist and asymmetric manner
 for the last two.  This means that, in case $i)$, 
instead of 
self-dual two-forms, we can just think of two-forms, without a self-duality 
condition. 
In lattice gauge theory, the implementation of 
the self-duality condition in a manifestly gauge covariant fashion is 
problematic.   
For example, in continuum, we will have 
$Q \psi^{\mu \nu, +}  =  F^{\mu \nu, +} \equiv  
 F^{\mu \nu} + 
\half \epsilon^{\mu \nu \rho \sigma}F_{\rho \sigma} $ where  both of 
$\psi^{\mu \nu, +}$ and $ F^{\mu \nu, +} $ are  in the
self-dual ${\bf (3,1)}$ 
representation \citep{Witten:1988ze}. \footnote{This equation has another 
utility. In the topological field theory context, the fixed points of the 
$\CN=2$ supersymmetric action are described by BPST-instantons. 
A useful complex  generalization of 
the instanton equation in the $\CN=4$ SYM theory 
was obtained by studying  the fixed points of 
$\widetilde Q= Q + * Q^{(4)}$. 
The fixed points of the $\widetilde Q$-action yield 
$ {\cal \mybar F}^{(2)}  +  *{\cal F}^{(2)} =0$, or in components, 
$ {\cal \mybar F}_{\mu \nu}+  
 \half \epsilon_{\mu \nu \rho \sigma} {\cal F}^{\rho \sigma}=0$. This equation 
was  derived first in the context of lattice supersymmetry
 \citep{Unsal:2006qp} and later in the study of dualities in $\CN=4$ SYM
\citep{Kapustin:2006pk}. }
It is not clear how to
implement
the self-dual field strength in a gauge covariant way on the lattice 
and hence, the meaning of the left hand side 
(a self-dual Grassmann) is also unclear. 
This means, a gauge covariant  
implementation of 
the twists $ii)$ and   $iii)$ in a lattice formulation is unlikely. 
Furthermore,  the  $iii)$ case also involves   
double-valued representations of scalars and spinors, 
which are again in double-valued spinor representations of the lattice point 
group symmetry and 
do not have a natural mapping to a  lattice, 
unlike the $p$-form to $p$-cell mapping that has been used in
the constructions described in this Report.

The conclusion of these arguments seems to be that
supersymmetric lattices always correspond to twists which do 
not involve 
any self-dual field-strengths and in which all
the  fields live in  single-valued 
integer spin representations of the twisted rotation group. Furthermore,
the spinors (and supercharges) must decompose into $p$-form integer spins:
\begin{equation}
  Q_{\alpha, I} \oplus  \mybar Q_{\dot \alpha, I} \longrightarrow  Q^{(0)} 
\oplus      
  Q^{(1)}  \oplus  Q^{(2)}  \oplus  Q^{(3)}  \oplus  Q^{(4)} 
\end{equation}
as we in for example the SL-twist of $\cN=4$ YM. Another way of stating this
is that supersymmetric lattice theories must always contain
a sufficient number of fermions to saturate
one or more single \KD fields.

\subsection{Matrix model regularizations} 
\label{matrixm}

In this section we briefly discuss an alternative to
the lattice constructions we have been describing but one which
shares many of the same features - the matrix model regularization 
of  supersymmetric gauge theories.  
This approach is  independent of the orbifolding/deconstruction and 
twisting approaches. The main utility of this approach is that it can be used 
to construct  a manifestly supersymmetric matrix  regularization for 
certain twisted supersymmetric gauge theories formulated on curved 
backgrounds, such as $S^2$ or $S^2 \times \mathbb R$, which are not accessible 
by the techniques described so far.

It is well-known that  global  scalar supersymmetry may be carried  
to curved spaces if a 
twisted version of the supersymmetry algebra is used \citep{Witten:1988ze}.  
On curved space, there are no covariantly constant spinors, hence global 
supersymmetry cannot be achieved in any naive way. 
On the other hand,  covariantly 
constant scalars exist in curved spaces thanks to the twisting 
procedure.  Indeed,  a mass deformation of the type IIB matrix model
provides a  matrix model  regularization 
for a twisted theory on a curved background -- a  two-sphere $S^2$.  
A remarkable 
feature of this construction is that both the regularized theory and 
continuum theory respect the same set of scalar supersymmetries. 
Instead of discussing an example on a curved background  (see \citep{Unsal:2008kx} for such an example),  which necessitates introducing additional notation, 
we will highlight  the main points of the 
matrix model regularization by employing  the already established 
notation of \S.\ref{sec:10}.

\subsubsection{A deformed $\CQ=1$ matrix model  for $\CN=4$ SYM in $d=4$} 
The type IIB matrix model possesses  $\CQ=16$ supersymmetries and a
$SO(10)_R$ 
global R-symmetry and $U(N^2k)$ gauge symmetry.    
We first construct a $\CQ = 1$ supersymmetry and 
$U(1)^5$ global symmetry  preserving deformation of the type IIB matrix model. 
This model  serves as a
nonperturbative regularization for $\CN = 4$ SYM  
theory in four Euclidean dimensions. As opposed to the orbifold projections 
where one starts with $U(N^4k)$ and projects out by $Z_N^4$ to obtain a $U(k)$  
lattice gauge theory on an $N^4$ lattice,  the  deformed $U(N^2k)$  
matrix model is itself a rewriting of a $U(k)$ gauge theory on
$N^4$ lattice, without any projections.  As a consequence, the latter formulation is not precisely local, however, 
this non-locality can be pushed to the 
cut-off scale by a judicious choice of the deformation parameter.

The deformed matrix model action with $\CQ=1$ exact supersymmetry is given by  
\begin{eqnarray}
S^{\rm DMM} &=& \frac{\Tr}{g^2}  \bigg[ \int \;  d \theta \left( 
-\frac{1}{2} {\bf \Lambda} {\partial}_{\theta} {\bf \Lambda }  
- {\bf\Lambda }[ \mybar z_{m} , 
{\bf Z}_{m} ] + 
\frac{ 1}{2}
{\bf \Xi}_{mn }{\bf E}^{mn}
\right) \cr \bigg.
&+& \bigg.
  \frac{\sqrt 2 }{8}  \epsilon^{mnpqr} {\bf \Xi}_{mn} 
( e^{-i( \Phi_{pq} + \Phi_{pr}) /2} \mybar z_{p}  {\bf \Xi}_{qr}  -  e^{+i( \Phi_{pq} + \Phi_{pr}) /2}
{\bf \Xi}_{qr}  \mybar z_{p } )
\bigg] \qquad \qquad
\eqn{matrixN=4}
\end{eqnarray}
where the $\CQ=1$ supersymmetric matrix multiplets are 
    \begin{equation}
\begin{aligned}
& {\bf \Lambda} = \lambda  -i\theta  d  \ ,\\
& {\bfz}^{m} = z^{m}  + \sqrt{2}\,\theta \,
  \psi^{m} , \qquad  \mybar z_{m},  \qquad m=1, \ldots,5 \\ 
 &  {\bf \Xi}_{mn}= \xi_{mn} -  
2\theta\,\, \mybar E_{mn} \, . 
\end{aligned}
\eqn{superfields}
\end{equation}
The  $\mybar z_{m}$ is a supersymmetry  singlet, and hence a multiplet on its 
own right. 
The fermi multiplet  ${\bf \Xi}_{mn}$ is anti-symmetric in its indices.  
 The holomorphic ${\bf E}^{mn}$ functions        are the analogs of the 
derivative of the 
 superpotential $\epsilon^{mnp}\frac{\partial W( {\bf Z})}{\partial \bfZ^p}$  
and   given by
\begin{eqnarray}
&{\bf E}_{mn} ( {\bf Z})&=  e^{-i\Phi_{mn}/2} {\bf Z}^{m} {\bf Z}^{n} - 
 e^{+i\Phi_{mn}/2}
{\bf Z}^{n} 
 {\bf Z}^{m} , \cr \cr
& { \mybar E}_{mn} ( {\mybar z})& =  e^{-i\Phi_{mn}/2} {\mybar  z}_{m} 
{\mybar z}_{n} -  e^{+i\Phi_{mn}/2}
{\mybar  z}_{n} 
 {\mybar  z}_{m}  \; .
 \eqn{Efunc}
\end{eqnarray}
The result \Eq{matrixN=4} is the $\CQ=1$ supersymmetry preserving 
deformed matrix model formulation of the target $\CN=4$ SYM theory. 
For $[\Phi_{mn}]=0$,  it is simply a rewriting of the 
$\CQ=16$ theory in terms of $\CQ=1$ superfields.

We choose the gauge group of the deformed matrix model as 
 $U(N^2 k)$ 
and a convenient choice  of deformation (flux)  matrix     with a local 
  continuum limit is    
\begin{eqnarray}
[\Phi_{mn}]= \left[ \begin{array}{cc|cc|c}
     & +\frac{2 \pi}{N}  & &  &- \frac{2 \pi}{N}  \\
     -\frac{2 \pi}{N}  & & &  &  +\frac{2 \pi}{N}   \\
\hline       &&& +\frac{2 \pi}{N} & -\frac{2 \pi}{N}  \\
      &&     -\frac{2 \pi}{N}  & &     +\frac{2 \pi}{N}  \\     
      \hline  
         +\frac{2 \pi}{N}  &  -  \frac{2 \pi}{N} &   +\frac{2 \pi}{N}  & -
 \frac{2 \pi}{N}     &   
 \end{array} \right]
\eqn{fluxdef}
\end{eqnarray}
With this choice of the flux matrix,  we may use the background solution to 
form a basis for a lattice theory on an $N^4$ lattice. (For details, see 
 \citep{Unsal:2005us}.) 
Splitting the background and fluctuations of the matrix field in
 \Eq{matrixN=4} formally as
\begin{equation} 
U(N^2k) \longrightarrow \underbrace{U(N^2)}_{T^4 \;  \rm background} \otimes \underbrace{U(k)}_{\rm gauge  \; fluctuations}  
\end{equation}
we obtain the  $\CQ=1$
lattice gauge theory action of \Eq{d4lat} {\it except}
with a modified (non-local) $\star$ product of lattice superfields.
The exact $\CQ=1$ supersymmetry of the deformed model is same as the  exact 
lattice supersymmetry of the lattice formulation. 
The $\star$-product  (which is more commonly known as Moyal $\star$-product) is encoded into a kernel $K( {\bf j -n, \; k-n})$  
\begin{eqnarray}
\Psi_{1}({\bfn}) \star  \Psi_{2}({\bfn}) =  &&
\sum_{{\bf j}, {\bf k}}
\Psi_{1}({\bf j}) \;K( {\bf j -n, \; k-n})    \Psi_{2}({\bf k} ) \;\; \cr
\equiv&& 
\sum_{{\bf j}, {\bf k}}
\Psi_{1}({\bf j})        \left(   \frac{1}{N^4}     e^{-\frac{4 \pi i} {N^2 \theta'} \;  ({\bf j}- \bfn) \wedge    ({\bf k}- \bfn) }  \right)
 \Psi_{2}({\bf k} ) \;\;
\eqn{kernel}
\end{eqnarray}
In this formula
$\theta'= 2/ N$ is a dimensionless non-commutativity parameter on the lattice, 
and $\wedge$ is the usual skew-product.

The resulting model corresponds to a 
$U(k)$ lattice gauge theory on a $N^4$ lattice.   
The  hypercubic  lattice and 
$A_4^*$  lattice are special points in its moduli space and were
examined in  \S.\ref{sec:10b.1} and \S.\ref{sec:10b.2}.   Distinct from the discussion in \S.\ref{sec:10b.2}, there 
is now a dimensionful length scale which measures the non-locality of the 
kernel in \Eq{kernel}.  Restoring the  dimensions,  it  is equal to 
\begin{eqnarray}
\Theta = \frac{N^2 a^2 \theta'}{4\pi}
\end{eqnarray} 
The length scale associated with the non-locality of the $\star$-product is, 
\begin{eqnarray}
\ell_{\star} \sim \sqrt \Theta  \sim N a \sqrt{\theta'} \sim \sqrt N a, 
\end{eqnarray}
Compared to the box size, which is $L= N a$, we have
\begin{eqnarray}
\frac{\ell_{\star}}{L}  \sim  \sqrt{\theta'} \sim \frac{1}{\sqrt N} \rightarrow 0 \; .
\end{eqnarray}
This means, in the continuum limit where we take $N \rightarrow \infty$, 
the non-locality   of the matrix model action tends to zero  relative to the 
size of the box.

A few remarks are in order:  The  deformed matrix model  is a natural 
generalization of the  $\beta$-deformed 
$\CN=4$  SYM theory in $d=4$, which is used to deconstruct slightly fuzzy
 theories in  six dimensions \citep{Adams:2001ne}. By tuning $\theta'$ to be 
$O(1)$ in $N$ counting, we may also achieve a non-commutative $\CN=4$ SYM 
theory on $T^4$ or ${\mathbb R}^4$ as in the supersymmetric examples of 
Refs.\citep{Nishimura:2003tf,Unsal:2004cf}. In Refs.
\citep{Nishimura:2003tf,Unsal:2004cf}, such supersymmetric non-commutative 
theories were obtained by using orbifolds with discrete torsion, 
which is just a way of saying that
the orbifold projection matrices used in generating various dimensions 
commute with each other only up to a phase, which substitutes 
for the deformation 
matrix in \Eq{fluxdef}. The lack of a need to orbifold the matrix
model at all, provided the model was suitably deformed,
was  recognized later in  \citep{Unsal:2005us}. 

There has been
also some recent interesting progress in the non-supersymmetric 
version of the deformed matrix model, which is known as the 
twisted Eguchi-Kawai (TEK) model. Along the same lines as above, a $U(N^2k)$ 
TEK  model, at the classical level, produces a slightly non-commutative $U(k)$
Yang-Mills theory on four dimensional $N^4$ lattice. 
Recently, \citep{Azeyanagi:2008bk, Bietenholz:2006cz} showed that, there is an 
quantum mechanical instability in the bosonic  TEK model, and the relation to 
the lattice theory is spoiled. Ref. \citep{Azeyanagi:2008bk} argued that, 
in supersymmetric theories, or supersymmetric theories with softly broken supersymmetry, the analog of the instability that takes place in the pure TEK model 
is cured. Thus,   
the deformed matrix model shown in \Eq{matrixN=4} with appropriate choice 
of flux yields  a non-perturbatively stable $d=4$ 
non-commutative supersymmetric gauge theory according to the criteria of 
Ref. \citep{Azeyanagi:2008bk}.

\section{Lattice Supergravity?}
\label{sec:11}

We have seen that it is
possible to construct globally supersymmetric lattices
and that they have a lot of interesting mathematical structure.  For
example, the series of well prescribed mathematical steps  described
in this review could have been used to discover staggered fermions (if
the methods hadn't come along 30 years too late!).  One might
wonder though whether the power of the analytical approach used
here could
be harnessed to create a lattice for {\it local} supersymmetry, known
as supergravity.  It would be pretty nifty if we could construct a
lattice theory for quantum gravity by walking down a straight and
narrow algebraic path  without having to worry
about the meaning of geometry and spacetime!  In this section we briefly outline such an attempt which was {\it not} successful, in hope that it might inspire the reader towards something better \footnote{The material in 
this section  is unpublished work by D.B. Kaplan and  Michael Endres.}.

Consider $(2,2)$ supergravity in $d=2$ dimensions. It's action is
derived from $\CN=1$ supergravity in $d=4$ dimensions by erasing two
spacetime dimensions.   The particle content of the theory is a
graviton and  a spin
$\textstyle{\frac{3}{2}}$ gravitino; the action for the graviton is
the usual  Hilbert action, and for the gravitino, the Rarita-Schwinger
action.  The theory also has
lots of auxiliary fields required to make the theory manifestly
supersymmetric off-shell.   The idea we will follow will be to invent
``staggered'' gravitinos on the lattice. We  will then introduce staggered
vierbeins, and try to realize one exact supercharge on the lattice,
and then hope that the action has enough Lorentz symmetry and
supersymmetry to have the desired continuum limit.

\subsection{Staggered gravitinos}
\label{sec:11a}
Consider spin $3/2$ Majorana fermions in four dimensions. These are
self-conjugate Dirac spinors $\psi_m$ where $m$ is a   4-vector
index.  The
Rarita-Schwinger action is given by
\beq
\epsilon_{mnpq} \psi_m^T C \gamma_n\gamma_5\partial_p\psi_q\ .
\eeq
This possesses a gauge symmetry $\psi_m \to \psi_m + \partial_m\chi$,
where $\chi$ is an arbitrary Dirac  spinor.  Following the derivation
for staggered fermions, we construct a naive latticization of this
action:
\beq
\frac{1}{2a}\epsilon_{mnpq} \psi_m^T(\bfn) C
\gamma_n\gamma_5\left[\psi_q(\bfn+\hat p)-\psi_q(\bfn-\hat p)\right]\ .
\eeq
This lattice action also possesses a gauge symmetry, $\psi_m(\bfn) \to
\psi_m(\bfn) + (\chi(\bfn+\hat m) - \chi(\bfn-\hat m))/(2 a)$.  We now
substitute
\beq
\psi_m(\bfn) = \gamma_m\left(\gamma_1^{n_1}\cdots\gamma_4^{n_4}\right)\lambda (\bfn)
\eeq
which is easily shown to eliminate the Dirac structure in the action,
leaving us with four identical copies of the action for each spinor component
of $\lambda_m$. We can therefore choose $\lambda_m$ to be a
one-component fermion (with a four-vector index). The lattice then has
one of these four-vector fermions at each site and a simple action
involving lattice derivatives with signs that encode the spin $3/2$ structure.

In General Relativity the vector index on the gravitino lives in
curved spacetime, while the spinor index lives in the tangent space;
the way the two talk to each other is through the vierbein $e_m^a$,
where $m$ is a curved space index and $a$ is a tangent space index;
the vierbein is related to the metric by $e_{am} e^{a}_{n} = g_{mn}$ and
to Lorentz symmetry by $e_{am} e^m_b = \eta_{ab}$, where $\eta$ is the
usual flat (Minkowski or Euclidean) space metric.
The ease with which one can construct staggered spin $3/2$ fermions is
encouraging, but the fact that the curved space index does not play
any structural role on the lattice is disturbing, even though the
action couples the curved space index to the index of lattice
derivative operators.

\begin{figure}[t]
\centerline{\resizebox{8.0cm}{!}{%
 \includegraphics{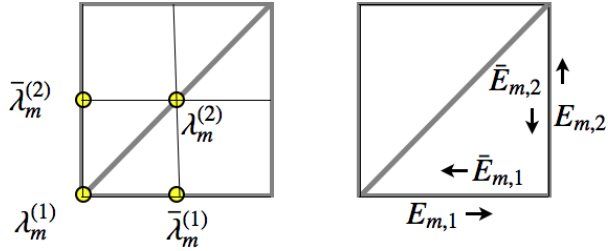} }}
\caption{Lattice assignments for the gravitino and vierbein for
  $(2,2)$ supergravity;  $m=1,2$ is a curved space index. }
\label{fig:vierbein}
\end{figure}

%

\begin{figure}[t]
\centerline{\resizebox{7.0cm}{!}{%
 \includegraphics{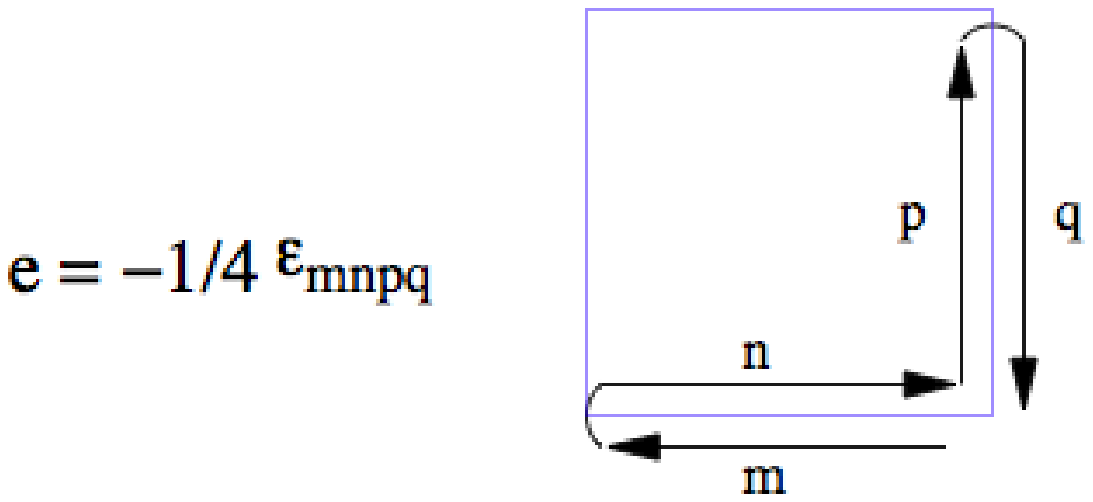} }}
\caption{A picture of the lattice operator equal to $e=\det e_m^a$
  in the continuum: A directed product of the $E$ fields defined of
  Fig.~\ref{fig:vierbein}, where the letters represent the curved
  space  indices of
  the $E$ variable, which are contracted by the $\epsilon$ tensor. }
\label{fig:e}
\end{figure}

Ignoring gathering confusion, one can try to construct a
lattice theory for $(2,2)$ supergravity in $d=2$.  The gravitino is
readily latticized following the staggering procedure, and the lattice
assignments are shown in Fig.~\ref{fig:vierbein}.  Pushing on, one can
latticize the  gravitino's supersymmetric partner, the vierbein.  Using
the structure of our $(2,2)$ lattice construction with matter fields
\citep{Endres:2006ic} as a guide, as well as the supersymmetry
transformations between vierbeins and gravitinos in $(2,2)$
supergravity, one can define
\beq
e_m^a \sigma_{a\alpha\dot\beta} \equiv \begin{pmatrix} E_{m,1} &
  E_{m,2}\cr -\bar E_{m,2} & \bar E_{m,1}\end{pmatrix}
\eeq
and assign the $E$ fields lattice positions shown in
Fig.~\ref{fig:vierbein}.  A heartening result is that various objects
needed in the supergravity action, such as $e\equiv\det e_m^a$ and $
(e_m^a)^{-1}$ are easily
constructed as local lattice operators.  For example, the determinant
$e$ is represented as a ``staple'' as shown in Fig.~\ref{fig:e}.

%

Nevertheless, it seems difficult to understand how to
formulate the lattice covariant derivative in this theory. 
At this time it is an
open and compelling question: can lattice supersymmetry  give us new
insights into lattice supergravity, and therefore about quantum
gravity in general ?

\section{Conclusions, prospects and open problems}
\label{sec:12}

In this report  we have discussed some of the problems
facing efforts to discretize supersymmetric theories.
In general one faces fine tuning problems when one tries to
do this as the classical symmetry is generally entirely broken
under discretization. However, we stress that in dimensions  
less than four this is not necessarily disastrous -- such theories
possess only a finite set of U.V divergent diagrams which occur
at low orders in perturbation theory. In principle such diagrams
can be calculated in lattice perturbation theory
and appropriate counterterms constructed, which when
added to the lattice action, ensure that the resulting theory
flows automatically to the supersymmetric fixed point
in the continuum limit. 
In general non-supersymmetric discretizations may offer computational
benefits such as positive definite determinants over lattice
actions with exact supersymmetry.

That said, we have spent the bulk of this review discussing
new ideas on how to put supersymmetric theories on the
lattice in a way which guarantees a subset of the
full supersymmetry is preserved at non-zero lattice
spacing \footnote{Approaches have also been pioneered based on twisting which claim to 
preserve
{\it all} supercharges on the lattice -- 
see \citep{D'Adda:2007ee,D'Adda:2007ax,Nagata:2008xk,Nagata:2008zz,Nagata:2007mz,Arianos:2008ai}.
These approaches have been examined in \citep{Bruckmann_nc,Bruckmann_crit, Unsal:2008kx}. 
Large  discrete chiral and space-time  symmetries  of these lattice theories  
is emphasized in  \citep{Unsal:2008kx}.}.
The approach only works for theories with a number of supercharges
which is an integer multiple of $2^d$ if $d$ is the dimension of 
(Euclidean) spacetime. This includes quantum mechanics, the
two dimensional Wess-Zumino model, sigma models and a large class  of SYM
theories,   including the important case of ${\cal N}=4$ SYM in
four dimensions.

Two constructions have been described; direct discretization
of a twisted form of the theory and a construction based on
orbifolding a matrix theory. The former technique can be used for theories
both with and without gauge symmetry, the latter is a powerful technique for deriving 
lattices for supersymmetric Yang-Mills theories. Remarkably the two approaches can be explicitly
connected in the case of gauge theories and in that case have been shown to
be
precisely equivalent \citep{Catterall:2007kn,Damgaard:2008pa, Unsal:2006qp}.  
In general the fermions
and supercharges 
of these theories can be embedded into one or more \KD fields containing
integer spin fields. The mapping of these fields onto the lattice is then
very natural. The scalar components of these \KD fields map to site fields
and correspond
to supersymmetries that can be preserved on the lattice. Furthermore, it has been known
for a long time that \KD fields can be mapped into staggered fermion
fields at the level of free field theory which is one way of seeing that these
lattice supersymmetric models do not exhibit fermion doubling.

In four
dimensions there is a unique theory that can be treated this way --
${\cal N}=4$
SYM. The resulting lattice action, derived either from orbifolding or
twisting, is invariant under both lattice gauge transformations and a single scalar
supersymmetry and is free of fermion doubles. Understanding the renormalization structure of this
lattice theory is a pressing issue since it governs whether the lattice theory requires
additional fine tuning in order for it to yield the correct continuum limit.
One and two loop calculations of the counterterms in this model are
crucial in thsi respect and await the interested researcher.

All these approaches potentially
suffer from a complex fermion effective action and it is an open question how
well current Monte Carlo algorithms can handle the resulting system. Initial results, particularly
for thermal systems are quite encouraging but much more work needs to be done 
\citep{Catterall:2008dv}.

The lattices described in this report represent only a small
fraction of the continuum supersymmetric theories one would like to
study, and it would be interesting to see if somehow the techniques
could be extended to include, for example, supersymmetric QCD in four
dimensions. The extensions to systems with
fermions in the fundamental representation are very
interesting in this regard \citep{Matsuura:2008cf,Sugino:2008yp}.
Since numbers of quark  flavors other than four
cannot be represented by staggered fermions, it would also be interesting
to see if one could somehow implement domain wall fermions in lattice
supersymmetry and escape the flavor tyranny of
staggered/Dirac-K\"ahler fermions.
 
Lattice supersymmetry has seen a resurgence of activity in
recent years. After years in
the desert, it is 
delightful to contemplate the intricate structure of the supersymmetric
lattices described here and  how they evade all the challenging  obstacles
outlined earlier. We still have some hope that 
these lattices will not only eventually be useful for numerical
studies of extended SYM theories, but also that
their reach might be extended to  shed light on both 
phenomenologically more realistic supersymmetric theories and perhaps  
some restricted class of lattice supergravity theories.

\section{Acknowledgments}
D.B.K was supported in part by U.S.\ 
Department of Energy Grant  DE-FG02-00ER41132; 
M.\"U.  was supported by the U.S.\ Department of Energy Grant 
DE-AC02-76SF00515 and S.C by U.S.\ Department of
Energy Grant DE-FG02-85ER40237.

The authors would like to acknowledge stimulating conversations
with Alessandro d'Adda,  Andy Cohen, Poul Damgaard,  Joel Giedt, 
Noboru Kawamoto, Ami Katz,  So Matsuura, Erich Poppitz, 
Fumihiko Sugino and Toby Wiseman.

\bibliographystyle{elsarticle-num}
\bibliography{ref}

\end{document}